\documentclass[traditabstract]{aa}

\usepackage{graphicx}
\usepackage[intlimits,sumlimits]{amsmath}
\usepackage{deluxetable}
\usepackage{times,epsfig} 
\usepackage{amssymb}
\usepackage{mathrsfs}  
\usepackage{color}
\usepackage{natbib}
\bibpunct{(}{)}{;}{a}{}{,} 
\usepackage{caption}

\usepackage{amsmath}
\usepackage[english]{babel}
\usepackage{longtable}
\usepackage{url}
\usepackage{hyperref}

\newcommand{\beqa}{\begin{eqnarray}} 
\newcommand{\eeqa}{\end{eqnarray}}

\newcommand{\bsub}{\begin{subequations}}
\newcommand{\esub}{\end{subequations}}
\newcommand{\beal}{\begin{align}}
\newcommand{\ealn}{\end{align}}

\newcommand{\nosne}{34}
\newcommand{\nosnenir}{26}

\authorrunning{Stritzinger et al.}
\titlerunning{Colors and host reddening of SE~SNe.}
\begin{document}

\title{The Carnegie Supernova Project~I:\\ methods to estimate host-galaxy reddening of stripped-envelope supernovae\thanks{Based on observations collected at  Las Campanas Observatory.}}

\author{M.~D. Stritzinger\inst{1}
\and F. Taddia\inst{2}
\and C. R. Burns\inst{3}
\and M.~M. Phillips\inst{4}
\and M. Bersten\inst{5,6,7}
\and C. Contreras\inst{4}
\and G. Folatelli\inst{5}
\and S. Holmbo\inst{1}
\and E.~Y. Hsiao\inst{8}
\and P. Hoeflich\inst{8}
\and G. Leloudas\inst{9}
\and N. Morrell\inst{4}
\and J. Sollerman\inst{2}
\and N.~B. Suntzeff\inst{10}
}

\institute{Department of Physics and Astronomy, Aarhus University, Ny Munkegade 120, DK-8000 Aarhus C, Denmark\\ (\email{max@phys.au.dk})
\and
The Oskar Klein Centre, Department of Astronomy, Stockholm University, AlbaNova, 10691 Stockholm, Sweden
\and
Observatories of the Carnegie Institution for Science, 813 Santa Barbara St., Pasadena, CA 91101, USA
\and
Carnegie Observatories, Las Campanas Observatory, 
  Casilla 601, La Serena, Chile
  \and
  Facultad de Ciencias Astron\'{o}micas y Geof\'{i}sicas, Universidad
Nacional de La Plata, Paseo del Bosque S/N, B1900FWA La Plata, Argentina
\and
Instituto de Astrof\'isica de La Plata (IALP),
  CONICET, Argentina
  \and
  Kavli Institute for the Physics and Mathematics of
the Universe, Todai Institutes for Advanced Study, University of
Tokyo, 5-1-5 Kashiwanoha, Kashiwa, Chiba 277-8583, Japan
\and
Department of Physics, Florida State University, 77 Chieftain Way, Tallahassee, FL, 32306, USA
\and
Department of Particle Physics and Astrophysics, Weizmann Institute of Science, Rehovot 7610001, Israel
\and
George P. and Cynthia Woods Mitchell Institute for Fundamental Physics and Astronomy, Department of Physics and Astronomy, Texas A\&M University, College Station, TX 77843, USA  
}

\date{Received 22 March 2017 / Accepted XX XXXX 2017}

\abstract{We aim to improve upon contemporary methods to estimate host-galaxy reddening of stripped-envelope (SE) supernovae (SNe).
To this end  the {\em Carnegie Supernova Project} (CSP-I)  SE SNe photometry data release, consisting of nearly three dozen objects, is used to identify a minimally reddened sub-sample for each traditionally defined spectroscopic sub-types (i.e, SNe~IIb, SNe~Ib, SNe~Ic). 
Inspection of the optical and near-infrared (NIR)  colors 
   and  color evolution of the minimally reddened sub-samples reveals  a high degree of homogeneity,  particularly  between 0d to $+$20d relative to $B$-band maximum. 
This motivated the construction of intrinsic color-curve templates, which when compared to the colors of reddened SE SNe, yields an entire suite of optical and NIR color excess measurements.
Comparison of  optical/optical vs. optical/NIR color excess measurements indicates the majority of the  CSP-I SE SNe suffer relatively low amounts of reddening (i.e., $E(B-V)_{host} < 0.20$ mag) and
we find evidence for different  $R_{V}^{host}$ values among  different SE SN.
Fitting the color excess measurements of the seven   most  reddened (i.e., $E(B-V)_{host} >  0.20$ mag) objects with the Fitzpatrick (1999) reddening law model  provides robust estimates of  the host visual-extinction $A_{V}^{host}$ and $R_{V}^{host}$.
In the case of the SE SNe with relatively low amounts of reddening, a preferred value of  $R_{V}^{host}$ is adopted for each sub-type, resulting in estimates of  $A_{V}^{host}$ through Fitzpatrick (1999) reddening law model  fits to the observed color excess measurements. 
Our analysis suggests SE SNe  reside in galaxies characterized by a range
of dust properties. We also find evidence  SNe~Ic   are more likely to occur in  regions   characterized by larger   $R_{V}^{host}$ values compared to SNe IIb/Ib and they also tend to suffer more extinction.
These findings are consistent with work in the literature  suggesting SNe~Ic tend to occur in 
regions of on-going star formation.}

\maketitle

\section{Introduction}
\label{sec:intro}

Stripped-envelope (SE) core-collapse (CC) supernovae (SNe) mark the terminal endpoint in the lives of massive stars that have shed their hydrogen (Type IIb/Ib)  and possibly helium (Type Ic) layers over their evolutionary lifetimes. 
The majority of mass loss for single massive stars is 
 due to line-driven winds, while  massive stars   in binary systems the majority of mass loss is  likely due to  mass transfer to  the companion via Roche lobe overflow  and/or through common envelope evolution.
Recent studies of samples of SE SNe \citep{drout11,cano13,taddia15a,lyman16,prentice16} indicate
 Type~IIb/Ib/Ic exhibit  a range of  $^{56}$Ni mass ($M_{Ni}$) and 
 explosion energy ($E_{K}$), while ejecta mass ($M_{ej}$) estimates are found to span a  more narrow range. 
 For example, \citet{lyman16} found from an extended sample of SE SNe that key explosion parameters range as: 
 $<M_{\rm Ni}>~=~$0.14--0.29~M$_\odot$, $<M_{ej}>~=~$2.6--3.9~M$_\odot$ and  $<E_{K}>~=~$2.7--5.2$\times 10^{51}$~erg. Only the more extreme  SNe~Ic BL (broad-line) exhibit significantly higher values of $E_K$ and $M_{Ni}$.   
  
The  low ejecta masses typically inferred for the current samples of SE~SNe, along with the direct progenitor detection for a handful of events (e.g., SN~1993J, SN 2008ax, SN~2011dh, SN~2013df, iPTF13bvn, see \citealp{smartt15} and references therein) and the rates of the various sub-types \citep{smith11rates}, suggest their progenitors are linked to  binary star systems \citep[see][for a review]{yoon15}. 
However, single massive stars may account for the brightest and most energetic SE~SNe, including those associated with long-duration gamma-ray bursts and some superluminous supernovae. To date no progenitor detections exist of  any SN~Ic.

An important requirement to accurately estimate the luminosity and explosion parameters of any type of SN is an estimate of its reddening due to dust along the line-of-sight to the observer.
Assuming the extinction of any given SN can be defined by a dust law that has a wavelength dependence, the level of reddening experienced by two passbands will  differ.
 Reddening  is typically parameterized by the  $E(B-V)$ color excess and the  absorption coefficient, $R_V$.
 The absorption coefficient dictates the total-to-selective absorption and is  defined as $R_V = A_V / E(B-V)$.
 In this parameterization $E(B-V)$ sets the optical depth of the intervening dust and gas  and $R_V$ is dependent on the average dust grain size  \citep{weingartner01}.
In practice large values of $R_V$ lead to shallow extinction curves and are associated with larger than normal size dust grains, while low values of $R_V$ lead to steep extinction curves due to smaller than normal size dust grains.
 
Reddening is typically split between two components. The first component accounts for reddening due to dust within the Milky Way (defined by $R_V^{MW}$ and $E(B-V)_{MW}$) and the second component accounts  for reddening external to the Milky Way (defined by $R_V^{host}$ and  $E(B-V)_{host}$). 
Reddening along any particular line-of-sight in the Milky Way can be approximately corrected for using the reddening maps of \citet{schlafly11} and \citet{schlafly16}.
However,  accounting for reddening  external to our own galaxy  is significantly more challenging.  
In principle light traveling from an extragalactic source may be reddened  by dust located in its  immediate vicinity, dust located in the  host galaxy of the source, or potentially by dust encountered within the intergalactic medium. 
Typically the combined effect of reddening due to dust associated with these locations is lumped together and labeled as $E(B-V)_{host}$. 
In principle, this could lead to potential problems as dust at any of these locations could have  different properties, leading to significantly different values of $R_V$ and hence various levels of reddening.
Indeed even in the MW the dust in the bulge shows  significant variations of $R_V^{MW}$ \citep{nataf16}.

Efforts to devise robust methods to estimate $A_V^{host}~=~R_V^{host} \cdot E(B-V)_{host}$ for SE SNe
have largely  been hampered by a lack of well-observed and homogeneous samples, and as a result, the inability to identify well-defined intrinsic SN colors.
Typically, host extinction values of SE~SN are inferred from empirically-derived relations between the measured equivalent width of \ion{Na}{i}~D  (hereafter $EW_{\ion{Na}{i}~D}$) and $E(B-V)_{host}$, and assuming a typical Galactic reddening law characterised by $R_V = 3.1$.  
These relations often rely on calibrations determined from observations within the Milky Way \citep[e.g.,][]{munari97,poznanski12} and their use to determine reddening in the hosts of extragalactic SNe is suspect as dust properties are known to vary among galaxies \citep{poznanski11}.
Indeed, these relations are associated with significant scatter, providing extinction values  with uncertainties of  $\sim$68\% of  the inferred value of  $A_V$ \citep{phillips13}. 
 When combined with extinction estimates computed from Balmer decrement 
measurements of the immediate environment of a SN \citep[cf.][]{xiao12} and/or  other less direct indicators, such as the position within the respective host galaxy and/or the SN peak colors, one can arrive at an estimate of $A_V^{host}$, but with significant uncertainty.

\citet{drout11} constructed an  intrinsic $V-R$ color curve template from a small sample of SE SNe. 
With this template  they were then  able to define an intrinsic $V-R$ color at $+$10d relative to $V$-band maximum with a scatter of only a few tenths of a magnitude.
This intrinsic color allows for the inference of the $E(V-R)_{host}$ color excess for reddened objects, and when adopting a reddening law, the host extinction.
Unfortunately the photometry of the objects in the \citeauthor{drout11}  sample was computed from images obtained with various telescopes, which is known to lead to considerable calibration issues \citep[e.g.][]{stritzinger02}. 
They also adopted  host extinction values for some of the literature-based objects that were  estimated from different \ion{Na}{i}~D vs. $E(B-V)_{host}$ relations. 
A similar approach  was employed by \citet{taddia15a} to estimate the host-galaxy reddening for a sample of SE SNe, but in this case  a  $g-r$  intrinsic color template was constructed from photometry of a number of objects  from the literature. 

Ideally, one should identify intrinsic SE SN colors from photometry obtained on a single, stable and well-understood photometric system, in addition to expanding the analysis to include various color combinations extending over a range of wavelength in order to facilitate a robust  estimate of both $A_V^{host}$ {\em and} $R_V^{host}$.
    
In this paper we build upon the work of \citeauthor{drout11} and in doing so present methods to  infer  $E(B-V)_{host}$, $A_V^{host}$, and in some instances, a preferred value of $R_V^{host}$ for SE SNe.
To realize these aims we make use of high-quality, multi-color light curves of over thirty SE SNe  observed by the first phase of the {\em Carnegie Supernova Project} (CSP-I; \citealt{hamuy06}).
The full CSP-I sample of SE SNe is presented in a companion paper \citep{stritzinger17}\footnote{Published CSP-I photometry is readily available in electronic format at our Pasadena-based webpage: \href{http://csp.obs.carnegiescience.edu/}{http://csp.obs.carnegiescience.edu/}}, and consists of optical ($ugriBV$) photometry  of \nosne\ objects, with a subset of \nosnenir\ of these  having at least some near-infrared (NIR, i.e. $YJH$) photometry. 
The CSP-I sample is ideal for the present purpose as it has been obtained on a well-understood photometric system \citep{stritzinger11,krisciunas17}, and consists of photometry measured from high signal-to-noise images typically taken under excellent observing conditions. 
Furthermore, the CSP-I has put forth a considerable effort to define accurate 
local sequences of stars, calibrated relative to standard star fields observed over multiple photometric nights.
Combined with host-galaxy template-image subtraction of each science image, our SN magnitudes have been measured differentially to the local sequence, yielding  photometry with typical  (statistical and systematics) uncertainties ranging between 0.01 to 0.03~mag. 
 Analysis of the full the CSP-I photometric light curves and spectroscopy are  presented in additional companion papers by \citet{taddia17} and \citet{holmbo17}, respectively. 
 From these works we adopt  for each object (i) estimates  of the epoch of maximum  for each filtered light curve, (ii)  $K$-corrections, and (iii) the spectral classification  of each object. 
 
In Sect.~\ref{sec:color} we examine the optical and NIR colors of the CSP-I SE SNe sample. 
This is followed by  Sect.~\ref{sec:intrinsic} where we define intrinsic color-curve templates for the different spectroscopic SE SN sub-types, i.e., Type~IIb, Type~Ib, and Type~Ic. 
In Sect.~\ref{sec:extinction} the intrinsic color templates are compared to the observed colors of reddened SE SNe to  compute optical and NIR color excesses. These inferred color excesses are then used to derive  host-extinction  $A_V^{host}$ for all reddened SE SNe in our sample and to constrain $R_V^{host}$ for the most reddened objects. 
Finally in Sect.~\ref{sec:discussion}  the results obtained from various  techniques and color-excess combinations are compared and  discussed. 

\section{Color curves of CSP stripped-envelope supernovae}
\label{sec:color}

Shown in each of the sub-panels of Fig.~\ref{fig:colors} are  various optical and optical/NIR colors plotted vs. time relative to maximum light for the CSP-I SE SN sample. 
All colors have been corrected for Milky Way extinction
 using values listed in NED that originate from the \citet{schlafly11} re-calibration of the \citet{schlegel98} extinction maps of the Milky Way, assuming a \citet[][hereafter F99]{fitzpatrick99} reddening law characterized by $R_V^{MW} = 3.1$. 
The plotted SN colors also include time-dilation corrections and $K$ corrections (see \citealp{taddia17}). 
To facilitate the construction of optical/NIR colors the NIR light curves were first interpolated as described by \citet{taddia17} and \citet{stritzinger17},  and then evaluated on  the epochs that optical observations were obtained. 

The overall shape of the color evolution for each color combination is quite similar. 
In the epochs preceding maximum brightness the colors of SE SNe  are typically at their bluest values \citep[e.g.,][]{stritzinger02, galyam04}.
As the SN evolve through maximum light and the temperature of their photosphere drops, the colors evolve towards the red almost linearly. 
The increase in colors observed in  Fig.~\ref{fig:colors}  during this time can exceed over two magnitudes for nearly all color combinations except for the $V-i$ color curves which exhibit  a color change of  no more than 1.5 mag.
Upon reaching their maximum red color between two to three weeks past maximum, the color curves make an abrupt turn and evolve  back towards the blue, following a linear evolution with a shallow slope out to later times.

Turning to NIR colors, shown in Fig.~\ref{fig:NIRcolors} (from top to bottom) are the $Y-J$, $Y-H$, and $J-H$ color curves for the  CSP-I SE SN sample. 
Inspection of  the  $Y-J$ and $Y-H$ colors reveals an evolution similar to the optical and optical/NIR colors shown in Fig.~\ref{fig:colors}, albeit with more complex behaviour.
At early phases the colors are blue, and as the SN evolve over the next fortnight their colors move towards  the red. This is  followed by a turnover to the blue that is significantly more prevalent than in the case of the optical and optical/NIR color curves. 
The $J-H$ color curves, on the other hand,  show blue SN colors at early times, which then evolve consistently towards the red over the entire duration of our photometric coverage.

To obtain a continuous representation of the color evolution the optical and optical/NIR color curves of each SN are fit with an analytic function.
The color curves shown in Fig.~\ref{fig:colors} largely mimic the color evolution of normal SNe~Ia. 
We therefore opted to fit each color curve with the same analytical function employed  by \citet{burns14} to study the color properties of the CSP-I sample of SNe~Ia. 
The \citet{burns14} analytical function is nearly linear with positive slope at early times, subsequently reaches a maximum value, and then makes a transitions back to a linear function with negative slope.
The functional form is as follows:
\begin{eqnarray}
y(t) = &\frac{\left(s_{0}+s_{1}\right)}{2}t+ \frac{\tau}{2}\left(s_{1}-s_{0}\right)\ln\left[\cosh\left(\frac{t-t_{1}}{\tau}\right)\right]
   \nonumber \\
   & +c+f(t,t_{0}),
   \label{eq:gifit}
\end{eqnarray}

\noindent where $s_0$ is the initial slope, $s_1$ is the final slope, $\tau$ is the length scale over which the transition occurs, and $c$ sets the overall normalization of the function.  
The location of the color maximum is defined as $t_{max}=t_1+\tau~{\rm tanh}^{-1}((s_0+s_1)/(s_0-s_1))$.
In order to capture the color evolution at the earliest epochs,
 a second-order polynomial, $f(t,t_0)$,  for $t < t_0$ and equal to 0 for $t  > t_0$,  completes Eq.~\ref{eq:gifit}. 
 An example fit of Eq.~\ref{eq:gifit} to the $B-V$ color curve of the Type~IIb SN~2006T is shown in Fig.~\ref{fig:colorfit}. 
 The function provides an objective representation of the color evolution, and the  various fit parameters are well-determined for all optical and optical/NIR color combinations.    
  
For some objects the photometric coverage is poor and does not cover the pre-peak phase or does not extend  to the epoch(s) when maximum color is reached.
In these cases the polynomial describing the early color evolution is removed from Eq.~\ref{eq:gifit} before the observed colors are fit. 
 In some other instances the color coverage only extends over  several  days, and when so,  the colors are fit with a low-order polynomial.
    
\section{Intrinsic color-curve templates and temperature constraints from hydrodynamical explosion models}
\label{sec:intrinsic}
\subsection{Color-curve templates}

The construction of intrinsic color-curve templates requires the identification of a  sub-set of objects believed to suffer minimal to no reddening.
To identify such objects we implemented selection criteria based on several factors.
The main two criteria in identifying a minimally reddened SE SN is little to no \ion{Na}{i}~D absorption in  the optical spectrum  and relatively blue optical colors. 
If an object fulfilled these two criteria we then also sought confirmation by considering the location of the SN in its host galaxy and the host's inclination. 
Objects that fulfilled the first two criteria and were  located in the outskirts of their hosts and/or in minimally inclined galaxies are considered to suffer minimal to no reddening. 

In principle, $EW_{\ion{Na}{i}~D}$  can  provide an indication of the host-galaxy color excess \citep[e.g.,][]{munari97,poznanski12}, although the relations used are associated  with large scatter \citep{phillips13}.
With this caveat in mind, plotted in Fig.~\ref{fig:NaID_bmv} (left panel) is $EW_{\ion{Na}{i}~D}$ measured from near-maximum light  spectra for each member of our sample \citep{holmbo17} vs. the $B-V$ color at $+$10d relative to $V_{\rm max}$ (hereafter $B-V_{+10d}$). 
We report the  $EW_{\ion{Na}{i}~D}$ values in Table~\ref{tab:NaID} and note no time variations in the measured $EW_{\ion{Na}{i}~D}$ values are found for  objects with multiple epochs of spectra available.   
For objects with minimal to no $EW_{\ion{Na}{i}~D}$ a 3$\sigma$ upper limit was computed based on the noise of the spectrum in the region of the \ion{Na}{i}~D line. These limits are plotted as arrows in Fig.~\ref{fig:NaID_bmv}. 
The two quantities exhibit a linear trend accompanied with  scatter.
 The best fit linear function between the $EW_{\ion{Na}{i}~D}$ and  $B-V_{+10d}$ (including the upper limits) is 
 given by 
 \begin{equation} 
 EW_{\ion{Na}{i}~D} = 1.568(\pm0.119)\cdot(B-V_{+10d}) - 0.709(\pm0.220).
 \end{equation}

The fact that those SE SNe with minimal to no $EW_{\ion{Na}{i}~D}$ are also the bluest objects within our sample is encouraging in the perspective of using SE SN colors to estimate host extinction. 
The  small region in the lower left corner of Fig.~\ref{fig:NaID_bmv} (left panel) defined by  two dashed lines ($B-V_{+10d}~<$~1.05~mag,  $EW_{\ion{Na}{i}~D}~<$~0.4~\AA) contains the objects characterized by both blue color and  minimal to no $EW_{\ion{Na}{i}~D}$.
These objects can be used to construct intrinsic color-curve templates for each of the SE SN  sub-types.  
Specifically, contained within the region are three objects for each spectroscopic sub-type and they correspond to
the Type~IIb SN~2005Q, SN~2008aq, and SN~2009Z; the
Type~Ib SN~2004ew (this SN is actually not observed at +10d but its color at epochs earlier than +20d are as blue as those of the other two unreddened SNe~Ib), SN~2007Y, and SN~2007kj; and the
Type~Ic SN~2004fe, SN~2005em, and SN~2008hh. 
These SNe are also marked in the right-panel of 
 Fig.~\ref{fig:NaID_bmv}, where plotted are the de-projected and normalized SN distances from their respective host galaxy centers vs. their respective host galaxy inclinations (cf. \citealp{taddia17}). 
 For eight of the nine objects we have the necessary host-galaxy information  to place them in the right panel of  Fig.~\ref{fig:NaID_bmv} (black circles). 
 Each of these SNe appear relatively far from the center of their host, and none of their respective hosts are significantly inclined.
 This provides an additional measure of confidence these objects  suffer minimal host reddening. 

With a minimally-reddened subsample  identified,  intrinsic color-curve templates for each of three canonical SE SN sub-types was constructed.  
An example is provided in Figure~\ref{fig:templates}  where the $B-V$ color curves of each of the  minimally-reddened Type~IIb, Type~Ib and Type~Ic SNe (hereafter SNe~IIb, SNe~Ib, SNe~Ic) are plotted, along with the intrinsic $B-V$ color templates derived from fitting simultaneously the colors of each SE SN sub-type with Eq.~\ref{eq:gifit}.
Also shown is a comparison  between each of the three template color curves. 
 The templates are slightly different among the three sub-types, with the SNe~Ic showing redder colors during the 20 days post maximum and bluer colors after the peak of $B-V$. This is most likely due to differences in the spectral features. 

Plotted in Fig.~\ref{fig:spectra_unreddened} are visual-wavelength spectra for 8 of the 9 objects selected as representative of the intrinsic color evolution of the  three sub-types. 
This includes post maximum spectra of the Type~IIb SN~2005Q, SN~2008aq and SN~2009Z, the Type~Ib SN~2007Y and SN~2007kj, and the Type~Ic SN~2004fe, SN~2005em and SN~2008hh.  
The spectra of the various spectral types are  broadly similar though clear differences between line strengths and line ratios are apparent between the different SN sub-types, particular around the location of H$\alpha$ and the \ion{He}{i} features, with the latter feature being absent in  SNe~Ic \citep[see][]{holmbo17}.

We next proceed to construct  intrinsic color-curve templates  extending over various optical and optical/NIR  color combinations for each SE SN sub-type. 
Plotted in  Fig.~\ref{fig:all_templates} are the intrinsic color-curve templates for eight color combinations: $u-V$, $B-V$, $g-V$, $V-r$, $V-i$, $V-Y$, $V-J$, and $V-H$, along with their accompanying 1$\sigma$ uncertainties computed by adding in quadrature the photometry errors. 
Color-curve templates were constructed by fitting  Eq.~\ref{eq:gifit} to the various color combinations of each of the SE SN subtypes.
In general  between 0d and $+$20d the different SE SN sub-types exhibit similar --though not identical-- colors and temporal morphology, while at earlier and later epochs color differences are more pronounced.
The average dispersion of our $V-X$ color templates between 0d and $+$20d is in the range 0.02--0.15~mag for SNe~Ib, 0.01--0.14~mag for SNe~IIb, 0.02--0.09~mag for SNe~Ic. The intrinsic $V-X$ colors at +10~d since $V_{max}$ are given in Table~\ref{tab:colors_10d}.  \citet{drout11} report a $V-R$ color at +10 d since $V_{max}$ of 0.26$\pm$0.06 mag, which is very close to our SN~IIb value for $V-r$ at that phase,  and as well is consistent within the errors with the $V-r$ colors of the SN~Ib and SN~Ic minimally reddened sub-samples at the same phase. 
The intrinsic $V-X$ color templates for each subtype as well as the $B-X$, $g-X$ and $r-X$ templates (see Sect.~\ref{sec:othercolors}). 
\footnote{The intrinsic color-curve templates are available in electronic format on our CSP Pasadena-based webpage: \href{http://csp.obs.carnegiescience.edu/}{http://csp.obs.carnegiescience.edu/}}

\subsection{Minimal post maximum color dispersion}

SE~SNe  potentially arise from multiple progenitors ranging from single to binary star systems and do exhibit a  (modest) range of inferred key explosion parameters.
Nevertheless, the low dispersion in colors around $+$10d
as demonstrated by \citet{drout11}, \citet{taddia15a}, and the present CSP-I sample is observed and an explanation is warranted if we are to 
 define and employ intrinsic colors to estimate host-galaxy reddening. 
For guidance  we turn to  hydrodynamical explosion models.
In \citet{taddia17} a grid of hydrodynamical models are presented \citep[see][for details on the code]{bersten11} that reproduces the bolometric light curves and velocity evolution  of our SE SNe sample. 
The explosion models represent  He-rich progenitor  stars that cover a range of mass, yielding SNe  with $M_{Ni}$ values  ranging between 0.034 and 1.8~$M_{\odot}$,  $M_{ej}$ values ranging  between 1.0 and 6.2~$M_{\odot}$, and explosion energies ranging between $0.6-4\times10^{51}$~erg.
Plotted as diamonds in the left panel of Fig.~\ref{fig:explanation_drout}  are the  $+$10d photospheric temperatures for our series of models  vs.  the ratio between each model's $M_{Ni}$ and $M_{ej}$ values.
The parameters clearly correlate and are well fit  by a  power law as indicated with a dashed black line. 
 Also shown  in Fig.~\ref{fig:explanation_drout} (left panel) is the  range  of the $M_{Ni}$ to $M_{ej}$ ratio inferred from simple analytical model fits to the bolometric light curves of several dozen SE SNe  \citep{lyman16}.
These regions are defined by the median and standard deviation values of the $M_{Ni}$ to $M_{ej}$ ratio for each sub-type  \citep[][see their Table~4]{lyman16}, and when compared to the hydrodynamical modeling, imply for each sub-type a rather narrow range in the $+$10d photospheric temperature (e.g., for SN~Ib we find $0.027<~M_{Ni}/M_{ej}~<0.076$). 

Assuming  SE SNe can be treated as black-body sources, we compute the $+$10d $B-V$ color for each model and plot it versus the corresponding $M_{Ni}$ to $M_{ej}$ ratio in the right panel of Fig.~\ref{fig:explanation_drout}.
The parameters show a trend that is well fit by a power law function. 
Over-plotted as vertical lines is the range of the  $M_{Ni}$ to $M_{ej}$ ratio as inferred by \citet{lyman16}, while the empty dotted points represent the corresponding  mean $B-V$ color value for each sub-type. 
Taking the standard deviation of the $B-V$ colors for each sub-type in the allowed range of the $M_{Ni}$ to $M_{ej}$ ratio, the uncertainty in the intrinsic $B-V$ color is estimated to be $\sigma_{B-V} =  0.09, 0.09, 0.05$ and 0.09 mag for SNe~IIb, Ib, Ic and Ic-BL, respectively. 
In the case of  SNe~Ic-BL, given the dearth of available models, $\sigma_{B-V}$ was taken to be half the difference between the $B-V$ values of the best fit function (black dashed curve) at the two extremes (magenta dashed lines) of the $M_{Ni}$ to $M_{ej}$ ratio.

If we consider the standard deviation of $B-V$ without distinguishing among SN sub-types, we obtain $\sigma_{B-V} =0.19$~mag. This suggests it is prudent to consider different intrinsic colors for each SE SN sub-type. 
Note that the average uncertainty for our three $B-V$ color templates between 0 and $+$20d is 0.02--0.03~mag and that the uncertainty quoted for the $+$10d $V-R$ color by \citet{drout11} is 0.06~mag, which is consistent with our estimates of $\sigma_{B-V}$ for each SE SN sub-type. 
This implies that the intrinsic color-curve templates can provide $E(B-V)_{host}$ color excess estimates with a systematic accuracy of $\leq$0.1 mag. 
Finally, we note that recently \citet{dessart16} has  published an extended set of SE SN models which also exhibit $+$10d $V-R$ colors characterized by minimal dispersion. 

Given the clear correlation between the $M_{Ni}$ to $M_{ej}$ ratio and the $B-V$ color at +10d, we also checked to see if  the minimally-reddened objects identified in this section are  also characterized by a high  $M_{Ni}$ to $M_{ej}$ ratio as compared to the reddened  objects. 
If this were the case the minimally-reddened objects should appear blue due to a higher 
$M_{Ni}$ to $M_{ej}$ ratio and not for being minimally reddened. 
However,  in \citet{taddia17}
the minimally reddening objects  are found to cover the full phase space of the M$_{Ni}$to M$_{ej}$ ratio for each sub-type, suggesting the  color templates are representative of  the  intrinsic colors of  SE SNe.

\section{Host-galaxy color excess, host-galaxy extinction and constraints on $R_V$}
\label{sec:extinction}
With  intrinsic color-curve templates in hand  various color excesses can be inferred. 
In the following section we describe how this is done and how color excesses are used to   estimate 
the host-extinction  $A_V^{host}$, and constrain $R_V^{host}$ for the most heavily reddened objects.

\subsection{Color excess estimates via intrinsic color-curve templates}
\label{sec:colormini}

To compute the color excess for any particular optical or optical/NIR color combination the observed SN color curve (corrected for Galactic reddening)  is compared to its corresponding intrinsic color-curve template. 
As already noted, \citet{drout11} used $+$10d as their fiducial epoch to infer the $E(V-R)_{host}$ color excess. 
We on the other hand have elected to compute color excesses by taking the difference between the observed  and  intrinsic colors between 0d and $+20$d.
This temporal interval was chosen for several reasons.
Firstly, the majority of our light curves have their best time sampling in  the two to three weeks just after maximum. 
Secondly, during this period the different SE SN sub-types follow a similar evolution in most colors, and thirdly, the overall scatter in the various intrinsic colors is minimal (see Fig.~\ref{fig:all_templates}).
To confirm this is an appropriate manner to proceed we experimented with making templates extending   from 0d to $+$10d and from $+$10d to $+$20d.  In short, these two sets of templates give  results consistent with one another and with the results obtained using the 0d to $+$20d templates.

The technique used to infer the various $E(V-X{_\lambda})_{host}$ (where $X{_\lambda}= u$, $B$, $g$, $r$, $i$, $Y$, $J$, $H$)  color excess values is demonstrated in  
Fig.~\ref{fig:get_color_excess_06T}.
Here the observed $V-X{_\lambda}$ color curve of SN~2006T (yellow triangles) is fit with Eq.~\ref{eq:gifit} (black line) and this fit is compared  to the corresponding  SN~IIb  intrinsic-color template (green line). 
Specifically, the difference between the two curves in the interval between 0 and $+$20d (where the line representing the template is plotted more thick in Fig.~\ref{fig:get_color_excess_06T}) represents the  $E(V-X{_\lambda})_{host}$ color excess. 
 $E(V-X{_\lambda})_{host}$ is computed by taking a weighted average of the difference between the intrinsic color template and the observed colors of the reddened SNe. The adopted weights correspond to the uncertainties associated with the measured observed colors and the estimated epoch of maximum light.

Armed with three sets of intrinsic color-curve templates, one for each of the SE SN  sub-types, the various  $E(V-X{_\lambda})_{host}$ color-excess combinations are computed for the entire sample of reddened SE SNe. 
The results are  provided in Table~\ref{tab:EVMX} with accompanying uncertainties computed by adding in quadrature (i) the weighted standard deviation of the difference between the observed and template colors, (ii) the average color uncertainty of the template, (iii) the average uncertainty of the observed color, and (iv) the   uncertainty related to the temporal phase.
Below in Sect.~\ref{sec:othercolors} this approach is extended to estimate color excess values for the entire suite of $B-X{_\lambda}$, $g-X{_\lambda}$ and $r-X{_\lambda}$ color combinations. 
 
 \subsection{Optical-NIR color excesses}
 \label{sec:colorexcesses}
 
Assuming the dust contained within the host-galaxies of our SE SN sample follows a 
F99 reddening law,  a universal value of $R_V^{host}$ can be derived for each of the SE SN spectral sub-types by comparing the color excesses between the optical and NIR bands.
 Plotted in Fig.~\ref{fig:evmx_evmx} are the measured $E(B-V)_{host}$  color excess values  compared with the measured $E(V-X_{\lambda})_{host}$ color excess values (for $X_{\lambda} = i, Y, J, H$) of 18 SE SN. 
Over-plotted as lines are the best-fit linear functions for each of the SE SN sub-types where 
the corresponding  slopes have been converted  to the $R_V^{host}$ values reported in each panel. 
To convert the $E(V-i, V-Y, V-J, V-H)_{host}$ / $E(B-V)_{host}$  slopes to $R_V^{host}$, we adopt the method described by 
\citet{krisciunas06} and \citet{folatelli10}.
This requires the calculation of the appropriate $a_{\lambda}$ and $b_{\lambda}$ coefficients \citep[see][their Appendix B]{folatelli10}, which is done using a $+$10d spectral template constructed for each SE SN  sub-type \citep[see][]{holmbo17}. 
A summary of this work and a list of the resulting coefficients is provided Appendix~\ref{appendixB}. 

The best-fit $R_V^{host}$  value for each SE SN sub-type is found to be strongly dependent on the color-excess combination considered and this is particularly the case of the SNe~Ic.
As the majority of the reddened objects in the sample  suffer relatively moderate amounts of reddening only weak constraints can be placed on $R_V^{host}$.
Nevertheless,  the comparison of optical with NIR color excess measurements in Fig.~\ref{fig:evmx_evmx}  suggests  that the  various SE SN sub-types are  not well represented by a universal $R_V^{host}$ value, although this simple analysis suggests  SNe~Ic favor  
$R_V^{host}$ values  that are on average larger than those inferred form the  SNe~IIb and SNe~Ib.
These results are based on small number statistic and conclusions should be taken with some caution, however this type of analysis demonstrates the potential to understand the relationship between SE SN and dust reddening.

\subsection{Host-extinction $A_V^{host}$  from multi-color fits and preferred  values of $R_V^{host}$}
\label{sec:reddening}

The comparison of optical and NIR color excess values in Sect.~\ref{sec:colorexcesses} highlights the difficulty to fit  for $R_V^{host}$ separately for  objects suffering little-to-moderate amounts of reddening.
In principle accurate estimates of both $A_V^{host}$ and $R_V^{host}$ are best obtained from optical and NIR color excess measurements of SN  suffering appreciable levels of reddening, i.e.,
$E(B-V)_{host}$ $\gtrsim 0.20$ mag.
{Seven objects meet the criteria and have both optical and NIR photometry. These include: SN~2004gt (Ic),  SN~2005aw (Ic), SN~2006T (IIb), SN~2006ep (Ib),  SN~2007C (Ib), SN~2009bb (Ic-BL), and SN~2009dt (Ic).
The $E(V-X_{\lambda})_{host}$ color excess values of each of these SN are fit with  the F99 reddening law with $R_V^{host}$ set as a free parameter.
The best non-linear least-squares  fit for each object is plotted in Fig.~\ref{fig:EVMX_vs_lambda_onlyRV} and the corresponding $R_V^{host}$ and $A_V^{host}$ values  are listed in each sub-panel and in Table~\ref{tab:veryred}. 
 In general, the shape of the color excess  vs. wavelength relations for the most highly reddened SE SN exhibit similar morphology, and this holds irrespective of spectroscopic subtype.  
Figure~\ref{fig:EVMX_vs_lambda_onlyRV} is encouraging and demonstrates the importance of NIR photometry  to estimate the reddening parameter. As of today we are limited by small number statistics and an expanded sample is required to confidently determine the intrinsic properties of SE SNe at NIR wavelengths.

Among the  objects with significant reddening  a range of  $R_V^{host}$ values are obtained. 
This includes  an $R_V^{host} = 2.6^{+1.2}_{-0.5}$ for the Type~Ib SN~2007C,  an $R_V^{host} = 3.3^{+0.4}_{-0.3}$ for the Type~Ic BL SN~2009bb and an $R_V^{host} = 4.3^{+0.4}_{-0.3}$  
for the  normal Type~Ic SN~2009dt.
Interestingly, this  is  consistent with the findings in Sect.~\ref{sec:colorexcesses} suggesting SNe~Ib and SNe~IIb tend  to occur in  low  $R_V^{host}$ hosts compared to SNe~Ic.
The only SN~IIb with sufficient reddening to determine $R_V^{host}$
is SN~2006T, and as indicated in Fig.~\ref{fig:EVMX_vs_lambda_onlyRV}, it is found to exhibit the  low value of $R_V^{host}=1.1^{+0.2}_{-0.2}$. 
 
The $E(B-V)_{host}$ color excess values presented in 
Sect.~\ref{sec:colorexcesses} reveal that the majority of the objects in our sample  suffer relatively low amounts of reddening (i.e., $E(B-V)_{host} < 0.20$ mag).  
For these events it is not possible to obtain reliable estimates of $R_V^{host}$ from our fitting method as the associated error  is exceedingly large. 
This also holds true for  objects  lacking NIR photometry.
For these objects we therefore assume  their 
 host-galaxy dust properties are similar to 
those characterizing the dust of the hosts of the most reddened objects in each of the SE SN sub-type samples.
In doing so we adopt $R_V^{host} = 1.1$ for all SNe~IIb (as derived from SN~2006T), $R_V^{host} = 2.6$ for all SNe~Ib (as derived from SN~2007C), and $R_V^{host} = 4.3$ for all SNe~Ic (as derived from in SN~2009dt). 
With $R_V^{host}$ set for each SE SN sub-type  each of the low reddened  object's  optical and NIR $E(V-X_{\lambda})_{host}$ sequences  are fit with  the F99 reddening law. 

Plotted in Fig.~\ref{fig:EVMX_vs_lambda} are the $E(V-X_{\lambda})_{host}$ color excess values for the SE SNe suffering low extinction (where $X_{\lambda} = u, B, g, R, i, Y, J, H$) vs.  the effective wavelength of passband $X_{\lambda}$.\footnote{The method used to compute an  average effective wavelength for each of the CSP-I passbands is provided in the Appendix~\ref{appendixA}.} 
 Over-plotted as solid lines in each panel are the best-fit  F99 models for the $R_V^{host}$ values associated to each sub-type with the associated  1$\sigma$ uncertain indicated by the dash lines.
The corresponding best-fit values of $A_V^{host}$ are reported in each sub-panel and listed in Table~\ref{tab:EBMV}. 
From close inspection of the various panels one is able to see the  adopted $R_V^{host}$ values generally provide a good fit to the various color excess values plotted as a function of wavelength.

We next compute  the peak absolute magnitude cumulative distribution function (CDF)  for our sample. To do so CDFs are computed with and without the inclusion of host-extinction corrections. Plotted in  Fig.~\ref{fig:reduceddispersion} are the peak luminosity CDFs for each of the nine CSP-I bandpasses, which we fit assuming a kernel distribution to guide the eye (the best fits are shown by curves of the same color). 
Inspection the CDFs  of the blue bands (i.e., $u$, $B$, $g$)  reveals   that host-extinction corrected  CDFs (in blue) are  steeper compared to the uncorrected CDFs (in red),  indicating  their distributions are narrower. 
Following the discussion of  \citet{faran14}, this is an encouraging results as a reduction in the dispersion of peak luminosities of the sample population after application of reddening corrections suggests the validity of the extinction corrections.

We now examine the distribution of inferred extinction values for the different SE SN subtypes. 
Plotted in  Fig.~\ref{fig:AV} (left-hand panel) are the best-fit $A_V^{host}$ cumulative distribution functions (CDFs) based on the fits shown  in
 Fig.~\ref{fig:EVMX_vs_lambda_onlyRV} and Fig.~\ref{fig:EVMX_vs_lambda}.
Comparing the different CDFs suggests  that on average SNe~Ic are  characterized by  larger $A_V^{host}$ values than SNe~IIb and SNe~Ib. 
A Kolmogorv-Smirnov test indicates the significance of these 
differences is larger than 2$\sigma$ for the comparison between SNe~Ic and SNe~IIb (p-value~$=$~0.0013). This is true also when we assume $R_V=3.1$ for all the objects suffering low extinction, rather than different $R_V$ based on their subclass. This is demonstrated in the right-hand panel of Fig.~\ref{fig:AV}, where SNe~Ic and SNe~IIb are different in their $A_V$ distributions, with p-value~$=$~0.0465. 

\subsection{Host-extinction $A_V^{host}$ from Goobar model}
\label{G08}
Up to here we have assumed the color excess values of our SE SN sample are well reproduce by a F99 reddening law. 
However, close inspection of the $E(V-u)_{host}$ color excess values and the best fits shown in Fig.~\ref{fig:EVMX_vs_lambda_onlyRV} and Fig.~\ref{fig:EVMX_vs_lambda} reveal that in a number of cases (e.g., SN~2004gt, SN~2004gv, SN~2005aw, SN~2006ir,  SN~2006lc, and SN~2009dp) the $E(V-u)_{host}$ color excess values are not perfectly reproduced by the F99 law.
This inspired us to fit an alternative reddening law to the observed optical and NIR color excess values. In this instance we adopt the \citet[][hereafter G08]{goobar08} reddening law, which corresponds to a power law taking the functional form:
\begin{equation}
\label{equation:GO08}
 E(V-X)_{host} = A_V^{host} \cdot a(1-(\lambda/\lambda_V)^{-p}).
\end{equation}
\noindent Here $A_V^{host}$ is the host-dust extinction, $\lambda_V$ is the central wavelength of the $V$ filter, and $a$ and $p$ correspond to the power-law fit parameters.
The power-law function described by Eq.~\ref{equation:GO08} was originally crafted to  describe the effects of reddening of SN light produced by multiple scatters off a shell of dust located in the circumstellar environment of  SNe~Ia. 
To break the degeneracy between $A_V^{host}$ and $a$ we adopt $a = 1$  \cite[see][]{amanullah15}. As a caveat we note the inferred $A_V^{host}$ values from the power-law fits cannot be taken explicitly as the extinction in the $V$ band, and as well in such a model where part of the extinction is due to scattering of light off of CSM one can infer a large range of $R_V^{host}$ values. If such scattering is common among the SE SN population would then suggest SE SN subtypes do not favor a preferred $R_V^{host}$ value.

Plotted in  Fig.~\ref{fig:goobar}  are the best fit power-law functions (blue line) to 17 SE SNe where the G08 reddening law is found to  provide a  good match to the $E(V-X_{\lambda}$) color excess values. 
In each of the instances where the $E(V-u)_{host}$ color excess points are poorly fit by a F99, the G08 model provides a superior fits.

We now examine the $p$ values obtained from the best G08 model fits  of the seven  (see Fig.~\ref{fig:EVMX_vs_lambda_onlyRV}) most highly reddening objects.  Of these objects the best-fit $p$ values of SN~2005aw, SN~2006T, SN~2006ep, and SN~2009dt are found to be inconsistent with the expected $p$ values between  $-1.5$ to $-2.5$ \citep{goobar08}, while SN~2004gt, SN~2007C, and SN~2009bb are found to have best-fit $p$ value confidence intervals that overlap with the range between $-1.5$ and $-2.5$ (see Fig.~\ref{fig:goobar}). 
This suggests these later three objects might have CSM dust driving multiple scatterings of light. We note, however, that the best F99 model fits to the color excesses of these objects  also fit quite well (see Fig.~\ref{fig:EVMX_vs_lambda_onlyRV}).

\subsection{Reddening constraints from additional color excess combinations: $E(B-X_{\lambda})_{host}$, $E(g-X_{\lambda})_{host}$ and $E(r-X_{\lambda})_{host}$}
\label{sec:othercolors}
 
 Most of the current SN follow-up programs do not necessarily observe objects in both SDSS and Johnson passbands. Therefore, we want to consider alternatives to the $V-X_{\lambda}$ color combinations to infer $A_V^{host}$ and $R_V^{host}$. This will, for example, allow one to use our intrinsic color templates also for SNe only observed in the SDSS passbands and not in Johnson $B$ and $V$. Specifically,  in the following  the $B-X_{\lambda}$, $g-X_{\lambda}$, and $r-X_{\lambda}$  color combinations are considered. 

Following the procedure used to construct  $V-X_{\lambda}$  intrinsic color-curve templates, intrinsic color-curve templates are constructed for each of the  $B-X_{\lambda}$,  $g-X_{\lambda}$, and $r-X_{\lambda}$ color combinations.
These intrinsic color-curve templates are  used to infer color excess values of reddened SE SNe  relying on the methodology described in Sect.~\ref{sec:reddening}. 
 The results of this exercise are summarized in Table~\ref{tab:veryred} and Table~\ref{tab:EBMV}.
 
In the case of the seven most heavily reddened objects, best-fit $R_V^{host}$  and $A_V^{host}$ values are determined for each color combination. The best-fit results for these objects are listed in Table~\ref{tab:veryred}, as well as  the average $R_V^{host}$ and $A_V^{host}$ values obtained from each color combination. 
The average values are characterized by relatively small standard deviations ranging between 0.1--1.1 and 0.03--0.15 for $R_V^{host}$ and $A_V^{host}$, respectively.
 
Reported in Table \ref{tab:EBMV} are the best-fit $A_V^{host}$ values for each color combination of the 14 SE SN suffering relatively low reddening.
 To obtain the best-fit value a universal $R_V^{host}$ value was assumed for each SE SN sub-type as determined from the most reddened objects (see Sect.~\ref{tab:veryred}). 
In addition, the average of the best-fit $A_V^{host}$ values obtained from each of the four color combinations is also given.
 Comparing the $A_V^{host}$ values obtained from the four different color combinations reveals that for most of the objects the $A_V^{host}$ standard deviation is relatively small. 
 We therefore  conclude   that when applied to different color combinations  the method gives consistent results.

\section{Discussion}
\label{sec:discussion}

The CSP-I  SE SN light-curve sample has enabled us to devise improved methods to quantify host-galaxy dust extinction, and with the addition of our NIR photometry the added ability to constrain $R_V^{host}$.
We now examine the consistency of the results obtained from the methods presented in Section~\ref{sec:colormini} and Section~\ref{sec:reddening}.
Plotted in Fig.~\ref{fig:checkEBMV} are  comparisons between the best-fit $A_V^{host}$ values obtained from fits of the F99 law to each object's set of color excesses (see the last column of Tables~\ref{tab:veryred} and \ref{tab:EBMV}) and the $A_V^{host}$ values inferred from individual 
$E(V-X_{\lambda})_{host}$ color excess measurement and  the F99 law (see Table~\ref{tab:EVMX}).
In doing so these comparisons are made for each of the four color excess combinations:  $E(V-X_{\lambda})_{host}$ (top-left panel), $E(B-X_{\lambda})_{host}$ (top-right panel), $E(g-X_{\lambda})_{host}$ (top-left panel), and $E(r-X_{\lambda})_{host}$ (top-left panel).
Included in each panel's sub-plots is the root-mean-square (rms) difference between the two methods.
In general the  $A_V^{host}$ values obtained by the two methods are in good agreement, as indicated by the fact that the data follow the black lines in the plots.

We find that the the optical and NIR color combinations providing the smallest dispersion among the suite of $E(V-X_{\lambda})_{host}$, $E(B-X_{\lambda})_{host}$, $E(g-X_{\lambda})_{host}$, and $E(r-X_{\lambda})_{host}$ color combinations considered in this work are as follows. 
 In the case of $E(V-X_{\lambda})_{host}$ and time relative to $V_{max}$ the $V-r$ and $V-H$ colors  exhibit minimal dispersion in $A_V^{host}$, while  for  $E(B-X_{\lambda})_{host}$ and time relative to $B_{max}$ the $B-i$ and $B-H$ colors exhibit minimal dispersion.
 Moreover, for $E(g-X_{\lambda})_{host}$ and time relative to $g_{max}$ the $g-r$ and $g-H$ colors exhibit minimal dispersion in $A_V^{host}$, while for $E(r-X_{\lambda})_{host}$ and time relative to $r_{max}$ the  $r-g/V$ and $r-H$ colors exhibit minimal dispersion.
 Discrepancies between the two methods are largely due to different spectral lines affecting the $A_V^{host}$ measurement from a single $V/B/g/r-X_{\lambda}$ color combination more than the global $V/B/g/r-X_{\lambda}$  fit value.

We now compare the $A_V^{host}$ value derived for each object in the sample to  their $EW_{\ion{Na}{i}~D}$ measurement obtained from visual-wavelength spectra \citep{holmbo17}. 
Specifically, $A_V^{host}$ average values given in Table~\ref{tab:veryred} and Table~\ref{tab:EBMV} are plotted in Fig.~\ref{fig:NAID_EBMV} vs. $EW_{\ion{Na}{i}~D}$. 
Over-plotted in the figure  is the relation (solid blue line) found  between these  parameters by \citet{poznanski12}, as well as a linear  fit (dashed line) to the data assuming $EW_{\ion{Na}{i}~D} = 0$ when $A_V^{host}=0$.  The latter of these is given by $A_V^{host}[{\rm mag}]~=~0.78(\pm0.15) \cdot EW_{\ion{Na}{i}~D}[{\rm \AA}]$.
As expected  there is significant scatter between the two quantities \citep[cf.][]{poznanski11}.

Returning to the $A_V^{host}$ CDFs shown in Fig.~\ref{fig:AV}, it is interesting to note that the SNe~Ic consistently exhibit higher extinction values compared to  SNe~IIb and SNe~Ib, and this  holds independent of the assume value of $R_V^{host}$.
This is consistent with the findings of \citet{anderson15} whom  used pixel statistics based on $H_{\alpha}$ emission  to study the environments of SNe by tracing on-going ($<$ 16 Myr old) star formation within SNe host galaxies.
\citeauthor{anderson15} found   SNe~Ic show a statistically significant preference  for being located in regions of on-going star formation, while  SNe~Ib are located in  regions  that do not closely follow ongoing star formation. 
The implications are the progenitors of  SNe~Ic  have shorter lifetimes and are more massive compared to the progenitors of SNe~Ib.
Being preferentially located in on-going star forming regions could therefore explain  why SNe~Ic experience (on average) more dust extinction compared with SNe~Ib.
 As a caveat to the SNe~Ic, we stress this subtype contains a mixed bag of transient phenomena, and assuming intrinsic colors based on more normal examples may not be appropriate to apply to more extreme cases such as the broad-line SN~2009bb and SN~2009ca. Unfortunately, we are limited by small number statistics though this will hopefully change in the future.

The findings of Sect.~\ref{sec:extinction}  indicate  SE SNe appear to reside in host galaxies characterized by a range of  dust properties.
Although the sample size of heavily reddened objects is small,  hosts of SNe~Ic are found to be characterized by standard $R_V^{host}$ values, while SNe~IIb and SNe~Ib tend to favor  lower $R_V^{host}$ values.

Low $R_V^{host}$ values have been inferred from the study of some core-collapse SNe. For example, \citet{folatelli14} recently inferred reddening parameters of $R_V^{host} = 1.5$ and $E(B-V)_{host}  = 0.41$ mag for the flat-velocity Type~IIb SN~2010as. Additionally in the process of constructing a low red-shift Hubble diagram populated with SNe~IIP,  \citet{olivares10} obtained a global value of $R_V^{host} = 1.4$. 

Low $R_V^{host}$ values are also obtained from the study of 
 individual and samples of moderate-to-heavily reddened SNe~Ia, suggesting 
 $R_V^{host}$ values ranging from 1.1 \citep{tripp98} to 2.2 \citep{kessler09,guy10,mandel11,burns14} to somewhere  between  $R_V^{host} = 1-2$ \citep{folatelli10}. 
 Plotted in Fig.~\ref{RV_cdf} is the $R_V$ CDF  obtained for the most reddened SE SNe in our sample compared to the $R_V$ CDF   obtained from an extended sample of CSP-I SNe~Ia  characterized by $E(B-V)_{host} \geq 0.2$ mag \citep[see][]{burns14}. 
 In general, compared to reddened SNe~Ia, SE SNe tend to exhibit larger $R_{V}^{host}$ values.

As noted in Sect.~\ref{sec:reddening} a handful of objects exhibit  $E(V-u)_{host}$ color excess values not reproduced by the fits obtained using the  F99 reddening law. 
This motivated us to fit each set of color excess values of our sample with the CSM-motivated reddening law of G08 
(see Fig.~\ref{fig:goobar}). 
This model was originally introduced to explain the abnormally low $R_V^{host}$ values implied from the study of   some heavily reddened SN~Ia.
Although the G08 reddening law provides superior fits to the $u$ colors of some objects it does not necessarily imply this is due to dust in the SN's immediate circumstellar environment. 
For example, \citet{burns14} found for several highly reddened SNe~Ia  with abnormal blue colors that they were   spectroscopically  peculiar objects exhibiting high-velocity features. 
This implies the anomalous $V-u$ colors could be driven by the prominent \ion{Ca}{ii} H\&K feature rather than related to the circumstellar environment.  
Unfortunately, the spectral coverage of our sample does not allow us to investigate the nature of this spectral feature given the wavelength cut-off off at the blue end of the spectrum for the majority of  the objects occurs at $\sim$ 4000 \AA\ \citep[see][]{holmbo17}.
 
In summary, we have identified a minimally-reddened sample of SE SNe enabling us to define intrinsic color-curve templates for the  main spectroscopically defined SE SN sub-types. 
The minimally-reddened sample is used  to infer the color excess of the reddened objects, and in the case of heavily reddened objects, a constraint on $R_V^{host}$. 
This work demonstrates the potential of securely estimating the host reddening parameters of SE SNe and highlights the  added benefits of NIR photometry when it comes to  determining the  host-galaxy reddening properties.
In the  future significant numbers of SE SNe will be discovered, and it is our hope that this work will provide a basis for further efforts  to improve upon our ability to understand their origins. 

\begin{acknowledgements}

We thank R. Amanullah,  E. Baron, A. Goobar, and P. Mazzali for useful discussions. 
M. D. Stritzinger, F. Taddia, E. Hsiao and C. Contreras gratefully acknowledge  support provided by the Danish Agency for Science and Technology and Innovation realized through a Sapere Aude Level 2 grant.
M. D. Stritzinger acknowledges funding by a research grant (13261) from the VILLUM FONDEN and  the Instrument Centre for Danish Astrophysics (IDA). 
M. D.~Stritzinger conducted a portion of this research at the Aspen Center for Physics, which is supported by NSF grant PHY-1066293. 
F. Taddia and J. Sollerman gratefully acknowledge the support from the Knut and Alice Wallenberg Foundation. 
This material is also based upon work supported by NSF under 
grants AST--0306969, AST--0607438, AST--0908886, AST--1008343, AST-1613426, AST--1613455, and AST--1613472.
This research has made use of the NASA/IPAC Extragalactic Database (NED), which is operated by the Jet Propulsion Laboratory, California Institute of Technology, under contract with the National Aeronautics and Space Administration.
\end{acknowledgements}

\bibliographystyle{aa}

\clearpage

\begin{deluxetable}{cc}
\tablewidth{0pt}
\tabletypesize{\scriptsize}
\tablecaption{Equivalent width measurements of {\ion{Na}{i}~D}.\label{tab:NaID}}
\tablehead{
\colhead{SN}&
\colhead{$EW_{\ion{Na}{i}~D}$ [\AA]}}
\startdata
2004ex&    1.4040$\pm$0.5280    \\
2004ff&    1.4015$\pm$0.0513    \\
2004gq&    1.4094$\pm$0.0962    \\
2004gt&    1.0192$\pm$0.0900    \\
2004gv&    0.7400$\pm$0.1970    \\
2005aw&    1.5032$\pm$0.1134    \\
2005em&         $<$1.4170       \\
2006ba&    0.5962$\pm$0.0339    \\
2006bf&         0$\pm$1.1280    \\
2006ep&         0$\pm$0.8030    \\
2006ir&    1.3670$\pm$0.5360    \\
2006lc&    0.9910$\pm$0.3870    \\
2006T &    0.8917$\pm$0.1247    \\
2007ag&    0.5700$\pm$0.0700    \\
2007C &    2.1973$\pm$0.2761    \\
2007hn&    0.7870$\pm$0.2670    \\
2007kj&         $<$0.5800       \\
2007rz&    1.6979$\pm$0.1342    \\
2007Y &    0.3270$\pm$0.0620    \\
2008aq&    0.3031$\pm$0.0010    \\
2008gc&    0.5515$\pm$0.0845    \\

2009bb&    1.9901$\pm$0.0410    \\
2009dt&    1.5190$\pm$0.1250    \\
2009K &    1.8439$\pm$0.1264    \\
2009Z &         $<$0.1610       \\
2009ca&    0.2610$\pm$0.2170    \\
\hline
2004ew&         $<$0.4340       \\
2004fe&         $<$1.2530       \\
2005bj&    1.0080$\pm$0.5310    \\
2005Q &         $<$0.4852           \\
2006fo&    0.8365$\pm$0.1428    \\
2008hh&         $<$0.4320       \\
2009dp&    2.3620$\pm$0.1458    \\
\enddata                                                        
\tablecomments{Objects are listed chronologically in two groups, first those with optical {\em and} NIR photometry and then those with only optical photometry.}
\end{deluxetable}

\clearpage
\begin{deluxetable}{cccc}
\tablewidth{0pt}
\tabletypesize{\scriptsize}
\tablecaption{Intrinsic values of $V/r/B/g-X_{\lambda}^{host}$ colors at +10d since $V/r/B/g$ maximum.\label{tab:colors_10d}}
\tablehead{
\colhead{Color}&
\colhead{IIb}&
\colhead{Ib}&
\colhead{Ic}\\
\colhead{}&
\colhead{[mag]}&
\colhead{[mag]}&
\colhead{[mag]}}
\startdata
$(u-V)_{host}(+10d)$ &  2.111(0.107) & $\cdots$        & 1.957(0.075)\\
$(B-V)_{host}(+10d)$ &  0.875(0.022) & 0.920(0.049)  & 0.962(0.027)\\
$(g-V)_{host}(+10d)$ &  0.444(0.015) & 0.450(0.024)  & 0.510(0.017)\\
$(V-r)_{host}(+10d)$ &  0.258(0.011) & 0.278(0.028)  & 0.293(0.018)\\
$(V-i)_{host}(+10d)$ &  0.279(0.029) & 0.358(0.056)  & 0.271(0.028)\\
$(V-Y)_{host}(+10d)$ &  0.703(0.062) & 0.775(0.053)  & 0.707(0.065)\\
$(V-J)_{host}(+10d)$ &  0.801(0.023) & 0.905(0.159)  & 0.408(0.031)\\
$(V-H)_{host}(+10d)$ &  0.966(0.041) & 1.070(0.072)  & 0.696(0.026)\\
\hline
$(u-r)_{host}(+10d)$  & 2.505(0.140) & $\cdots$        & 2.309(0.075) \\
$(B-r)_{host}(+10d)$  & 1.171(0.031) & 1.316(0.039)  & 1.331(0.025) \\
$(g-r)_{host}(+10d)$  & 0.765(0.019) & 0.790(0.048)  & 0.851(0.018) \\
$(V-r)_{host}(+10d)$  & 0.256(0.019) & 0.308(0.020)  & 0.322(0.016) \\
$(r-i)_{host}(+10d)$  &-0.002(0.100) & 0.148(0.022)  & 0.042(0.015) \\
$(r-Y)_{host}(+10d)$  & 0.463(0.049) & 0.606(0.038)  & 0.651(0.045) \\
$(r-J)_{host}(+10d)$  & 0.593(0.016) & 0.658(0.043)  & 0.288(0.020) \\
$(r-H)_{host}(+10d)$  & 0.760(0.031) & 0.932(0.068)  & 0.621(0.021) \\
\hline
$(u-B)_{host}(+10d)$  & 0.879(0.163) & 1.589(0.078)  & 0.996(0.062)\\
$(B-g)_{host}(+10d)$  & 0.411(0.024) & 0.409(0.033)  & 0.441(0.018)\\
$(B-V)_{host}(+10d)$  & 0.833(0.029) & 0.818(0.040)  & 0.915(0.018)\\
$(B-r)_{host}(+10d)$  & 1.064(0.031) & 1.123(0.047)  & 1.148(0.031)\\
$(B-i)_{host}(+10d)$  & 1.026(0.092) & 1.107(0.074)  & 1.162(0.048)\\
$(B-Y)_{host}(+10d)$  & 1.461(0.071) & 1.500(0.053)  & 1.537(0.109)\\
$(B-J)_{host}(+10d)$  & 1.597(0.030) & 1.635(0.056)  & 1.238(0.074)\\
$(B-H)_{host}(+10d)$  & 1.706(0.044) & 1.811(0.063)  & 1.526(0.071)\\
\hline
$(u-g)_{host}(+10d)$ & 2.188(0.159) & 1.296(0.099)  & 1.440(0.070) \\
$(B-g)_{host}(+10d)$ & 0.436(0.031) & 0.427(0.028)  & 0.449(0.019) \\
$(g-V)_{host}(+10d)$ & 0.427(0.021) & 0.421(0.014)  & 0.503(0.012) \\
$(g-r)_{host}(+10d)$ & 0.686(0.038) & 0.663(0.030)  & 0.786(0.014) \\
$(g-i)_{host}(+10d)$ & 0.751(0.071) & 0.629(0.060)  & 0.778(0.026) \\
$(g-Y)_{host}(+10d)$ & 1.184(0.211) & 1.039(0.068)  & 1.284(0.050) \\
$(g-J)_{host}(+10d)$ & 1.259(0.274) & 1.185(0.022)  & 0.881(0.053) \\
$(g-H)_{host}(+10d)$ & 1.462(0.066) & 1.306(0.036)  & 1.179(0.051) \\
\enddata                                                        
\end{deluxetable}

\clearpage
\begin{deluxetable}{cccccccccc}
\tablewidth{0pt}
\rotate
\tabletypesize{\scriptsize}
\tablecaption{$E(V-X_{\lambda})_{host}$ color excesses inferred from post maximum colors.\label{tab:EVMX}}
\tablehead{
\colhead{SN}&
\colhead{spectral ID}&
\colhead{$E(V-u)_{host}$}&
\colhead{$E(V-B)_{host}$}&
\colhead{$E(V-g)_{host}$}&
\colhead{$E(V-r)_{host}$}&
\colhead{$E(V-i)_{host}$}&
\colhead{$E(V-Y)_{host}$}&
\colhead{$E(V-J)_{host}$}&
\colhead{$E(V-H)_{host}$}\\
\colhead{}&
\colhead{}&
\colhead{[mag]}&
\colhead{[mag]}&
\colhead{[mag]}&
\colhead{[mag]}&
\colhead{[mag]}&
\colhead{[mag]}&
\colhead{[mag]}&
\colhead{[mag]}}
\startdata
    2004ex &      IIb  &$-$0.749(0.161)  &$-$0.175(0.036)  &$-$0.080(0.025)  &0.055(0.020)  &0.129(0.037)  &0.191(0.070)  &0.106(0.033)  &0.124(0.050)   \\
    2004ff &      IIb  &$-$0.757(0.139)  &$-$0.151(0.034)  &$-$0.081(0.083)  &0.040(0.018)  &0.233(0.038)  &0.411(0.072)  &0.354(0.035)  &0.161(0.054)   \\
    2004gq &      Ib   &$-$0.041(0.229)  &$-$0.067(0.037)  &$-$0.026(0.022)  &0.035(0.023)  &0.175(0.048)  &0.364(0.063)  &0.398(0.158)  &0.282(0.082)   \\
    2004gt &      Ic   &$-$1.220(0.099)  &$-$0.237(0.032)  &$-$0.149(0.027)  &0.149(0.029)  &0.396(0.034)  &0.572(0.090)  &1.120(0.046)  &1.028(0.040)   \\
    2004gv &      Ib   &$-$0.174(0.156)  &$-$0.053(0.039)  &$-$0.031(0.022)  &0.053(0.024)  &0.039(0.047)  &0.016(0.058)  &0.031(0.149)  &0.028(0.075)   \\
    2006ba &      IIb  &$-$0.576(0.171)  &$-$0.088(0.078)  &$-$0.118(0.035)  &0.042(0.030)  &0.273(0.057)  &0.496(0.087)  &0.482(0.077)  &$\cdots$         \\
    2006bf &      IIb  &$\cdots$         &$-$0.350(0.112)  &$-$0.150(0.037)  &0.155(0.066)  &0.286(0.059)  &0.459(0.099)  &0.366(0.063)  &0.497(0.092)   \\
    2006ep &      $\cdots$Ib   &$-$0.186(0.306)  &$-$0.233(0.040)  &$-$0.128(0.023)  &0.098(0.025)  &0.212(0.051)  &0.412(0.065)  &0.447(0.160)  &0.462(0.182)   \\
    2006ir &      Ic   &$-$0.208(0.143)  &$-$0.022(0.048)  &$-$0.034(0.043)  &$-$0.003(0.034) &$-$0.035(0.044) &$-$0.043(0.100) &0.399(0.175)  &0.068(0.085)   \\
    2006lc &      Ib   &$-$1.478(0.274)  &$-$0.298(0.077)  &$-$0.173(0.039)  &0.195(0.039)  &0.346(0.074)  &$\cdots$         &$\cdots$         &$\cdots$       \\
    2006T  &      IIb  &$-$1.073(0.159)  &$-$0.277(0.035)  &$-$0.143(0.023)  &0.152(0.016)  &0.234(0.032)  &0.395(0.066)  &0.339(0.028)  &0.399(0.050)   \\
    2007ag &      Ic   &$\cdots$         &$-$0.191(0.142)  &$-$0.084(0.102)  &0.169(0.108)  &0.218(0.124)  &0.338(0.134)  &0.972(0.143)  &0.989(0.126)   \\
    2007C  &      Ib   &$-$1.106(0.412)  &$-$0.549(0.057)  &$-$0.313(0.038)  &0.311(0.037)  &0.532(0.089)  &0.999(0.190)  &1.018(0.178)  &1.019(0.195)   \\
    2007hn &      Ic   &$\cdots$         &$-$0.170(0.104)  &$-$0.054(0.102)  &0.044(0.102)  &0.036(0.104)  &0.304(0.071)  &0.622(0.060)  &$\cdots$         \\
    2007rz &      Ic   &$\cdots$         &$-$0.078(0.043)  & 0.002(0.023)  &0.159(0.048)  &0.385(0.100)  &0.500(0.163)  &0.889(0.172)  &0.816(0.201)   \\
    2009bb &      Ic-BL&$-$0.787(0.103)  &$-$0.540(0.027)  &$-$0.204(0.024)  &0.177(0.024)  &0.224(0.035)  &0.760(0.090)  &1.122(0.050)  &0.902(0.040)   \\
    2009dt &      Ic   &$\cdots$         &$-$0.490(0.044)  &$-$0.256(0.027)  &0.273(0.035)  &0.555(0.068)  &1.060(0.081)  &1.562(0.037)  &1.526(0.040)   \\
    2009K  &      IIb  &$-$0.677(0.165)  &$-$0.096(0.021)  &$-$0.083(0.019)  &0.136(0.020)  &0.200(0.040)  &0.425(0.058)  &0.368(0.037)  &0.337(0.067)   \\
    2005aw &      Ic   &$-$1.438(0.117)  &$-$0.496(0.088)  &$-$0.197(0.032)  &0.132(0.033)  &0.126(0.072)  &0.442(0.107)  &0.781(0.064)  &0.499(0.087)   \\
    2005bj &      IIb  &$-$41.225(0.134)  &$-$0.346(0.073)  &$-$0.199(0.083)  &0.077(0.038)  &0.117(0.085)  &$\cdots$         &$\cdots$         &$\cdots$       \\
    2009dp &      Ic   &$\cdots$         &$-$0.387(0.045)  &$-$0.065(0.044)  &0.106(0.048)  &0.246(0.061)  &$\cdots$         &$\cdots$         &$\cdots$       \\
\enddata                                                        
\tablecomments{$E(V-X_{\lambda})_{host}$ color excess values  are computed for all of the CSP SE SN except for the nine objects  considered to be minimally-reddened. 
Due to poor light curve coverage color excesses are not available for SN~2006fo, SN~2008gc, and SN~2009ca.} \end{deluxetable}

\clearpage
\begin{deluxetable}{cc|cc|cc|cc|cc|cc}
\rotate
\tablewidth{0pt}
\tabletypesize{\scriptsize}
\tablecaption{Best-fit $R_V^{host}$ and $A_V^{host}$ values for the  seven most reddened SE SNe with  eight color excess combinations. Fits are performed to $E(V-X_{\lambda})_{host}$, $E(r-X_{\lambda})_{host}$, $E(B-X_{\lambda})_{host}$ and  $E(g-X_{\lambda})_{host}$.\label{tab:veryred}}
\tablehead{
\colhead{SN}&
\colhead{spectral ID}&
\colhead{$R_V^{[E(V-X)_{host}]}$}&
\colhead{$A_V^{[E(V-X)_{host}]}$}&
\colhead{$R_V^{[E(r-X)_{host}]}$}&
\colhead{$A_V^{[E(r-X)_{host}]}$}&
\colhead{$R_V^{[E(B-X)_{host}]}$}&
\colhead{$A_V^{[E(B-X)_{host}]}$}&
\colhead{$R_V^{[E(g-X)_{host}]}$}&
\colhead{$A_V^{[E(g-X)_{host}]}$}&
\colhead{$<R_V^{host}>$}&
\colhead{$<A_V^{host}>$}\\
\colhead{}&
\colhead{}&
\colhead{}&
\colhead{[mag]}&
\colhead{}&
\colhead{[mag]}&
\colhead{}&
\colhead{[mag]}&
\colhead{}&
\colhead{[mag]}&
\colhead{}&
\colhead{[mag]}}
\startdata

2004gt  & Ic    &  2.6$^{+0.2}_{-0.2} $ & 1.24$^{+0.09}_{-0.10}$ & 2.9$^{+0.2}_{-0.1} $ & 1.15$^{+0.07}_{-0.08}$  & 2.5$^{+0.5}_{-0.3}$  & 0.95$^{+0.08}_{-0.08}$  &  2.0$^{+0.3}_{-0.2} $ & 0.93$^{+0.07}_{-0.08}$ &  2.5(0.4)& 1.07(0.15)  \\
2005aw  & Ic    &  1.2$^{+0.2}_{-0.1} $ & 0.71$^{+0.05}_{-0.05}$ & 1.4$^{+0.1}_{-0.1} $ & 0.69$^{+0.07}_{-0.07}$  & 1.2$^{+0.4}_{-0.3}$  & 0.63$^{+0.05}_{-0.06}$  &  1.7$^{+0.3}_{-0.3} $ & 0.58$^{+0.05}_{-0.05}$ &  1.4(0.2)& 0.65(0.06)  \\
2006ep  & Ib    &  4.8$^{+22.4}_{-2.3}$ & 0.60$^{+0.53}_{-0.50}$ & 5.5$^{+14.6}_{-2.1}$ & 0.56$^{+0.42}_{-0.42}$  & 3.7$^{+1.2}_{-0.7}$  & 0.62$^{+0.10}_{-0.12}$  &  6.3$^{+12.7}_{-2.8}$ & 0.62$^{+0.42}_{-0.39}$ &  5.1(1.1)& 0.60(0.03)  \\
2006T   & IIb   &  1.1$^{+0.2}_{-0.2} $ & 0.47$^{+0.06}_{-0.06}$ & 1.2$^{+0.2}_{-0.1} $ & 0.46$^{+0.07}_{-0.07}$  & 1.3$^{+0.4}_{-1.0}$  & 0.38$^{+0.16}_{-0.06}$  &  1.5$^{+0.2}_{-0.2} $ & 0.37$^{+0.03}_{-0.03}$ &  1.3(0.2)& 0.42(0.05)  \\
2007C   & Ib    &  2.6$^{+1.2}_{-0.5} $ & 1.36$^{+0.38}_{-0.38}$ & 2.3$^{+0.5}_{-0.3} $ & 1.21$^{+0.25}_{-0.27}$  & 2.4$^{+0.6}_{-0.5}$  & 1.43$^{+0.16}_{-0.15}$  &  2.4$^{+0.3}_{-0.3} $ & 1.32$^{+0.09}_{-0.10}$ &  2.4(0.1)& 1.33(0.09)  \\
2009bb  & Ic-BL &  3.3$^{+0.4}_{-0.3} $ & 1.23$^{+0.12}_{-0.13}$ & 2.8$^{+0.2}_{-0.2} $ & 1.08$^{+0.09}_{-0.09}$  & 4.1$^{+0.6}_{-0.4}$  & 1.32$^{+0.10}_{-0.12}$  &  3.4$^{+0.6}_{-0.4} $ & 1.06$^{+0.11}_{-0.12}$ &  3.4(0.5)& 1.17(0.12)  \\
2009dt  & Ic    &  4.3$^{+0.4}_{-0.3} $ & 1.89$^{+0.12}_{-0.15}$ & 3.5$^{+0.2}_{-0.1} $ & 1.63$^{+0.08}_{-0.08}$  & 4.0$^{+0.7}_{-0.4}$  & 1.87$^{+0.14}_{-0.18}$  &  4.6$^{+0.5}_{-0.3} $ & 1.78$^{+0.11}_{-0.13}$ &  4.1(0.5)& 1.79(0.12)  \\

    \enddata                                                        
\end{deluxetable}

\clearpage
\begin{deluxetable}{cc|cc|cc|cc|cc|c}
\rotate
\tablewidth{0pt}
\tabletypesize{\scriptsize}
\tablecaption{Best-fit $A_V^{host}$ values of 15 low reddened SE SN obtained from $E(V-X_{\lambda})_{host}$, $E(r-X_{\lambda})_{host}$, $E(B-X_{\lambda})_{host}$ and  $E(g-X_{\lambda})_{host}$ assuming a universal $R_V^{host}$ value for each SE SN sub-type.\label{tab:EBMV}}
\tablehead{
\colhead{SN}&
\colhead{type}&
\colhead{$R_V^{[E(V-X_{\lambda})_{host}]}$}&
\colhead{$A_V^{[E(V-X_{\lambda})_{host}]}$}& 
\colhead{$R_V^{[E(r-X_{\lambda})_{host}]}$}& 
\colhead{$A_V^{[E(r-X_{\lambda})_{host}]}$}& 
\colhead{$R_V^{[E(B-X_{\lambda})_{host}]}$}& 
\colhead{$A_V^{[E(B-X_{\lambda})_{host}]}$}&
\colhead{$R_V^{[E(g-X_{\lambda})_{host}]}$}&
\colhead{$A_V^{[E(g-X_{\lambda})_{host}]}$}& 
\colhead{$<A_V^{host}>$}\\
\colhead{}&
\colhead{}&
\colhead{}&
\colhead{[mag]}&
\colhead{}&
\colhead{[mag]}&
\colhead{}&
\colhead{[mag]}&
\colhead{}&
\colhead{[mag]}&
\colhead{[mag]}}
\startdata
2004ex  &      IIb   & 1.1 & 0.28$^{+0.05}_{-0.05}$ &   1.2 & 0.26$^{+0.05}_{-0.05}$ &  1.3 & 0.25$^{+0.03}_{-0.02}$ &        1.5 & 0.24$^{+0.03}_{-0.03}$   &   0.26(0.02)\\
2004ff  &      IIb   & 1.1 & 0.34$^{+0.04}_{-0.05}$ &   1.2 & 0.28$^{+0.05}_{-0.06}$ &  1.3 & 0.29$^{+0.04}_{-0.02}$ &        1.5 & 0.30$^{+0.03}_{-0.03}$   &   0.30(0.03)\\
2004gq  &      Ib    & 2.6 & 0.33$^{+0.11}_{-0.10}$ &   2.3 & 0.16$^{+0.09}_{-0.10}$ &  2.4 & 0.28$^{+0.03}_{-0.03}$ &        2.4 & 0.26$^{+0.10}_{-0.08}$   &   0.26(0.07)\\
2004gv  &      Ib    & 2.6 & 0.09$^{+0.08}_{-0.07}$ &   2.3 & 0.07$^{+0.08}_{-0.08}$ &  2.4 & 0.11$^{+0.03}_{-0.03}$ &        2.4 & 0.04$^{+0.10}_{-0.07}$   &   0.08(0.03)\\
2006ba  &      IIb   & 1.1 & 0.32$^{+0.06}_{-0.06}$ &   1.2 & 0.24$^{+0.07}_{-0.07}$ &  1.3 & 0.28$^{+0.04}_{-0.04}$ &        1.5 & 0.42$^{+0.05}_{-0.06}$   &   0.32(0.08)\\
2006bf  &      IIb   & 1.1 & 0.48$^{+0.05}_{-0.05}$ &   1.2 & 0.49$^{+0.07}_{-0.07}$ &  1.3 & 0.51$^{+0.06}_{-0.02}$ &        1.5 & 0.42$^{+0.05}_{-0.04}$   &   0.47(0.04)\\
2006ir  &      Ic    & 4.3 & 0.20$^{+0.10}_{-0.08}$ &   3.5 & 0.13$^{+0.09}_{-0.08}$ &  4.0 & 0.01$^{+0.09}_{-0.07}$ &        4.6 & 0.15$^{+0.09}_{-0.07}$   &   0.12(0.08)\\
2006lc  &      Ib    & 2.6 & 1.43$^{+0.23}_{-0.22}$ &   2.3 & 0.98$^{+0.09}_{-0.10}$ &  2.4 & 1.02$^{+0.11}_{-0.13}$ &        2.4 & 1.04$^{+0.13}_{-0.15}$   &   1.12(0.21)\\
2007ag  &      Ic    & 4.3 & 1.04$^{+0.11}_{-0.09}$ &   3.5 & 0.80$^{+0.11}_{-0.08}$ &  4.0 & 0.93$^{+0.14}_{-0.11}$ &        4.6 & 0.91$^{+0.10}_{-0.08}$   &   0.92(0.10)\\
2007hn  &      Ic    & 4.3 & 0.63$^{+0.08}_{-0.06}$ &   3.5 & 0.66$^{+0.20}_{-0.20}$ &  4.0 & 0.45$^{+0.08}_{-0.07}$ &        4.6 & 0.45$^{+0.08}_{-0.06}$   &   0.55(0.11)\\
2007rz  &      Ic    & 4.3 & 0.99$^{+0.15}_{-0.11}$ &   3.5 & 0.74$^{+0.10}_{-0.08}$ &  4.0 & 0.79$^{+0.13}_{-0.10}$ &        4.6 & 0.80$^{+0.12}_{-0.09}$   &   0.83(0.11)\\
2009K   &      IIb   & 1.1 & 0.34$^{+0.05}_{-0.05}$ &   1.2 & 0.32$^{+0.05}_{-0.05}$ &  1.3 & 0.02$^{+0.03}_{-0.03}$ &        1.5 & 0.07$^{+0.03}_{-0.03}$   &   0.19(0.17)\\
2005bj  &      IIb   & 1.1 & 0.51$^{+0.05}_{-0.05}$ &   1.2 & 0.50$^{+0.06}_{-0.07}$ &  1.3 & 0.47$^{+0.05}_{-0.06}$ &        1.5 & 0.55$^{+0.06}_{-0.07}$   &   0.51(0.03)\\
2009dp  &      Ic    & 4.3 & 0.96$^{+0.11}_{-0.12}$ &   3.5 & 0.81$^{+0.08}_{-0.09}$ &  4.0 & 1.29$^{+0.09}_{-0.10}$ &        4.6 & 0.69$^{+0.07}_{-0.09}$   &   0.94(0.26)\\
    \enddata                                                        
\end{deluxetable}

\clearpage
\begin{figure*}
\includegraphics[width=8cm]{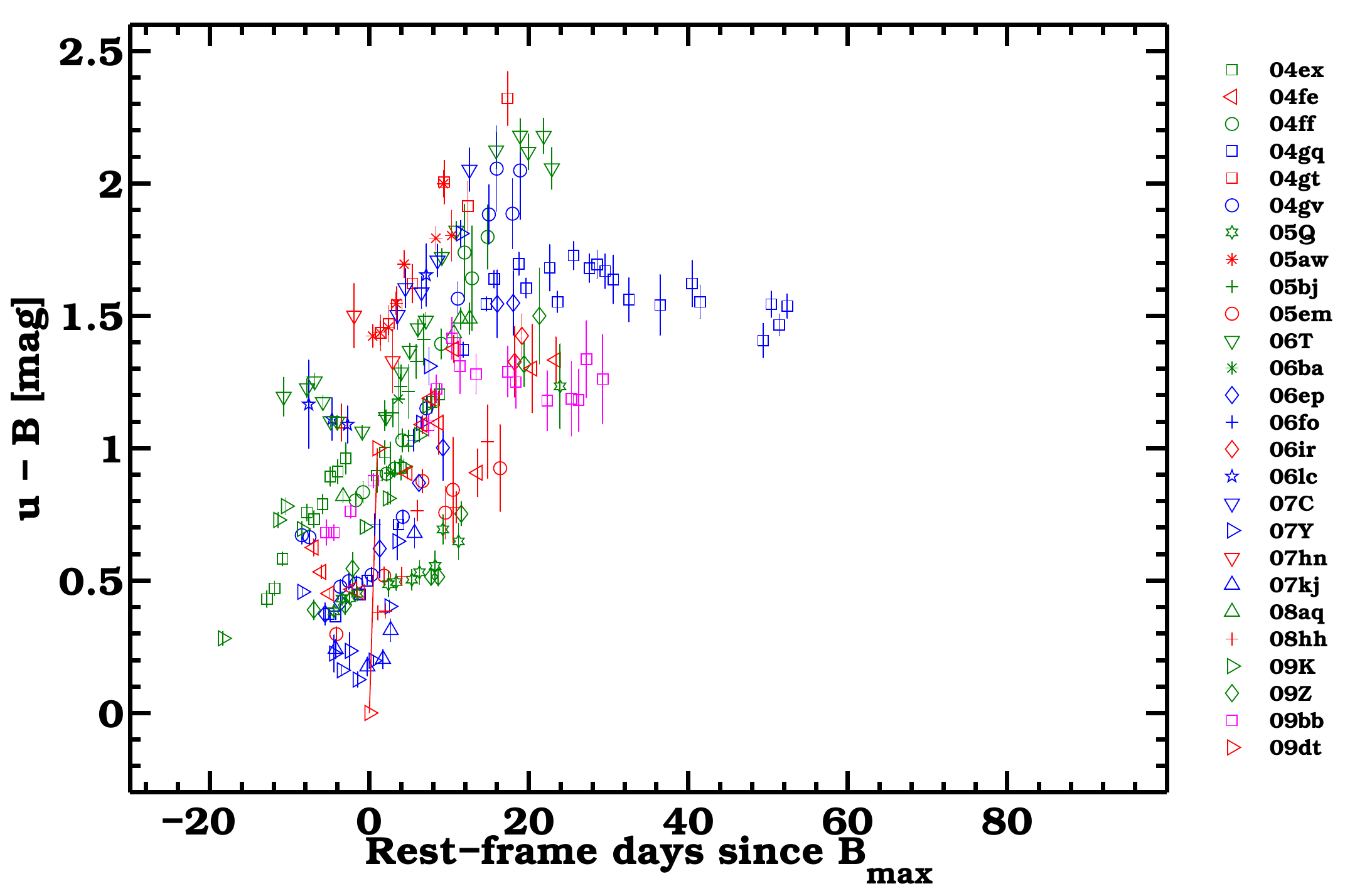}
\includegraphics[width=8cm]{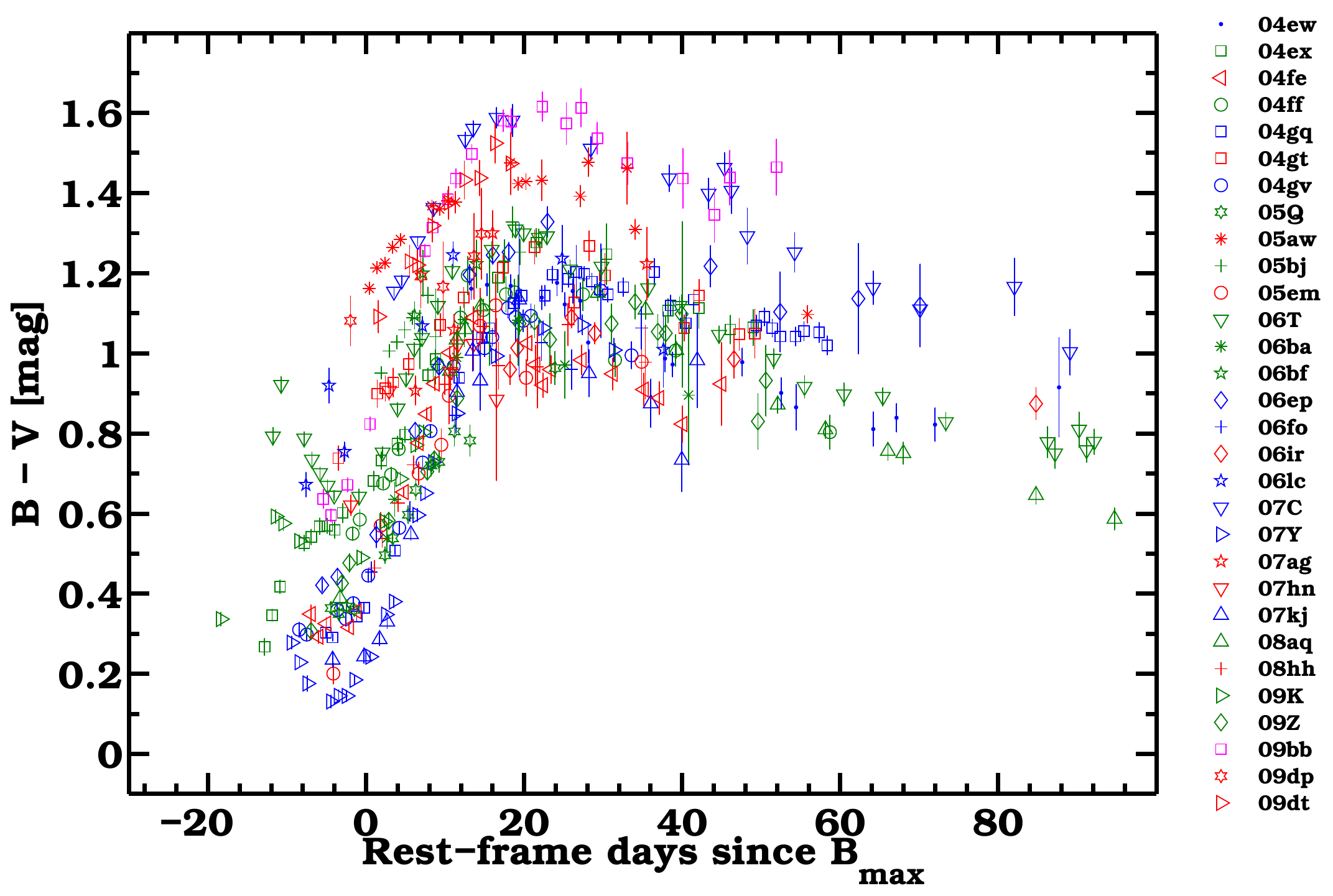}\\
\includegraphics[width=8cm]{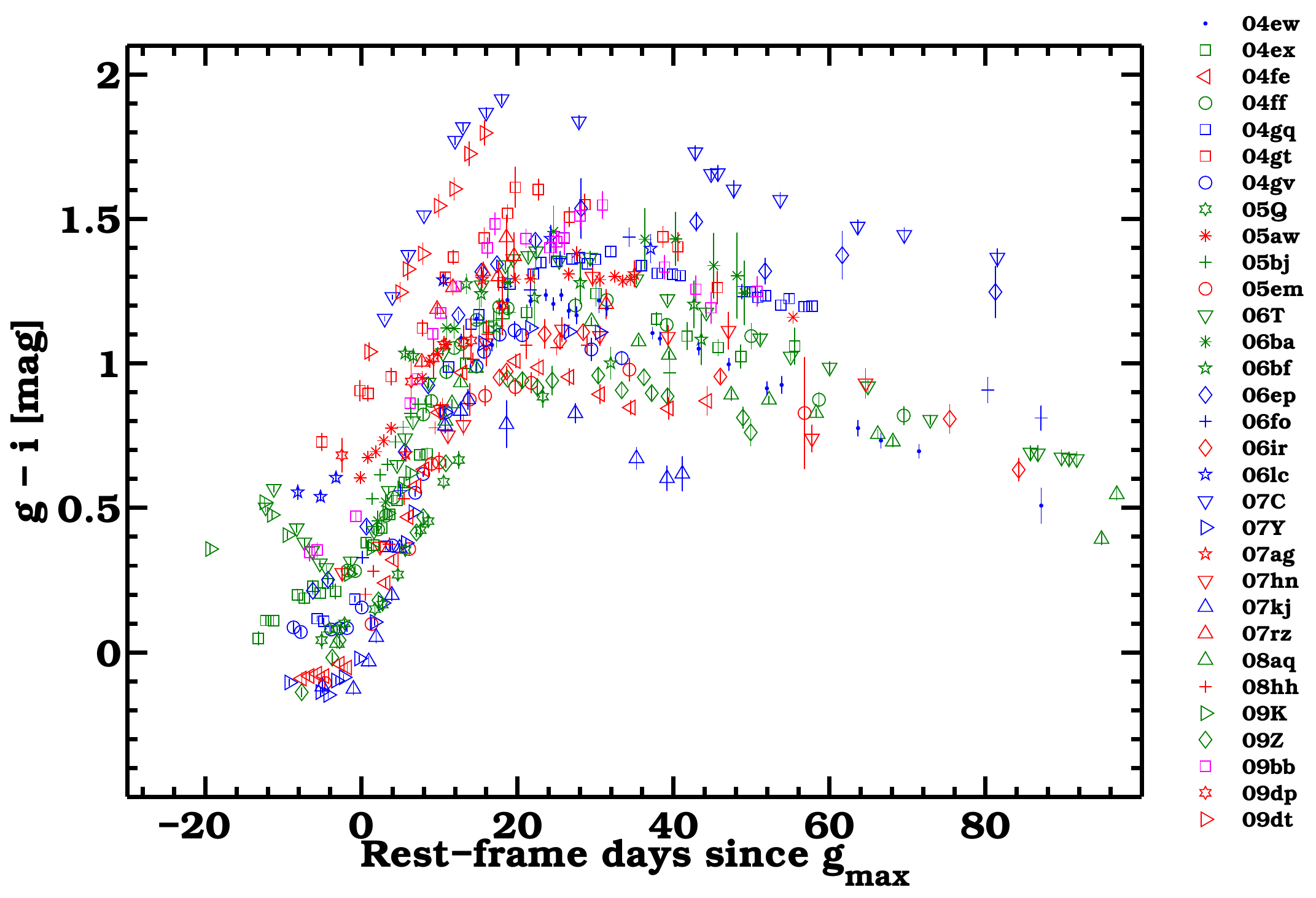}
\includegraphics[width=8cm]{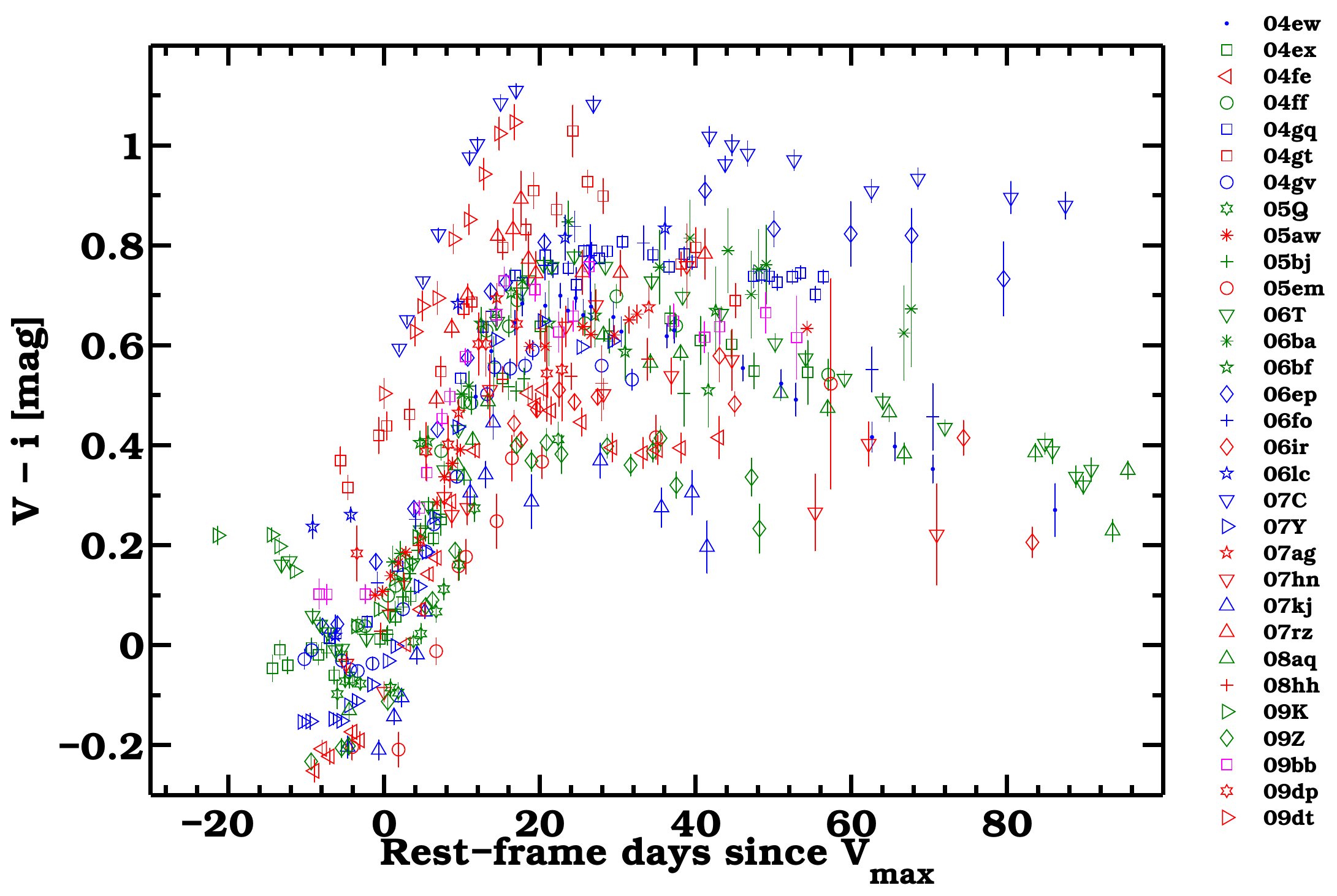}\\
\includegraphics[width=8cm]{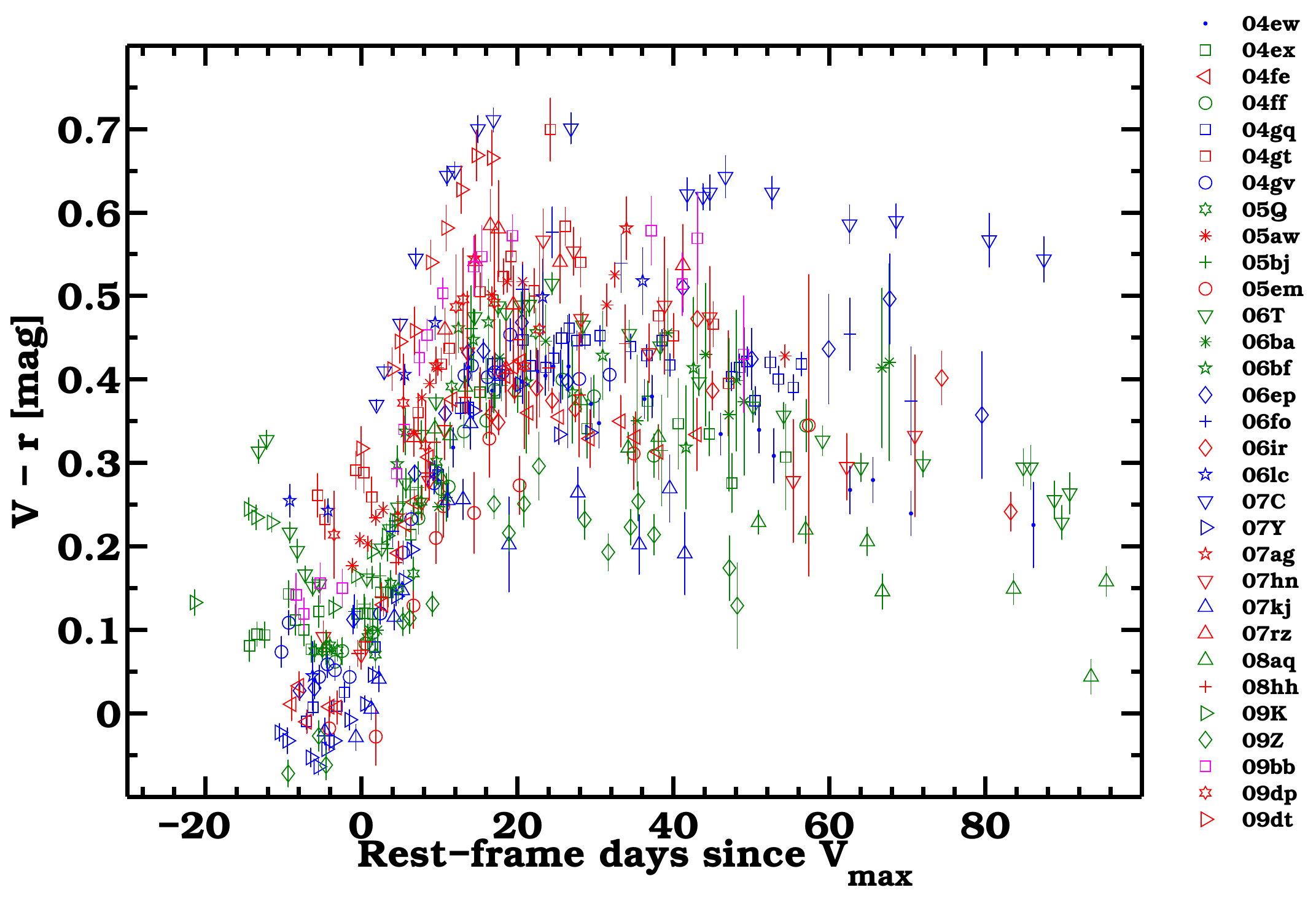}
\includegraphics[width=8cm]{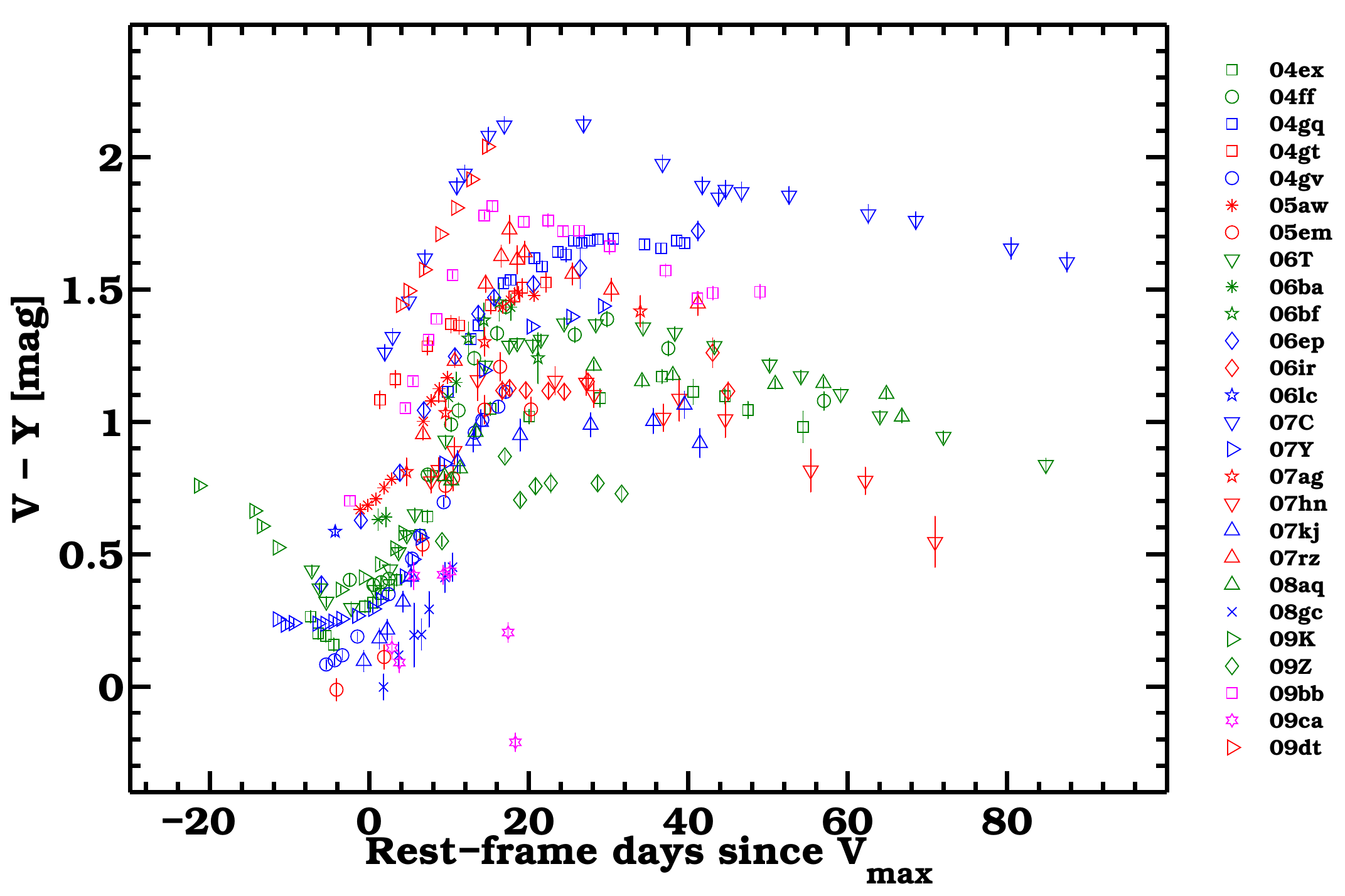}\\
\includegraphics[width=8cm]{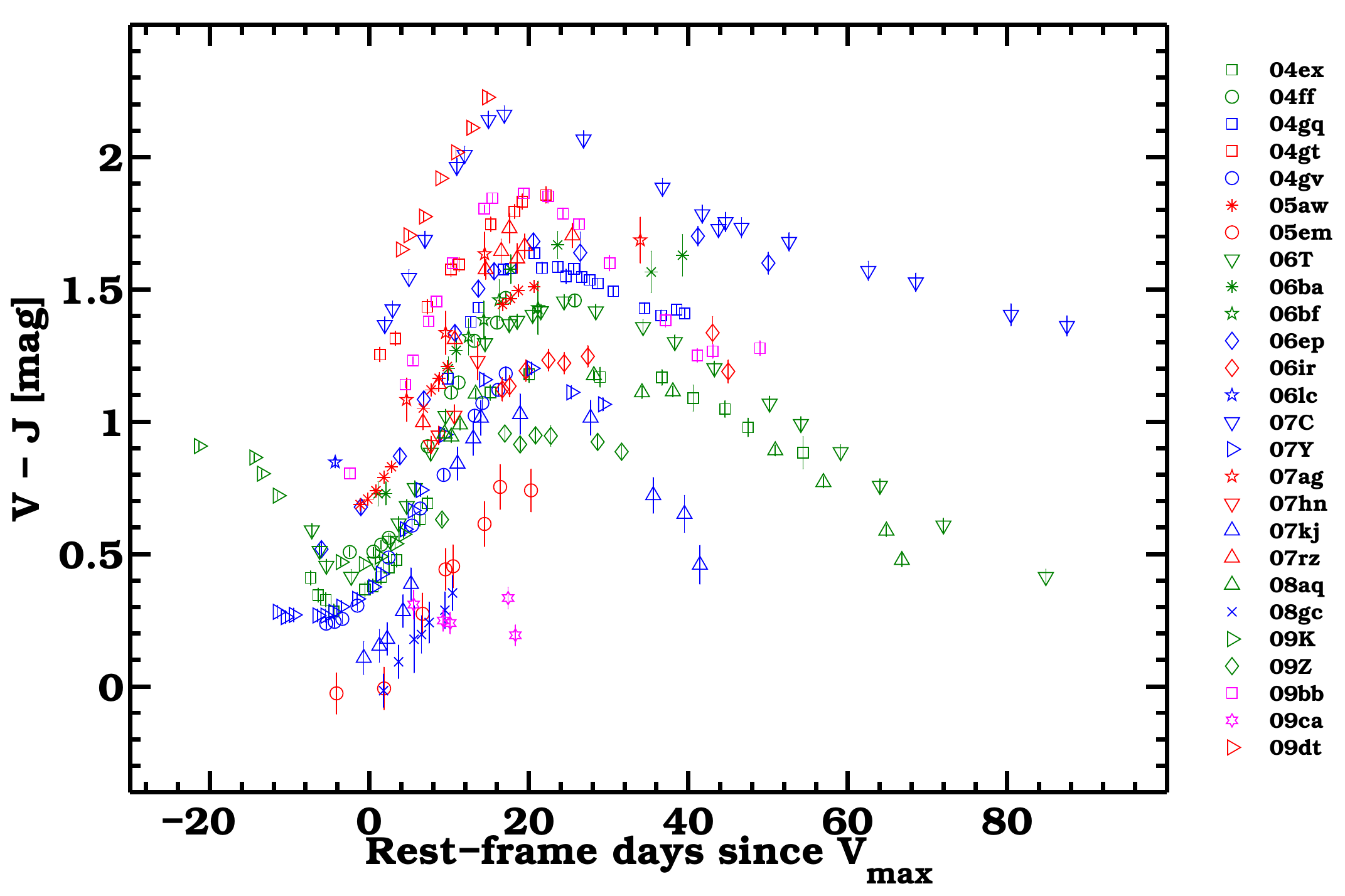}
\includegraphics[width=8cm]{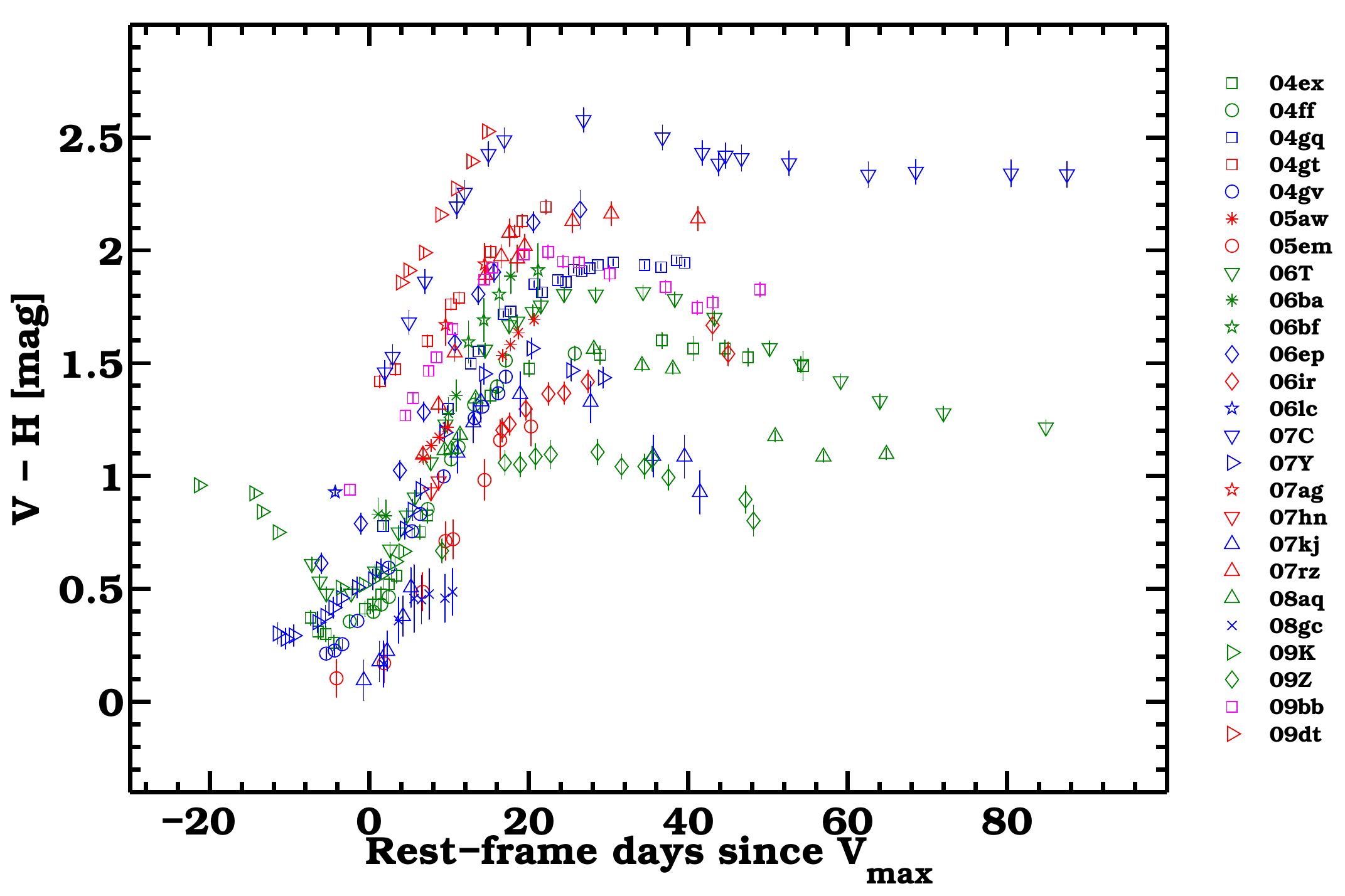}\\
\caption{Various optical and optical/NIR colors plotted as a function of days relative to maximum for the CSP SE SN sample. Symbols are color coded here and largely throughout this manuscript based on the different sub-types. Specifically, green, blue, red and magenta correspond to SNe~IIb, Ib, Ic and Ic-BL, respectively.\label{fig:colors}}
\end{figure*}

\clearpage
\begin{figure*}[h]
\centering
\includegraphics[height=8cm]{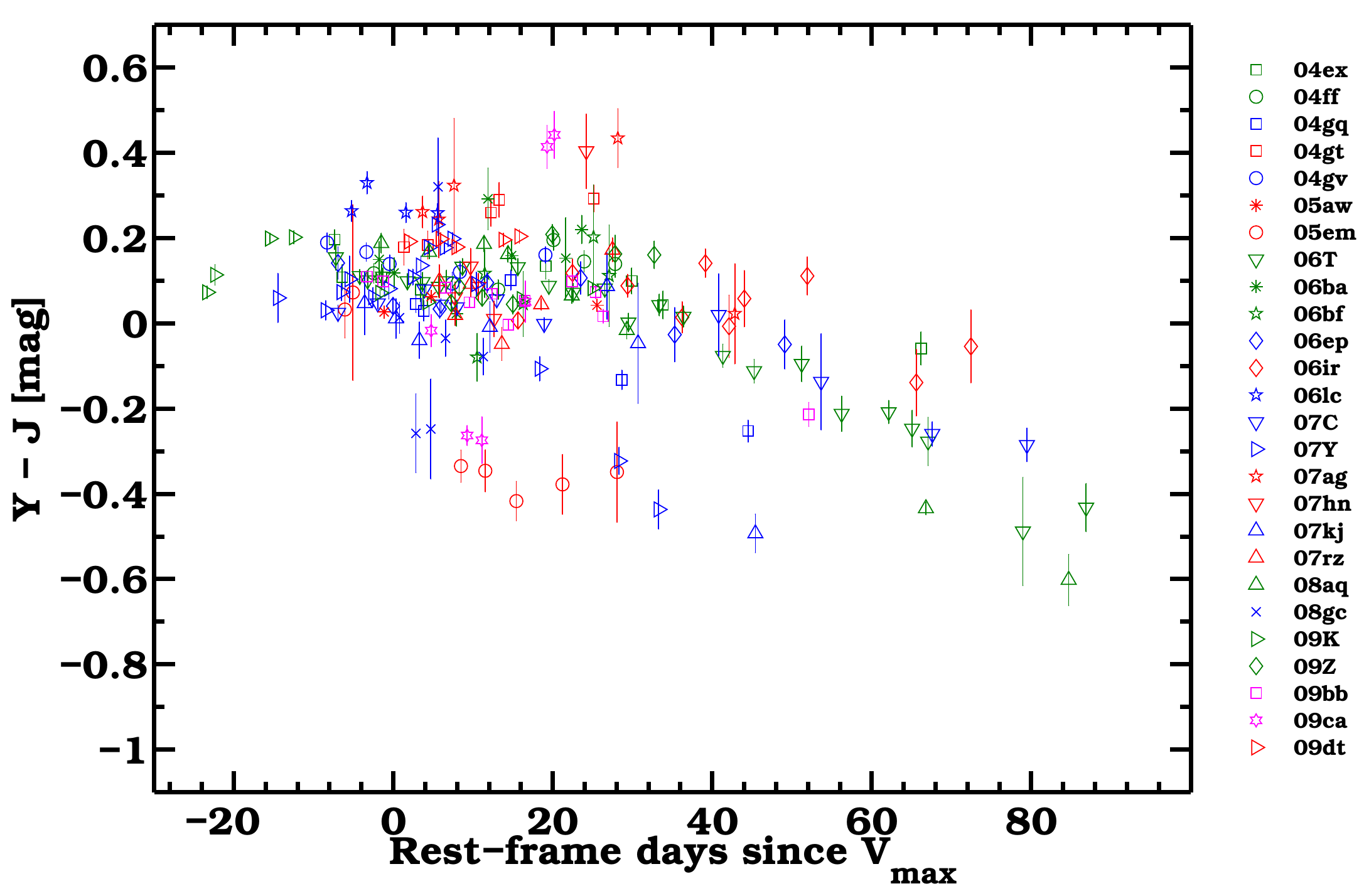}\\
\includegraphics[height=8cm]{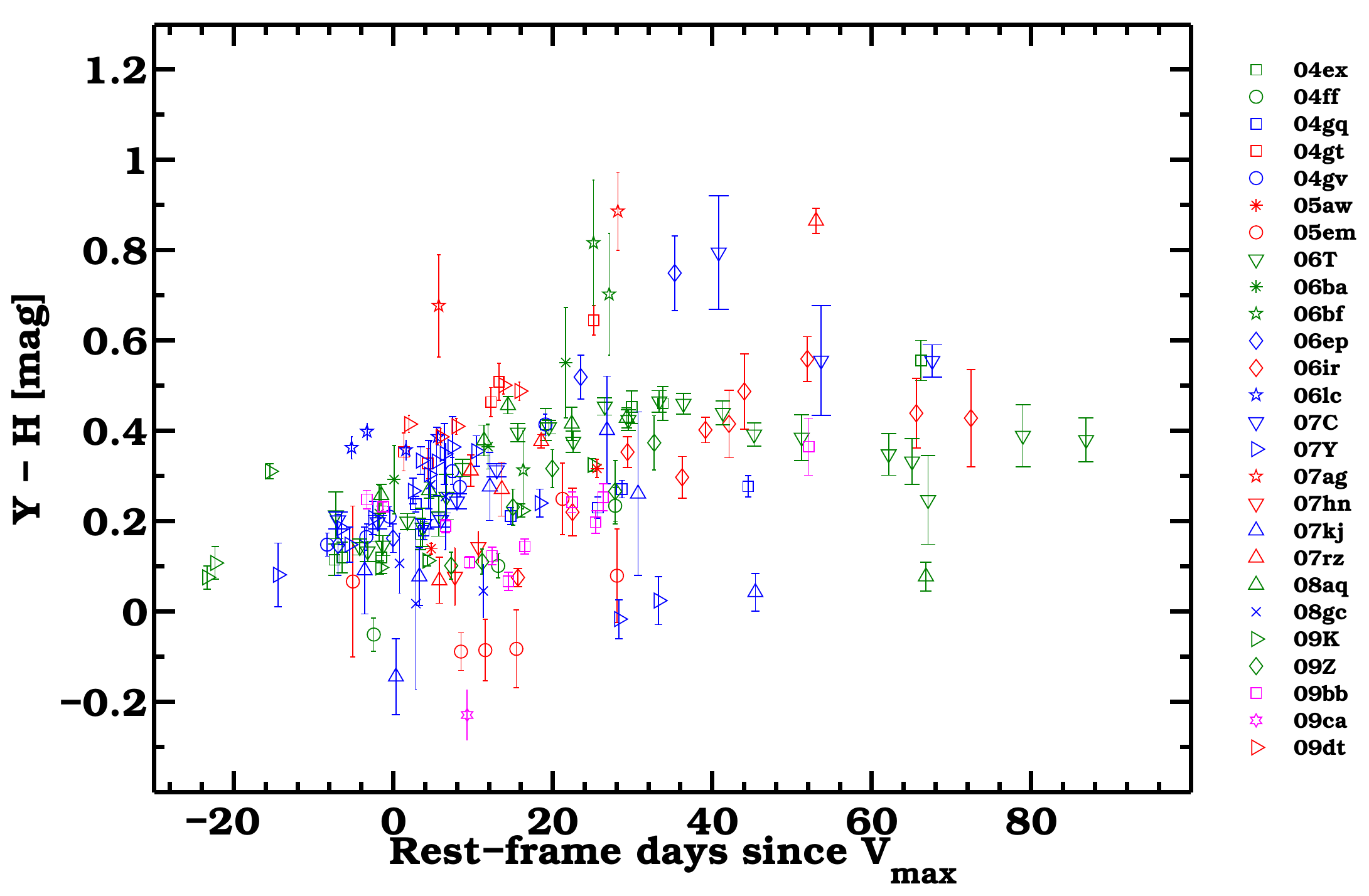}\\
\includegraphics[height=8cm]{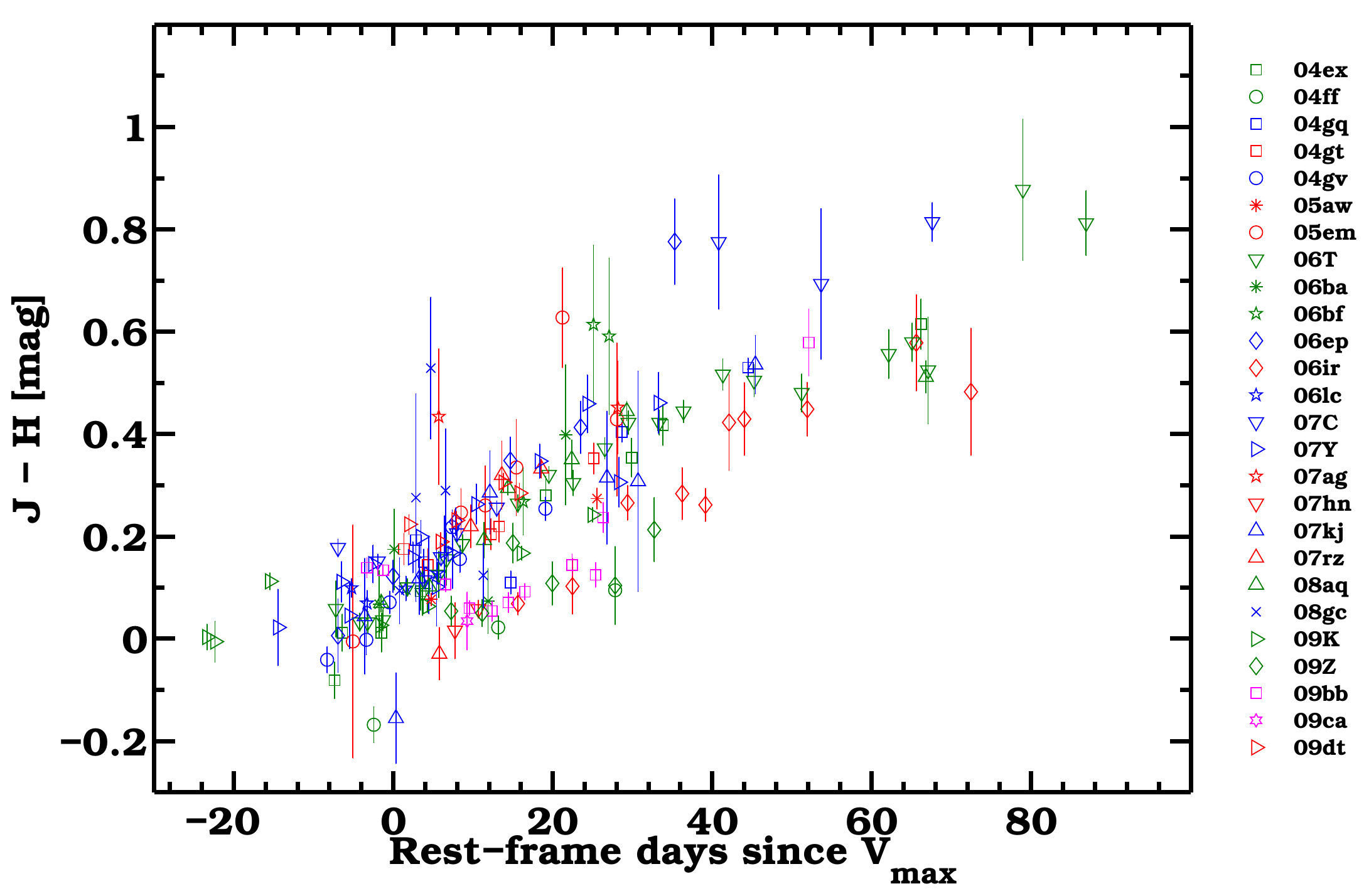}
\caption[]{NIR colors for our sample of SE SNe.\label{fig:NIRcolors}}
\end{figure*}

\clearpage
\begin{figure*}[h]
\centering
\includegraphics[height=8cm]{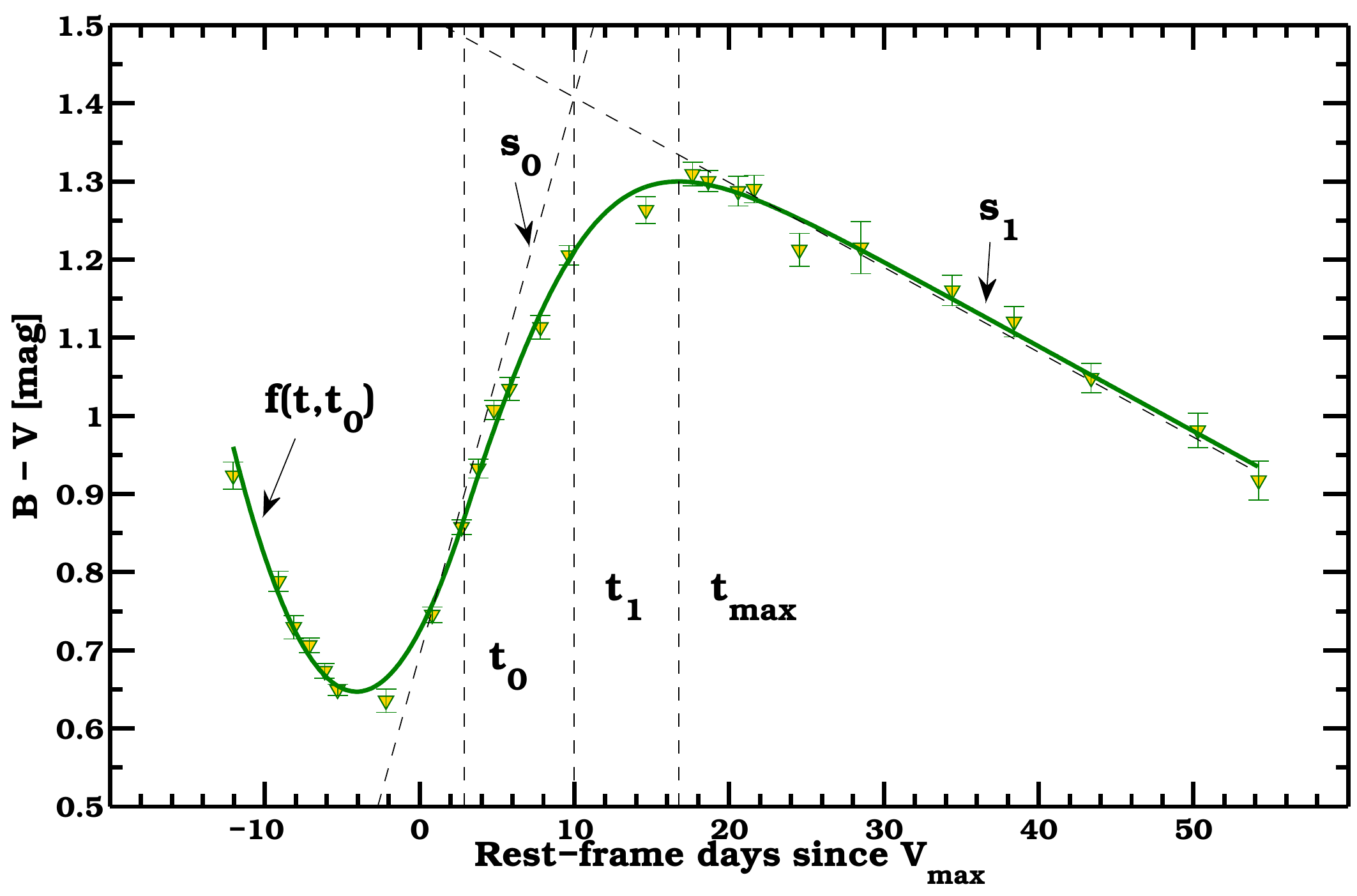}
\caption[]{{\em (Top panel)} An example of the $B-V$ color curve of SN~2006T fit with the analytical function described by  Eq.~\ref{eq:gifit}.\label{fig:colorfit}}
\end{figure*}

\clearpage
\begin{figure*}[h]
\begin{center}$
\begin{array}{cc}
\includegraphics[width=9cm,]{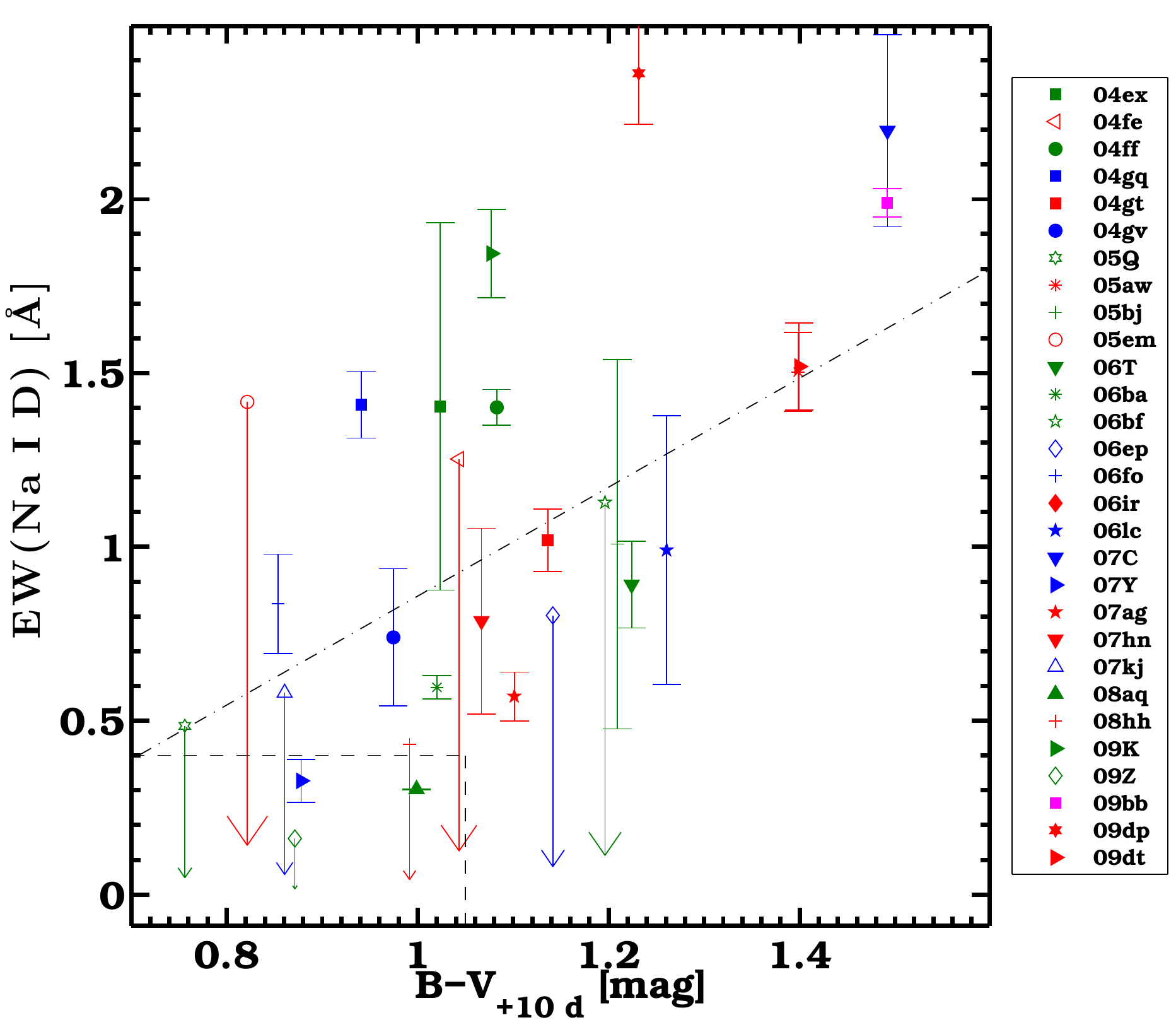}&
\includegraphics[width=9cm]{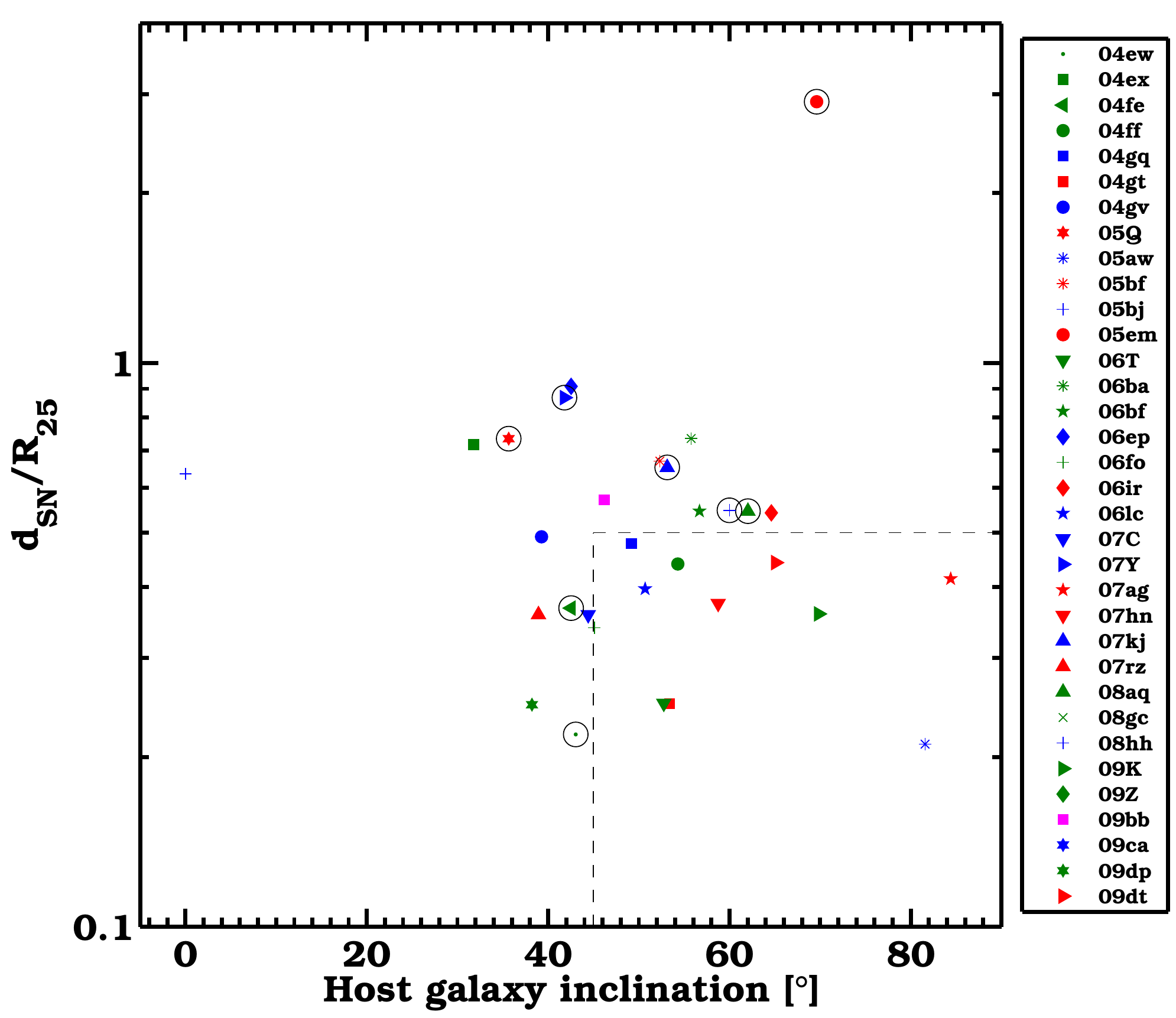}\\
\end{array}$
\end{center}
\caption[]{{\em (Left panel)} \ion{Na}{i}~D EW versus the galactic-extinction corrected $B-V$ at $+$10d for the CSP sample of SE SN. Arrows indicate upper limits. The best linear fit is shown by a dashed-dotted line. The area in the low-left corner contained within the borders of the square dashed box includes 9 objects (SN~2004aw is not reported because it was not observed at +10d but its color within +20d is as blue as that of the other 8 events) whose host extinction is considered to be negligible. {\em (Right panel)} De-projected and normalized SN offset from the host center versus host galaxy inclination. The objects suspected of suffering low extinction are circled in black, and they never appear in both strongly inclined galaxies and at locations close to their host center (i.e., they always sit outside the box delimited by dashed lines in the figure). Symbols are color coded based on the different sub-types: green, blue, red and magenta for SNe~IIb, Ib, Ic and Ic-BL, respectively. \label{fig:NaID_bmv}}
\end{figure*}

\clearpage
\begin{figure*}
\centering
$\begin{array}{cc}
\includegraphics[width=9cm]{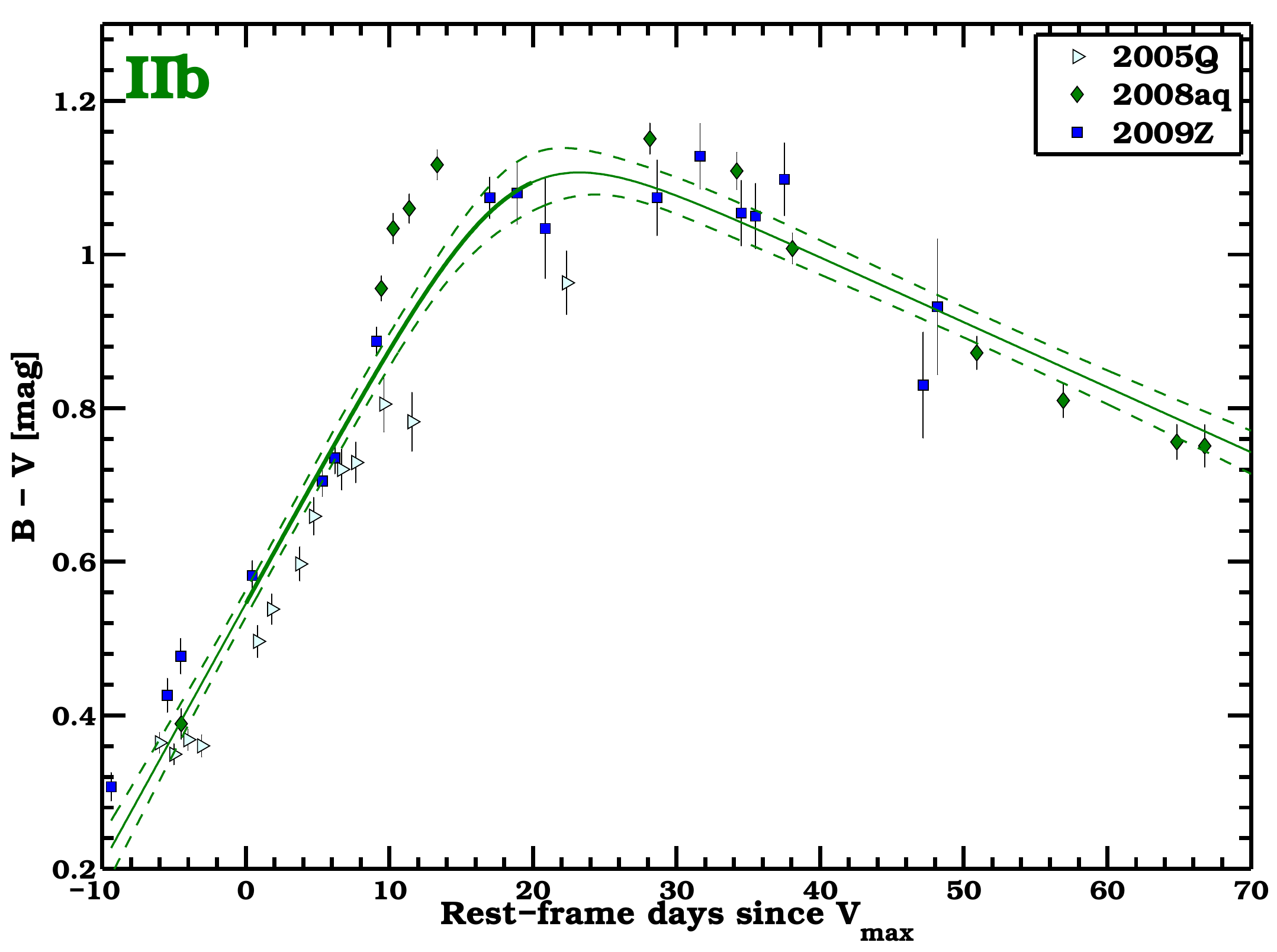}&
\includegraphics[width=9cm]{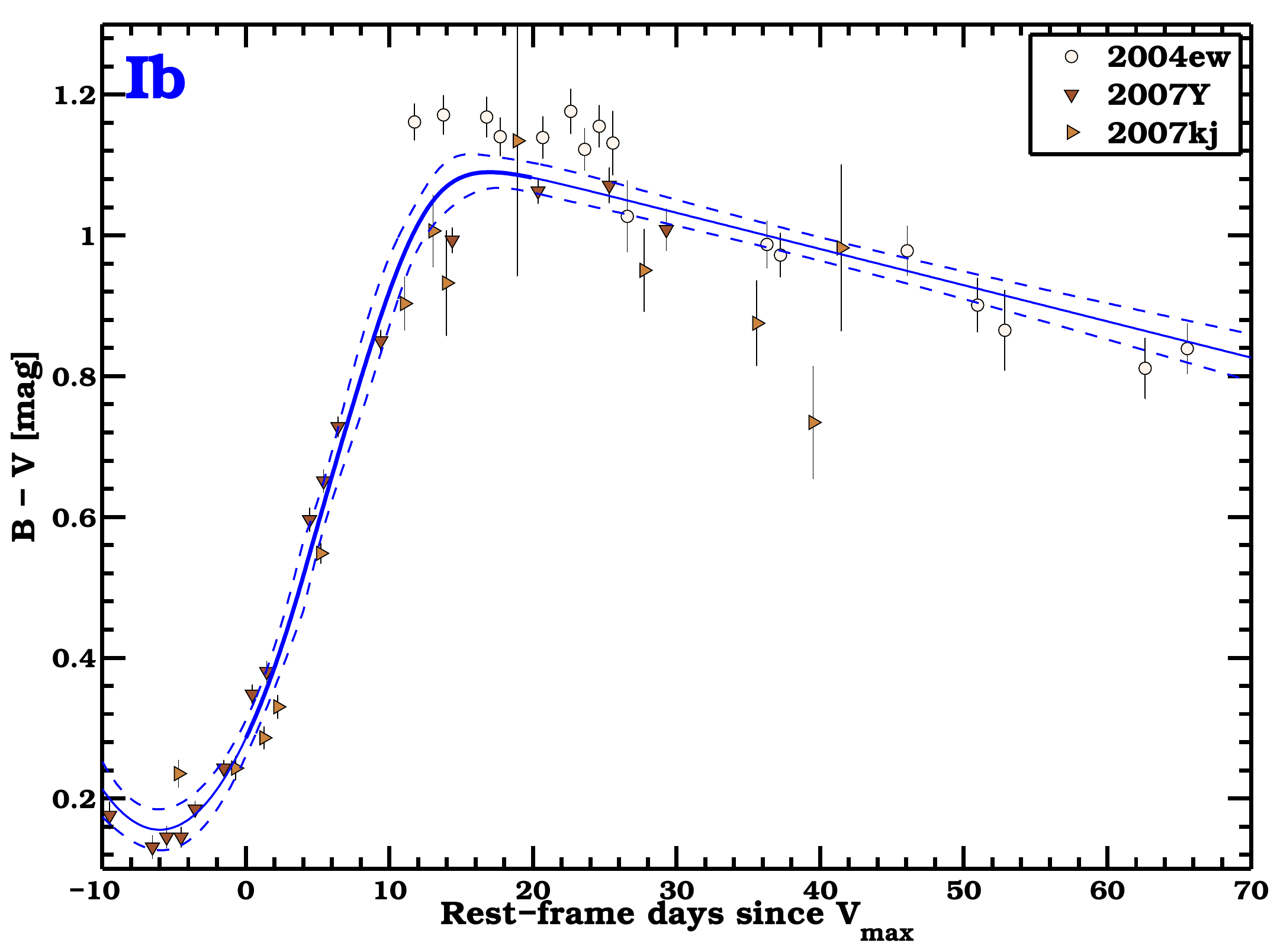}\\
\includegraphics[width=9cm]{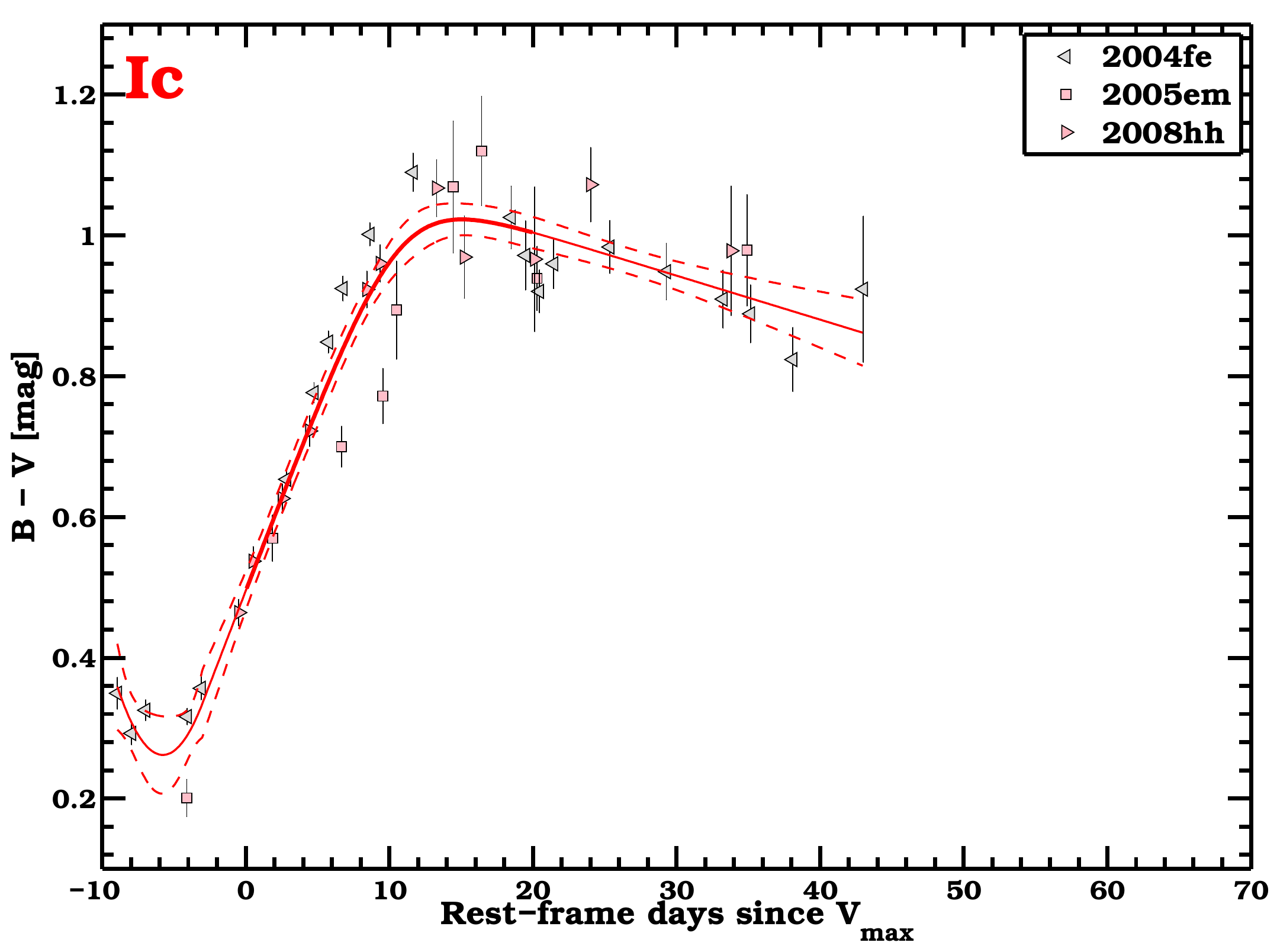}&
\includegraphics[width=9cm]{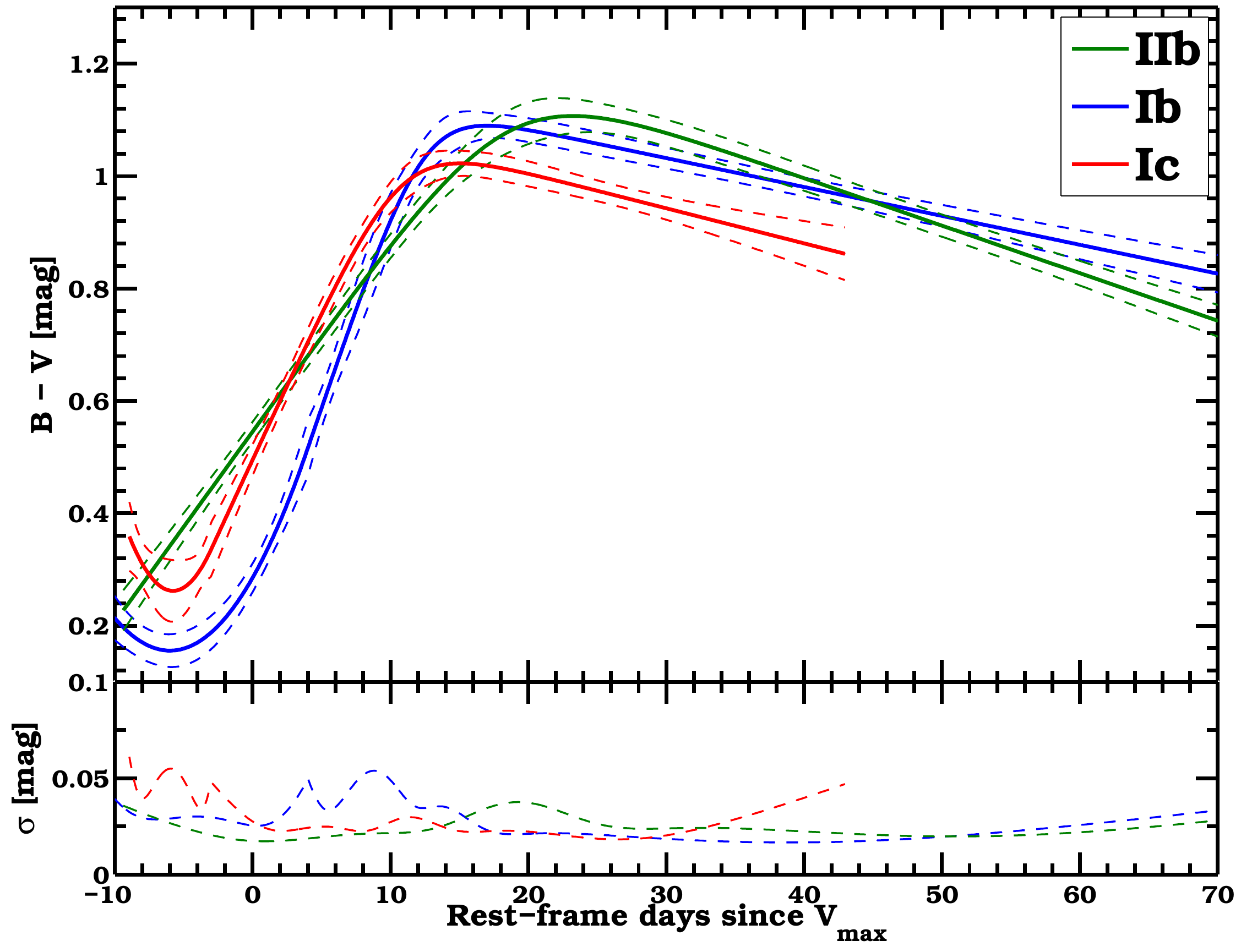}\\
\end{array}$
\caption{Intrinsic $B-V$  color-curve templates for SNe~IIb, Ib and Ic, as obtained from the best fit (solid lines) of Eq.~\ref{eq:gifit} to the observed colors of nine minimally-reddened  objects. Dashed lines are the associated 1$\sigma$ uncertainties. The three templates are over-plotted in the bottom-right panel, showing that there are differences among the SE SN sub-types at all phases. The template uncertainties (highlighted in the bottom-right sub-panel) are lower in the first month after maximum, and then increase over time. \label{fig:templates}}
\end{figure*}

\clearpage
\begin{figure*}
\centering
\includegraphics[width=12cm]{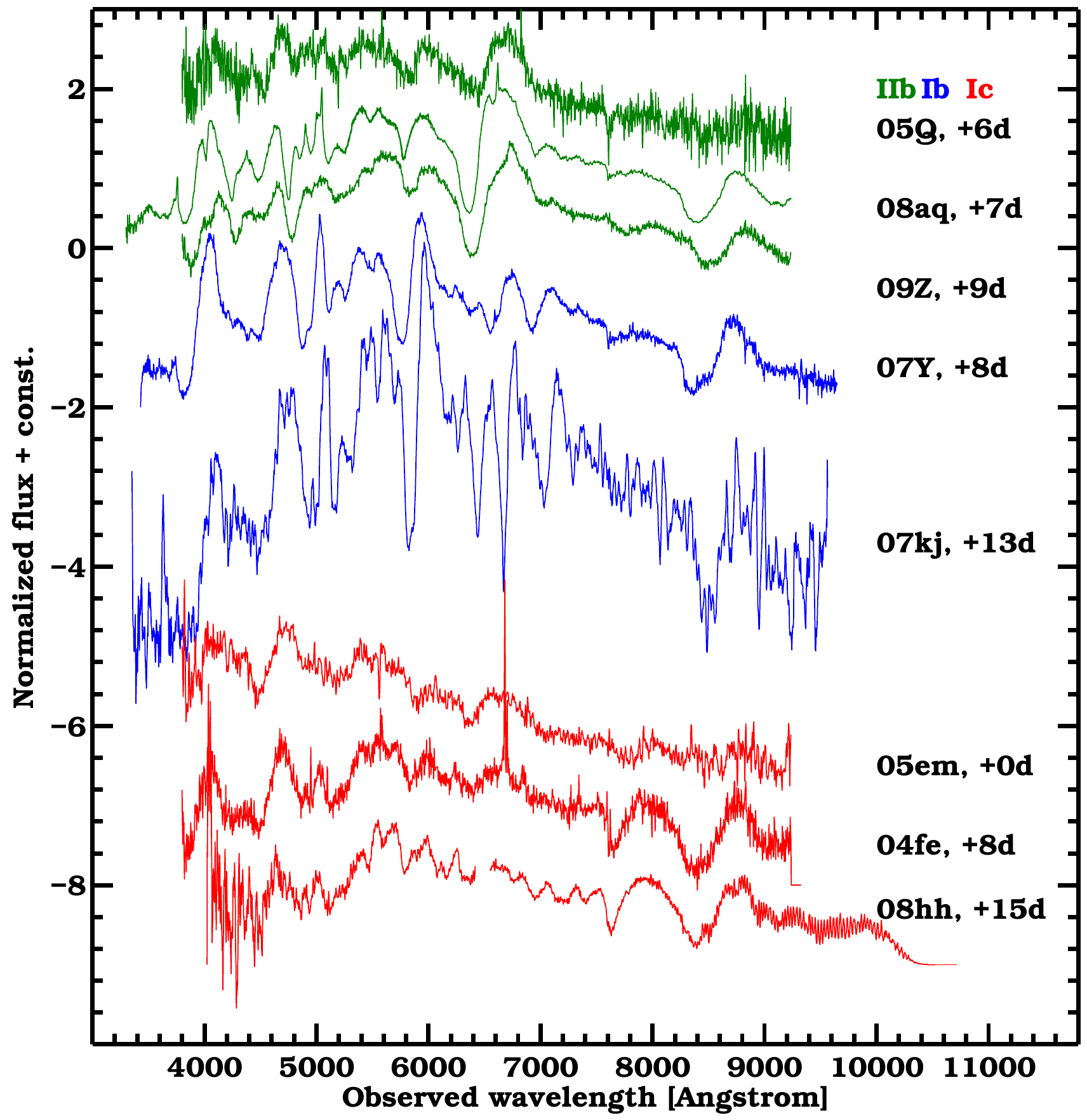}
\caption{Post $V$-band maximum  visual-wavelength spectroscopy of eight of the nine minimally-reddened objects identified in the CSP-I SE SNe sample. Overall the spectra are similar for each individual sub-type, through  differences between line strengths and line ratios are apparent, particular at the location of H$\alpha$ and the \ion{He}{i} features which evolve considerably as a function of phase. \label{fig:spectra_unreddened}}
\end{figure*}

\clearpage
\begin{figure*}
\centering
\includegraphics[width=16cm]{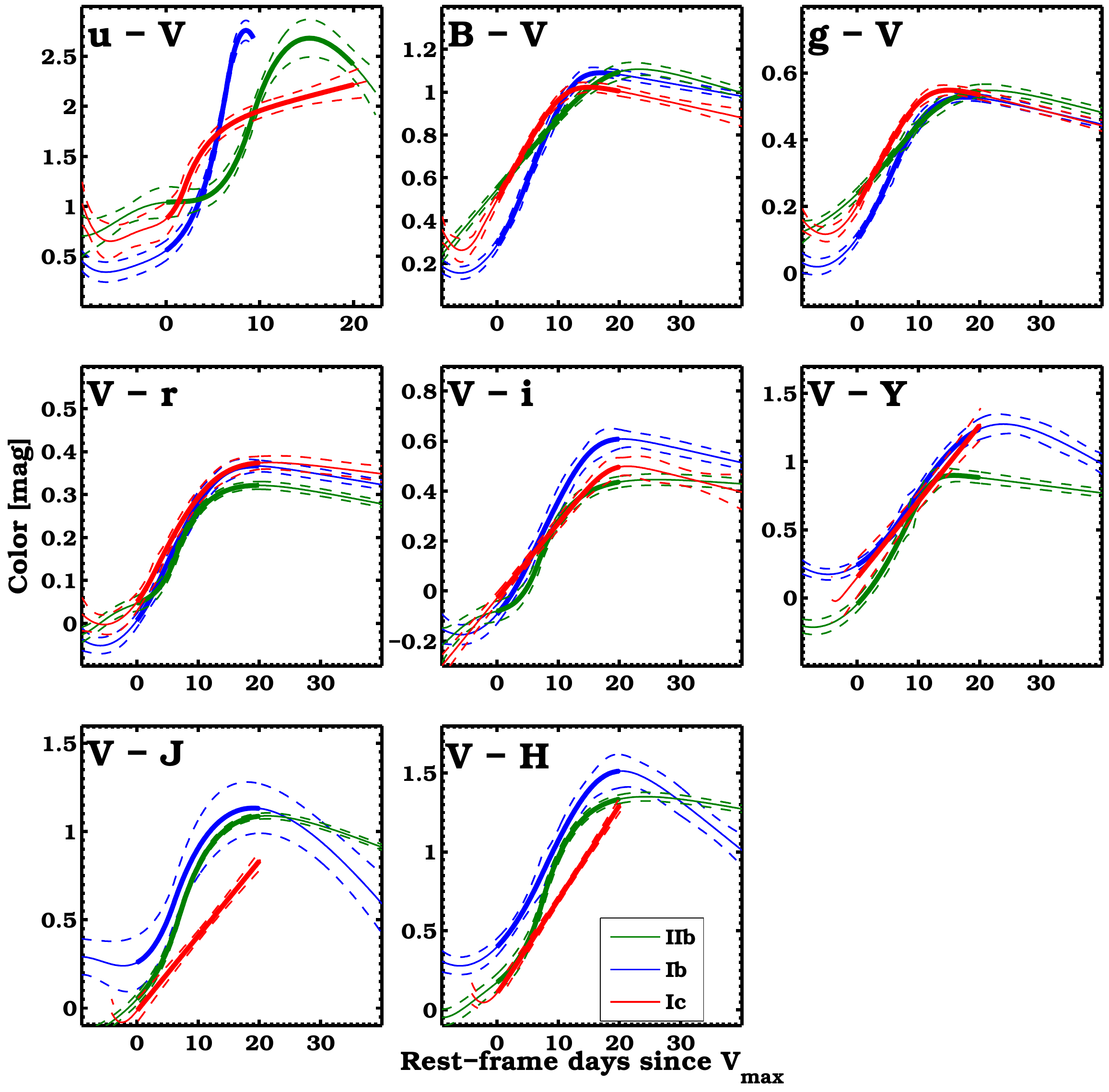}
\caption{Intrinsic color-curves templates for SNe~IIb, Ib, and Ic for eight different color combinations. Each template is constructed by averaging the best fit of Eq.~\ref{eq:gifit} to the observed colors of minimally-reddened  SE SNe. Dashed lines are the associated 1$\sigma$ uncertainties. 
Note that the  thick portion of the color curves corresponds to the epochs used to infer color excesses. \label{fig:all_templates}}
\end{figure*}

\clearpage
\begin{figure*}
\centering
$\begin{array}{cc}
\includegraphics[height=6cm]{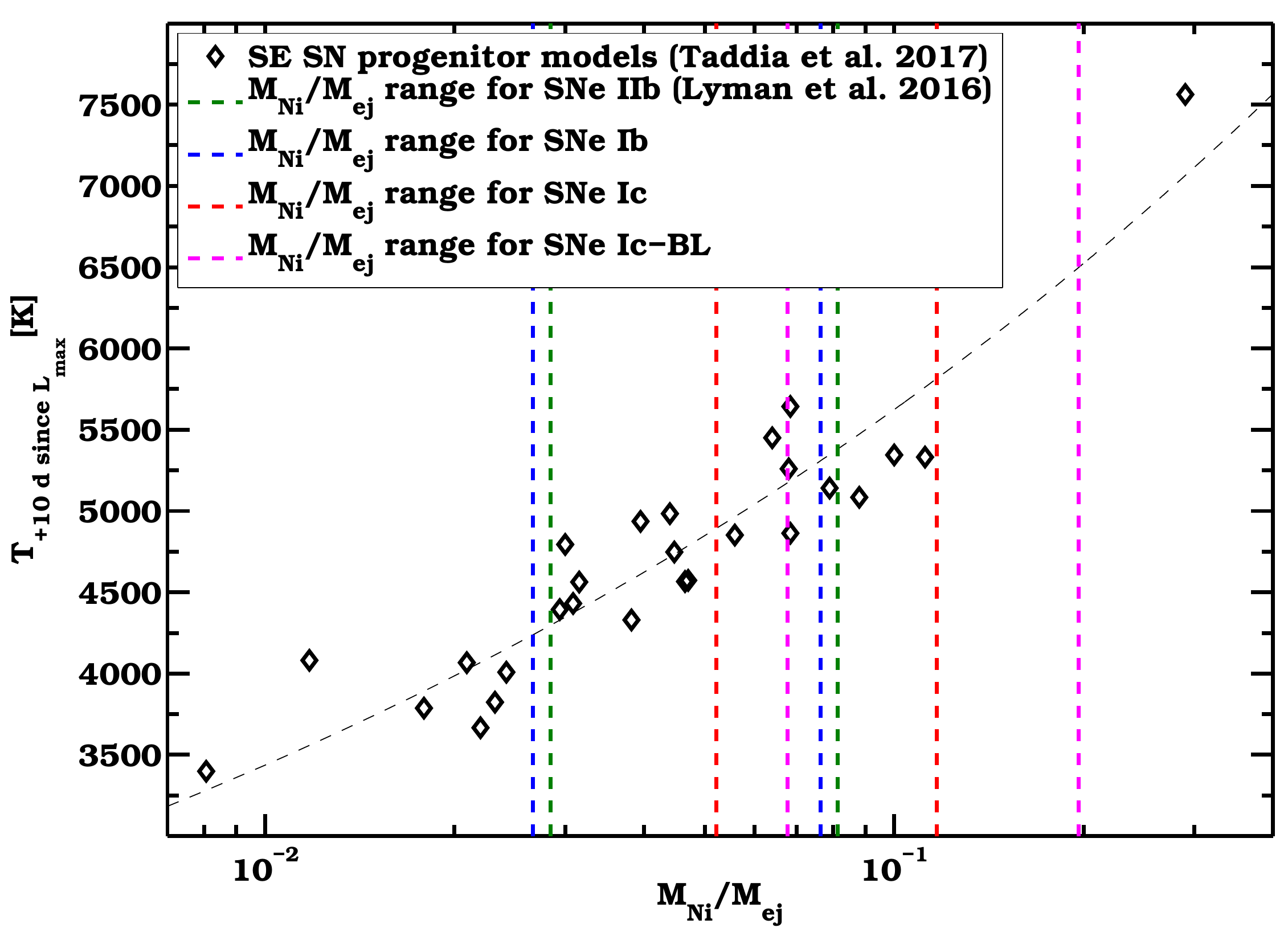}&
\includegraphics[height=6cm]{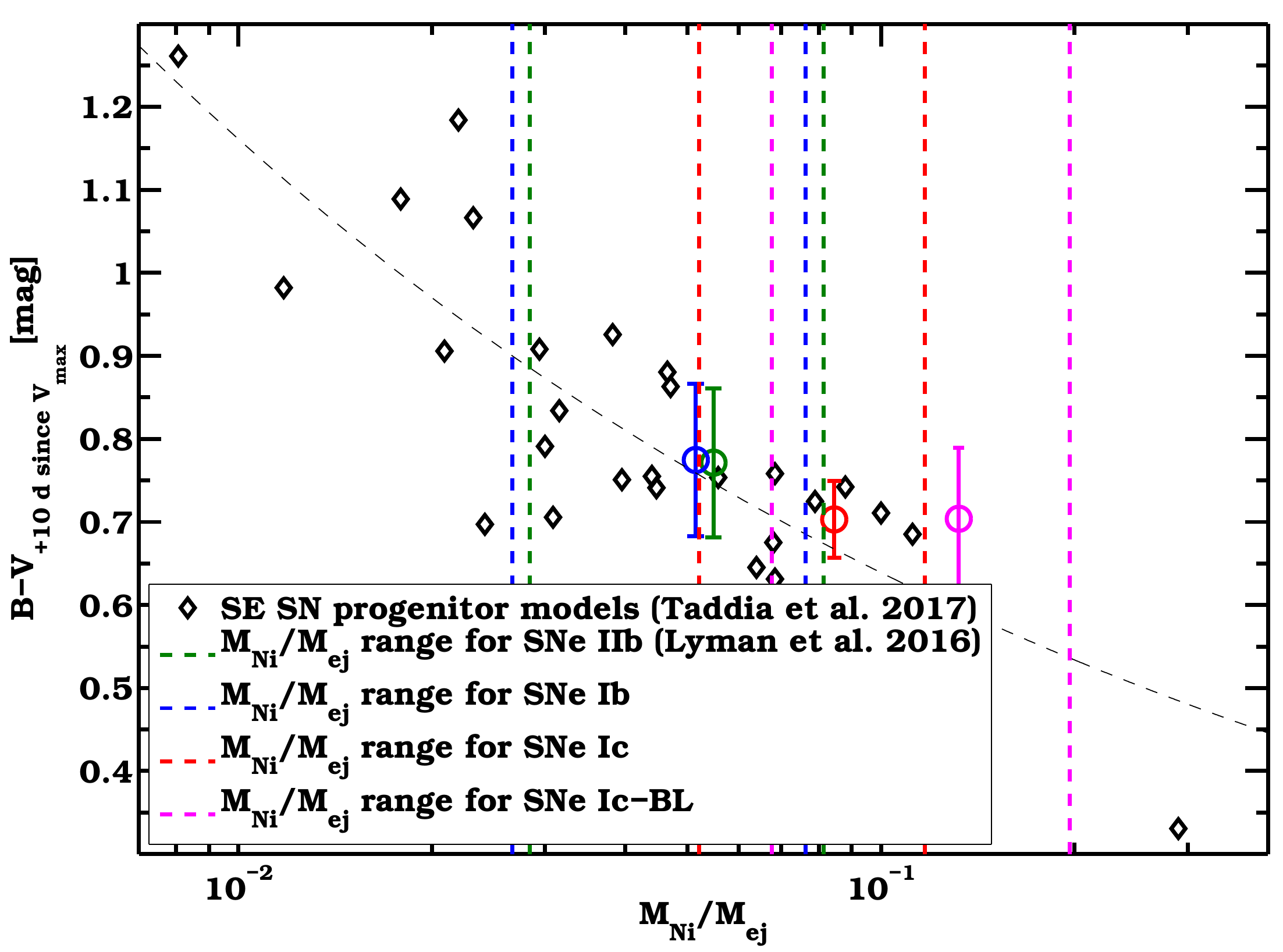}
\end{array}$
\caption{\label{fig:explanation_drout}{\em (Left panel)}  Plotted as diamonds is the temperature at $+$10d calculated from an extended set of hydrodynamical explosion models that best-fit the bolometric light curves of our sample \citep[see][]{taddia17}  vs. the ratio of the $^{56}$Ni mass and $M_{ej}$ for each model.
The  models indicate  SE SNe with higher $^{56}$Ni to $M_{ej}$ ratios  tend to exhibit higher temperatures, and the correlation between these parameters is well described by a power-law  fit (black dashed line). 
Dashed vertical lines  indicate the range of the $^{56}$Ni mass to  $M_{ej}$ ratio  inferred  by \citet{lyman16} based on a large observational sample of SE SNe.
The range of the $^{56}$Ni mass to  $M_{ej}$ ratio for the each SE SN sub-type is relatively narrow, and implies a narrow range of $\sim$ 1000 degrees in temperature at $+$10d. 
{\em (Right panel)} Inferred $B-V$ color at $+$10d determined from hydrodynamical models, plotted versus  the ratio between $^{56}$Ni mass and $M_{ej}$. 
The parameters are fit with a power law function plotted as black dashed line.
Dashed vertical lines indicate the range in parameter space inferred by \citet{lyman16} for the various SE SN sub-types, while color points are mean values  obtained from the explosion models contained within each of the  indicated regions with errorbars  corresponding to 1$\sigma$ uncertainty. This plot indicates a rather narrow  range of intrinsic $B-V$ colors are expected for the span of 
  $^{56}$Ni to $M_{ej}$ ratios observed in the various SE SN sub-types.
  $B-V$ color uncertainties are found to be lower than 0.1 mag for each SE SN sub-type.}
\end{figure*}

\clearpage
\begin{figure*}
\centering
\includegraphics[width=16cm]{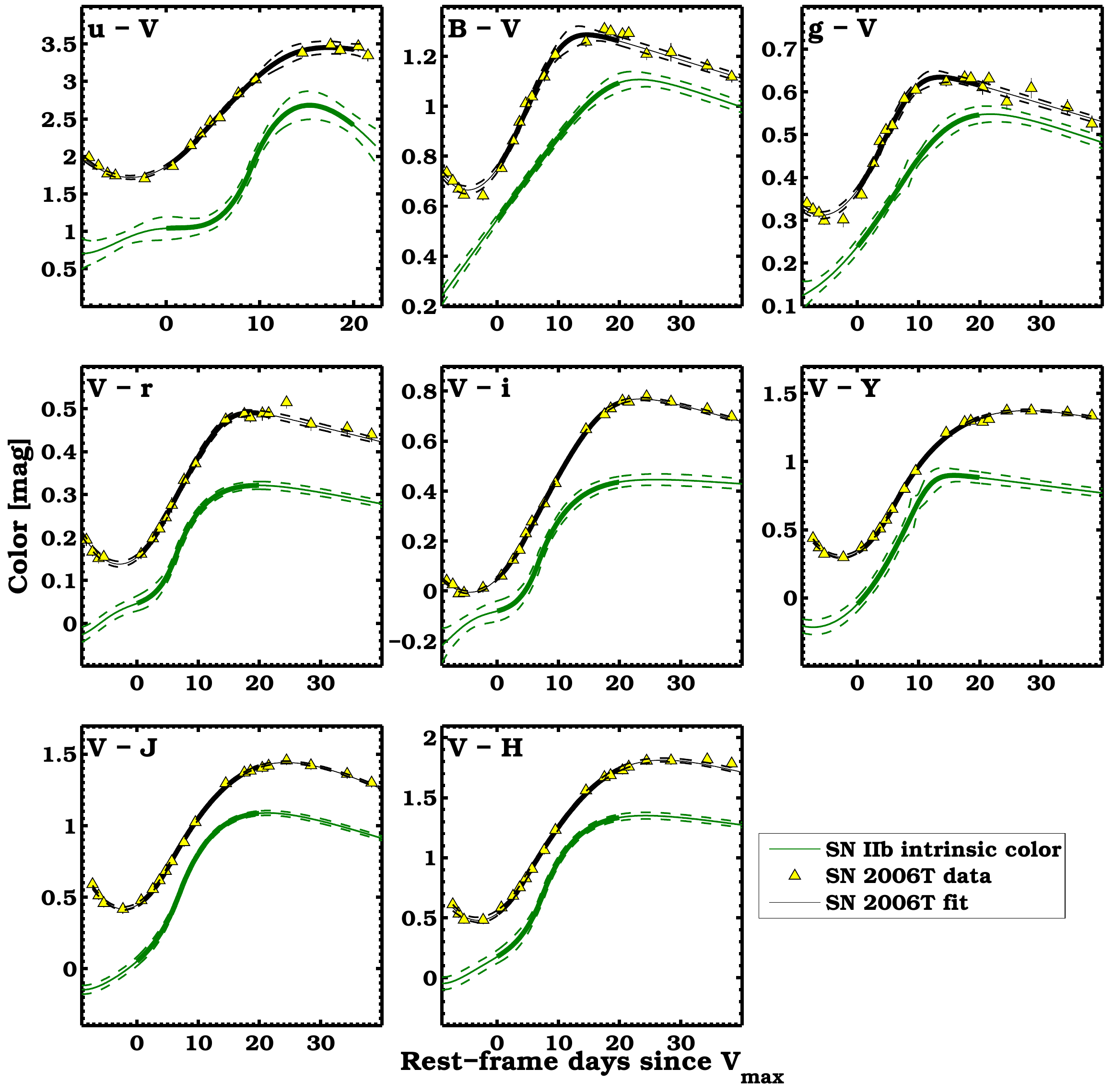}
\caption{Observed color curves (yellow triangles) of the Type~IIb SN~2006T  compared to the SNe~IIb intrinsic color-curves templates (green lines) for eight different color combinations. 
The color excesses are computed for  the various color combination by taking the difference between the 
the observed and  intrinsic colors  (black lines) between 0d and $+$20d (thick lines). \label{fig:get_color_excess_06T}}
\end{figure*}

\clearpage
\begin{figure*}
\centering
\includegraphics[width=14.0cm]{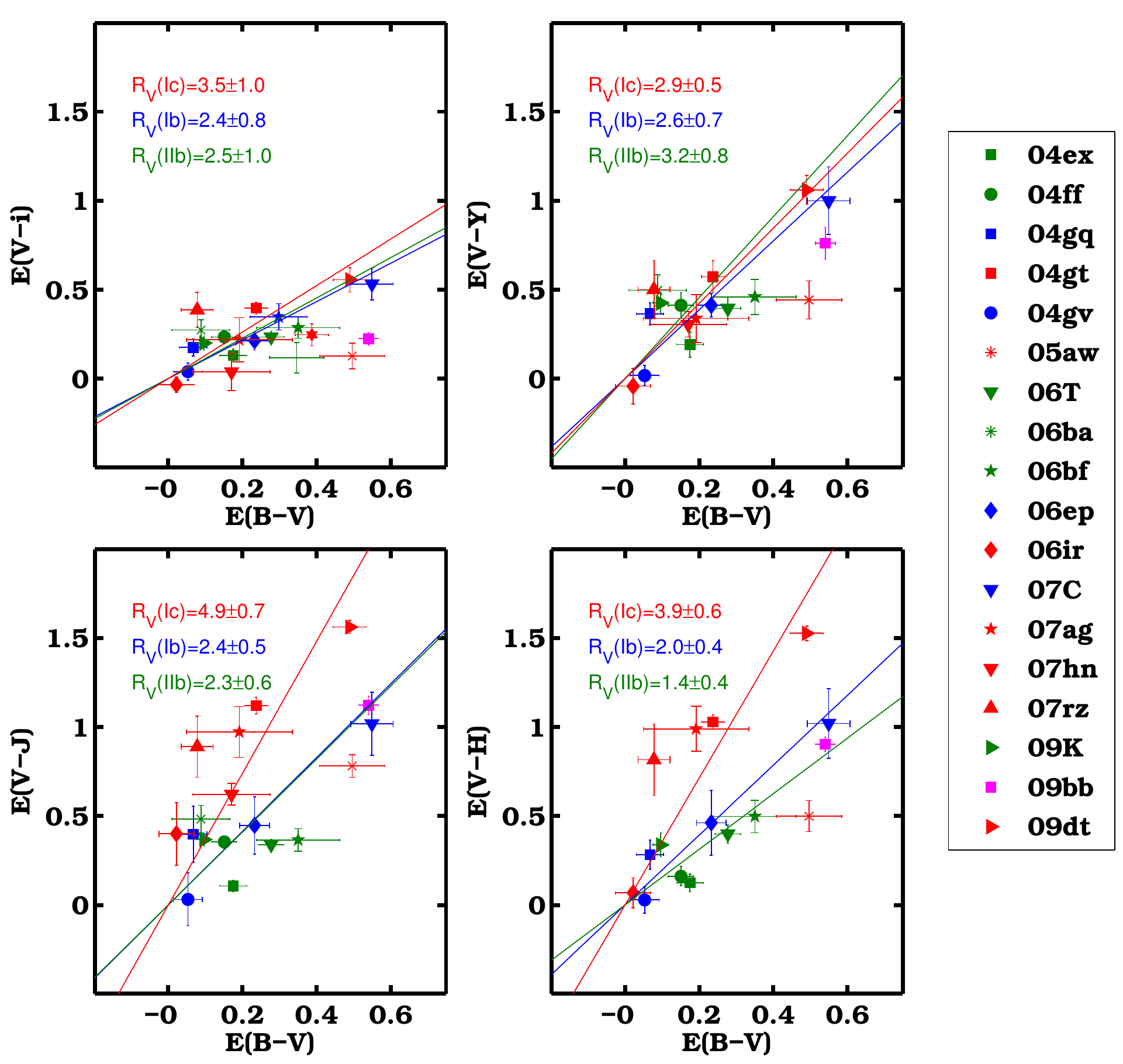}
\caption[]{Comparison of $E(B-V)_{host}$ color excesses with $E(V-X)_{host}$ (for X = $i$, $Y$, $J$, $H$) color excesses  of 18 reddened SE SN. 
Color excess values are computed by taking the difference between the intrinsic color-curve templates and the observed colors of the reddened SE SNe from 0d and $+$20d.
Over-plotted as solid lines are the best $R_V^{host}$ fits to each of the SE SN sub-types. 
Symbols and lines are color coded with green, blue, red and magenta corresponding to SNe~IIb, Ib, Ic and Ic-BL, respectively. \label{fig:evmx_evmx}}
\end{figure*}

\clearpage
\begin{figure*}[h]
\centering
\includegraphics[width=13cm]{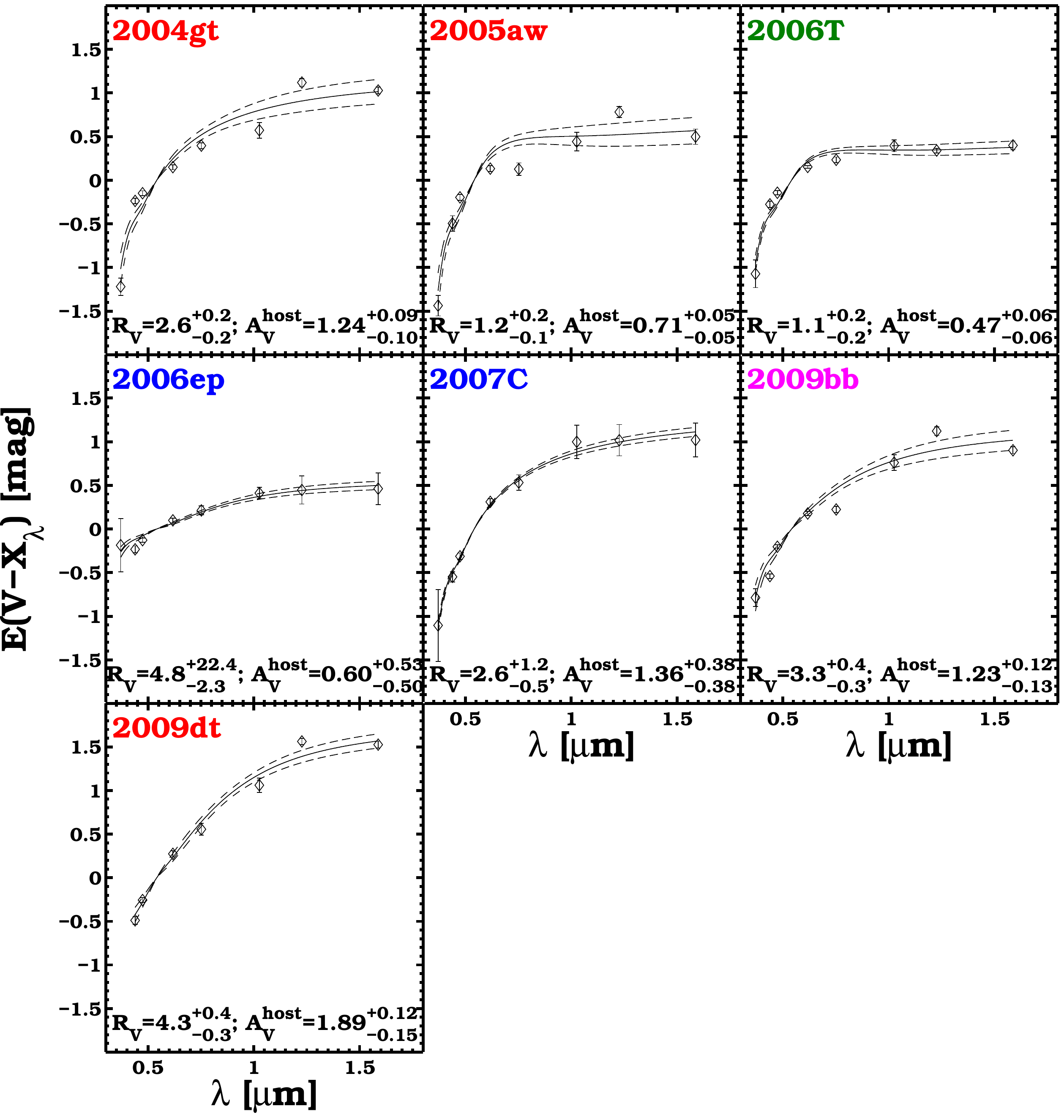}
\caption[]{$E(V-X_{\lambda})_{host}$ as a function of wavelength for seven objects whose $E(V-B)_{host} < -0.20$~mag and with 8 different $E(V-X_{\lambda})_{host}$ combinations extending from $E(V-u)_{host}$ to $E(V-H)_{host}$ (except for SN~2009dt where $u-V$ is missing).  Diamonds correspond to $E(V-X_{\lambda})_{host}$ values obtained by comparison with the intrinsic color-curve templates for each object's spectroscopic sub-type. Solid lines represent the best reddening law fit to 
each series of data and the corresponding  1$\sigma$ uncertainty of the fit is indicated with dash lines. 
The SN names are color coded with respect to their spectroscopic sub-type, i.e. SNe~IIb in green, SNe~Ib in blue, SNe~Ic in red, and SNe~Ic-BL in magenta.}
\label{fig:EVMX_vs_lambda_onlyRV}
\end{figure*}

\clearpage
\begin{figure*}[h]
\centering
\includegraphics[width=16cm]{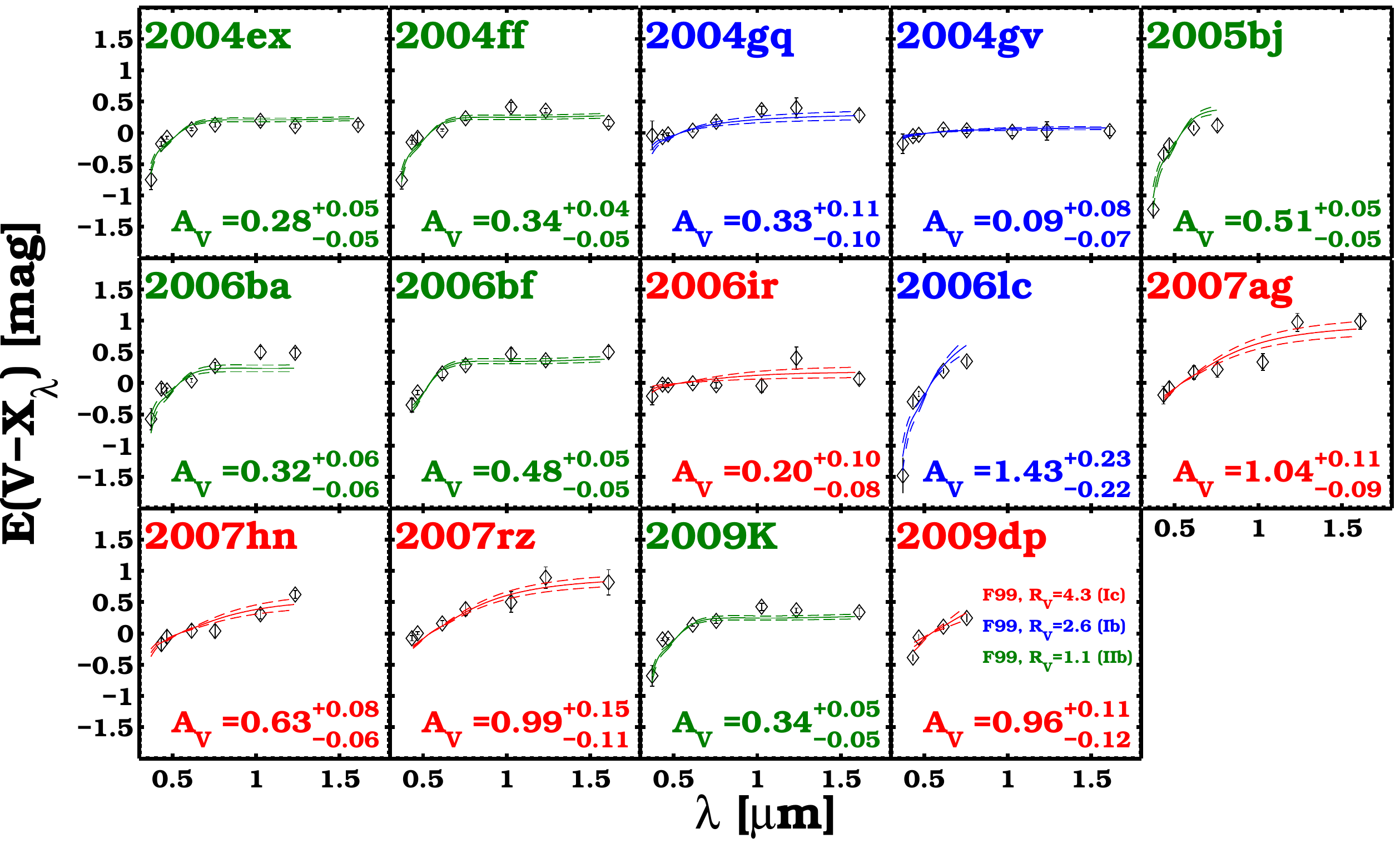}
\caption[]{{$E(V-X_{\lambda})_{host}$ plotted as a function of wavelength for 14 SE SNe  affected by $0<E(B-V)_{host}<0.2$ or with a limited set of $E(V-X_{\lambda})_{host}$ combinations. Note that the nine minimally-reddened objects are not included and that due to limited photometric coverage,  SN~2006fo, SN~2008gc, and SN~2009ca 
are also omitted. Black diamonds indicate the $E(V-X_{\lambda})_{host}$ color excess values obtained by taking the difference between the intrinsic color-curve template and the observed  colors of each  reddened SE SN from 0d to $+$20d.  
Solid lines correspond to the best reddening law fit to each series of data and the corresponding 1$\sigma$ uncertainty of the fit is indicated with dashed lines. 
Shown in red is the best fit for $R_V^{host} = 4.3$, which is assumed for SNe~Ic. Blue lines correspond to the best fit for $R_V^{host} = 2.6$, which is assumed for SNe~Ib.  Green lines correspond to the best fit for $R_V^{host} = 1.1$, which is assumed for SNe~IIb.  The corresponding best-fit $A_V^{host}$ values are reported in each sub-panel. The SNe names are color coded with respect to their spectroscopic sub-type, i.e, SNe~IIb in green, SNe~Ib in blue, SNe~Ic in red, and SNe~Ic-BL in magenta.}\label{fig:EVMX_vs_lambda}}
\end{figure*}

\clearpage
\begin{figure*}[h]
\centering
\includegraphics[width=14cm]{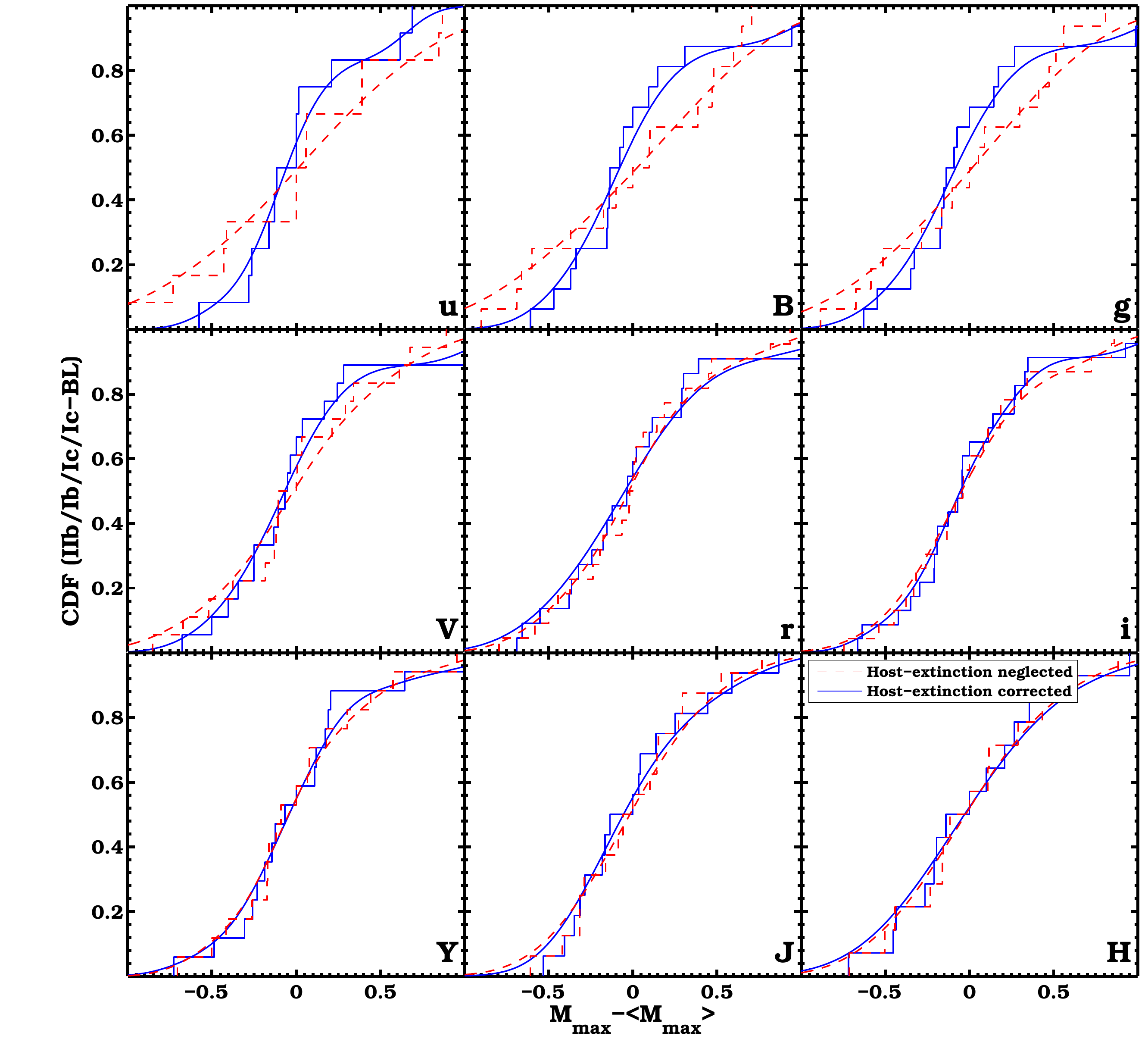}
\caption[]{Cumulative distribution of peak absolute magnitudes of the CSP-I SE SN sample with (blue) and without (dashed red) host-extinction corrections. The actual distributions are shown by segmented lines, their best fit assuming kernel distributions are shown by curves of the same color.
\label{fig:reduceddispersion}}
\end{figure*}

\clearpage
\begin{figure*}[h]
\centering
$\begin{array}{cc}
\includegraphics[width=9cm]{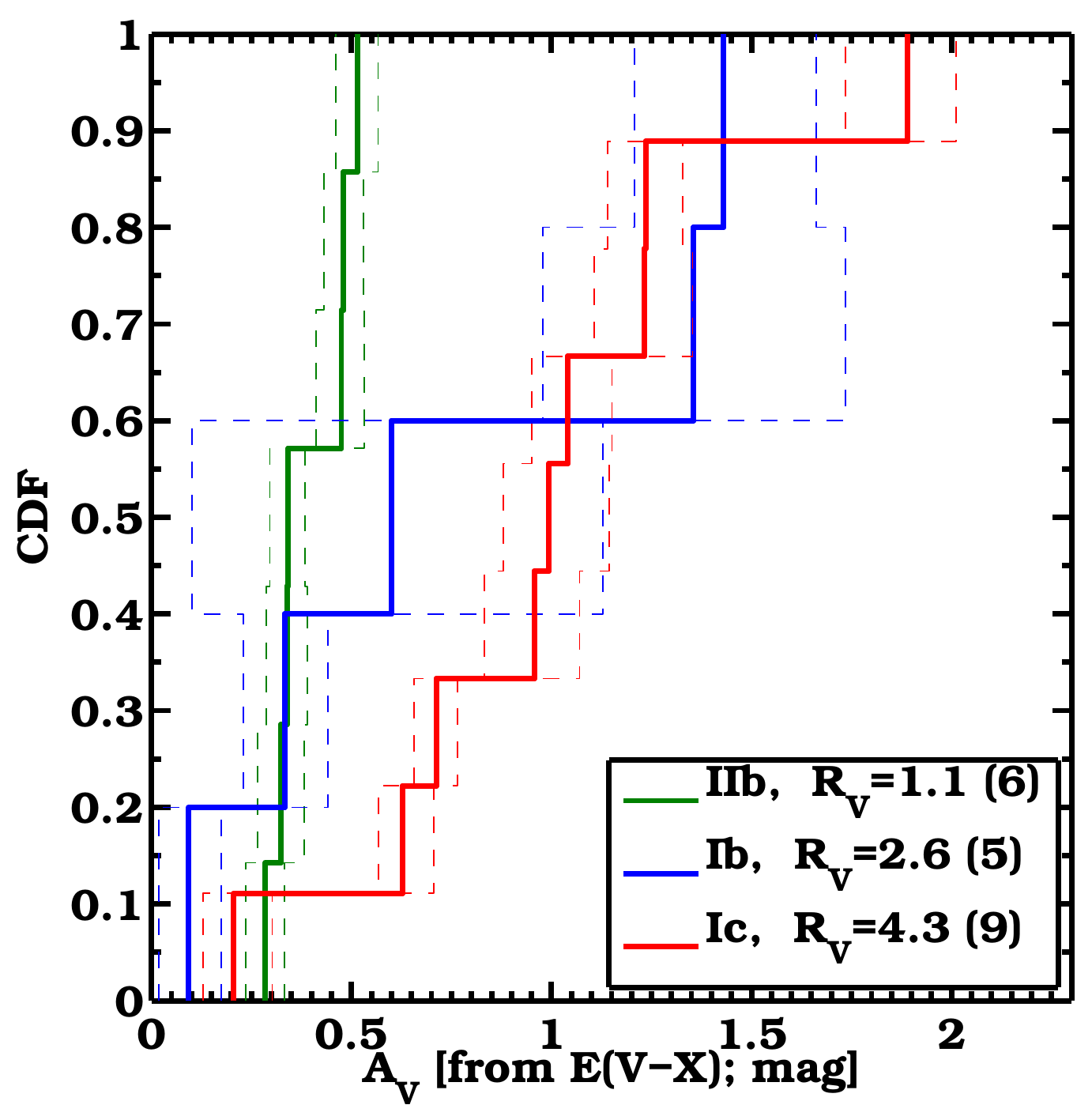}&
\includegraphics[width=9cm]{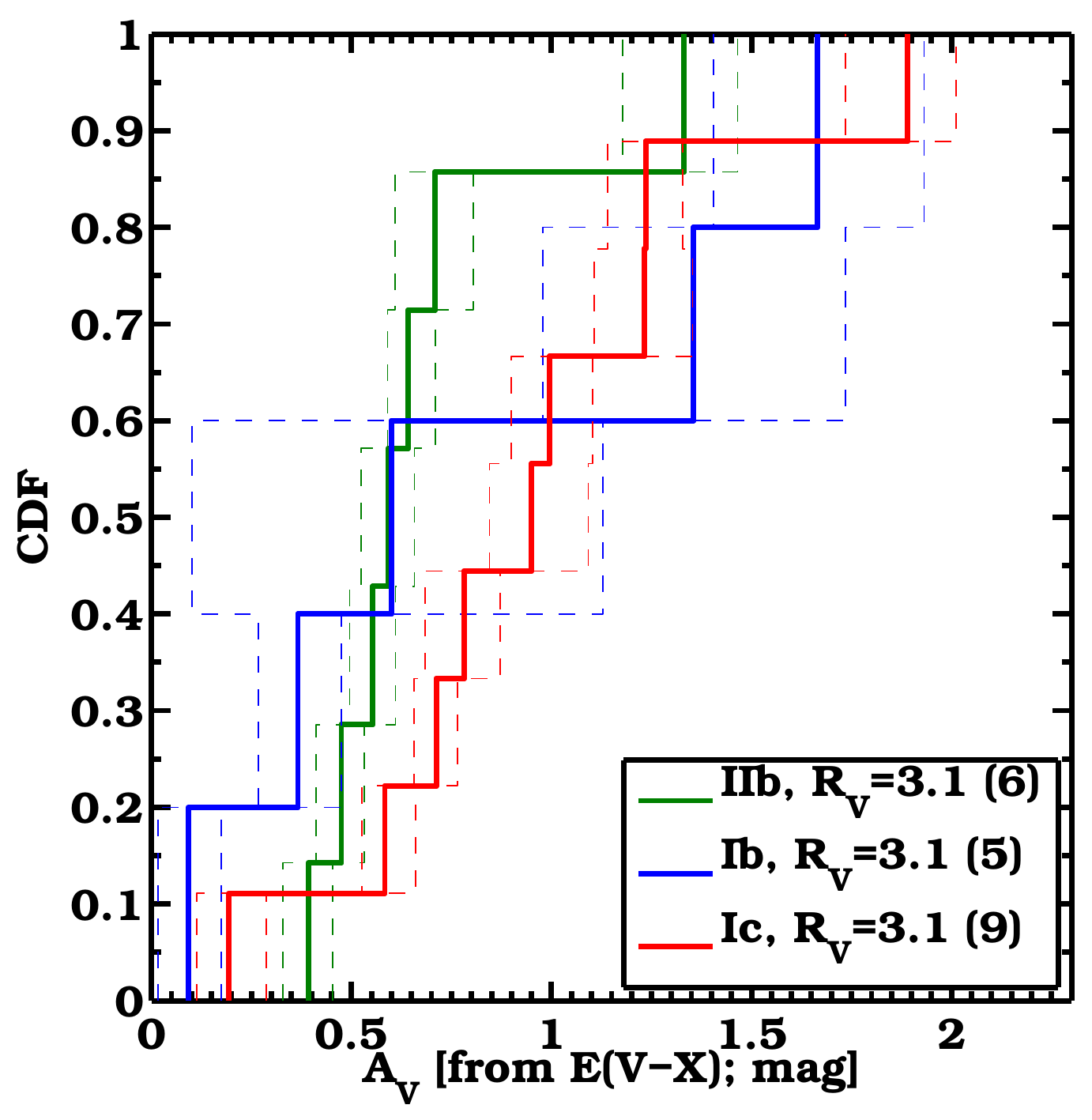}
\end{array}$
\caption[]{(Right-hand panel) Visual-extinction $A_V^{host}$ cumulative distributions for the different SE SN sub-types of the sample. Solid lines represent the $A_V^{host}$ values, dashed lines their uncertainties. The values of $A_V^{host}$ are those listed in Fig.~\ref{fig:EVMX_vs_lambda_onlyRV} and Fig.~\ref{fig:EVMX_vs_lambda}. SNe~Ic suffer more extinction than SNe~IIb and SNe~Ib. (Left-hand panel) Same as in the other panel, this time assuming $R_V=3.1$ for all the objects suffering low extinction. The differences between SNe IIb, Ib and Ic still hold, with SNe~Ic being the most reddened.  
\label{fig:AV}}
\end{figure*}

\clearpage
\begin{figure*}[h]
\centering
\includegraphics[width=16cm]{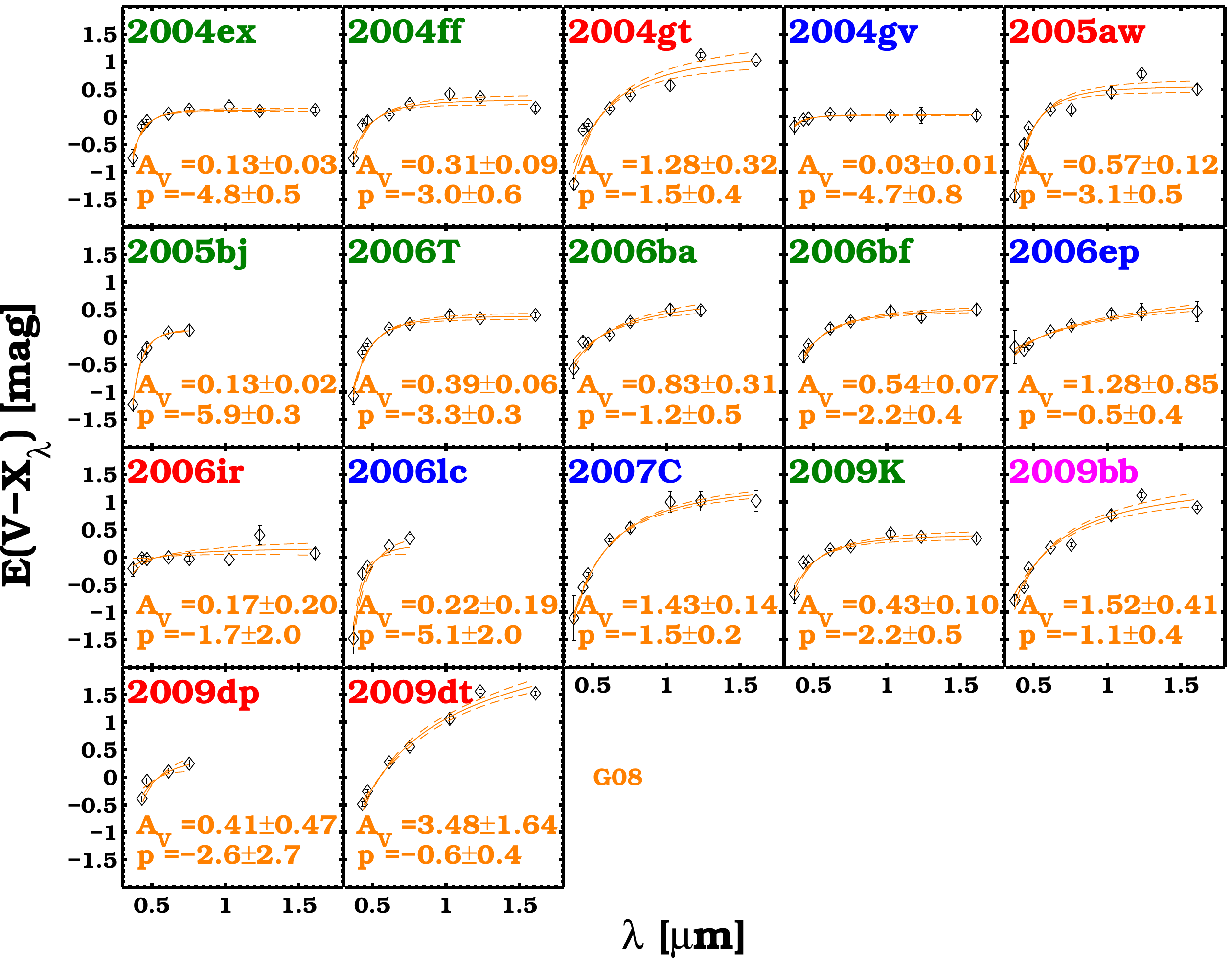}
\caption[]{{$E(V-X_{\lambda})_{host}$ plotted as a function of wavelength for 17 reddened SE SN.
The full sample was culled down to these objects due to omission of  the nine minimally-reddened objects, SN~2006fo, SN~2008gc, and SN~2009ca 
 due to poor photometric coverage, and also SN~2004gq, SN~2007hn, SN~2007ag, and SN~2007rz  as the G08 reddening law provides a poor fit. 
Black diamonds indicate the $E(V-X_{\lambda})_{host}$ color excess values obtained by minimizing the difference between the intrinsic color-curve template and the observed  colors of each  spectroscopic sub-type. 
Solid orange lines correspond to the best G08 reddening law fit to each series of data and the corresponding 1$\sigma$ uncertainty of the fit is indicated with dashed lines. The corresponding best-fit $A_V^{host}$ and $p$ values are reported in each sub-panel. The SN names are color coded with respect to their spectroscopic sub-type, i.e, SNe~IIb in green, SNe~Ib in blue, SNe~Ic in red, and SNe~Ic-BL in magenta.}\label{fig:goobar}}
\end{figure*}

\clearpage
\begin{figure*}[h]
\centering
$\begin{array}{cc}
\includegraphics[width=8cm]{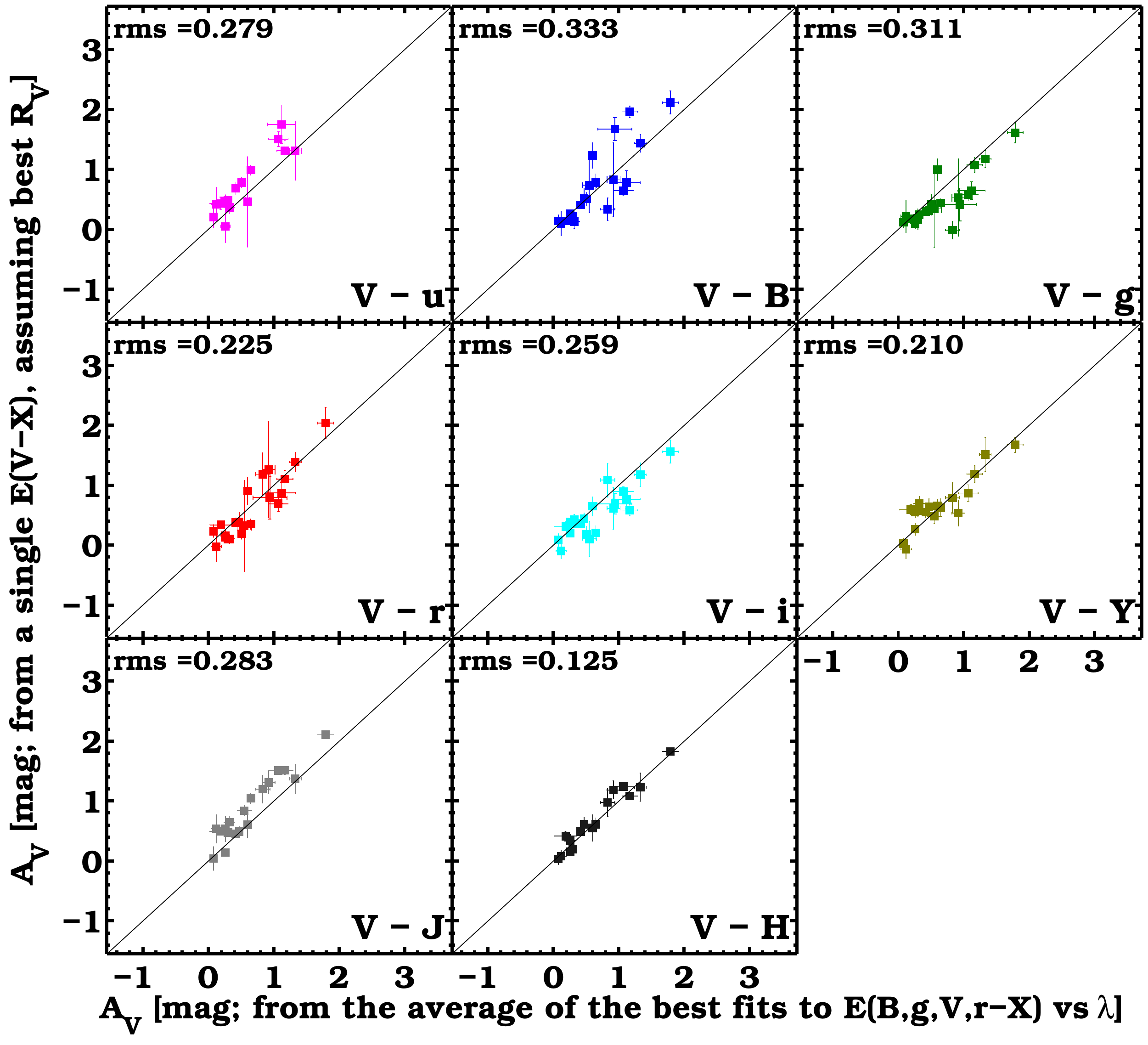}&
\includegraphics[width=8cm]{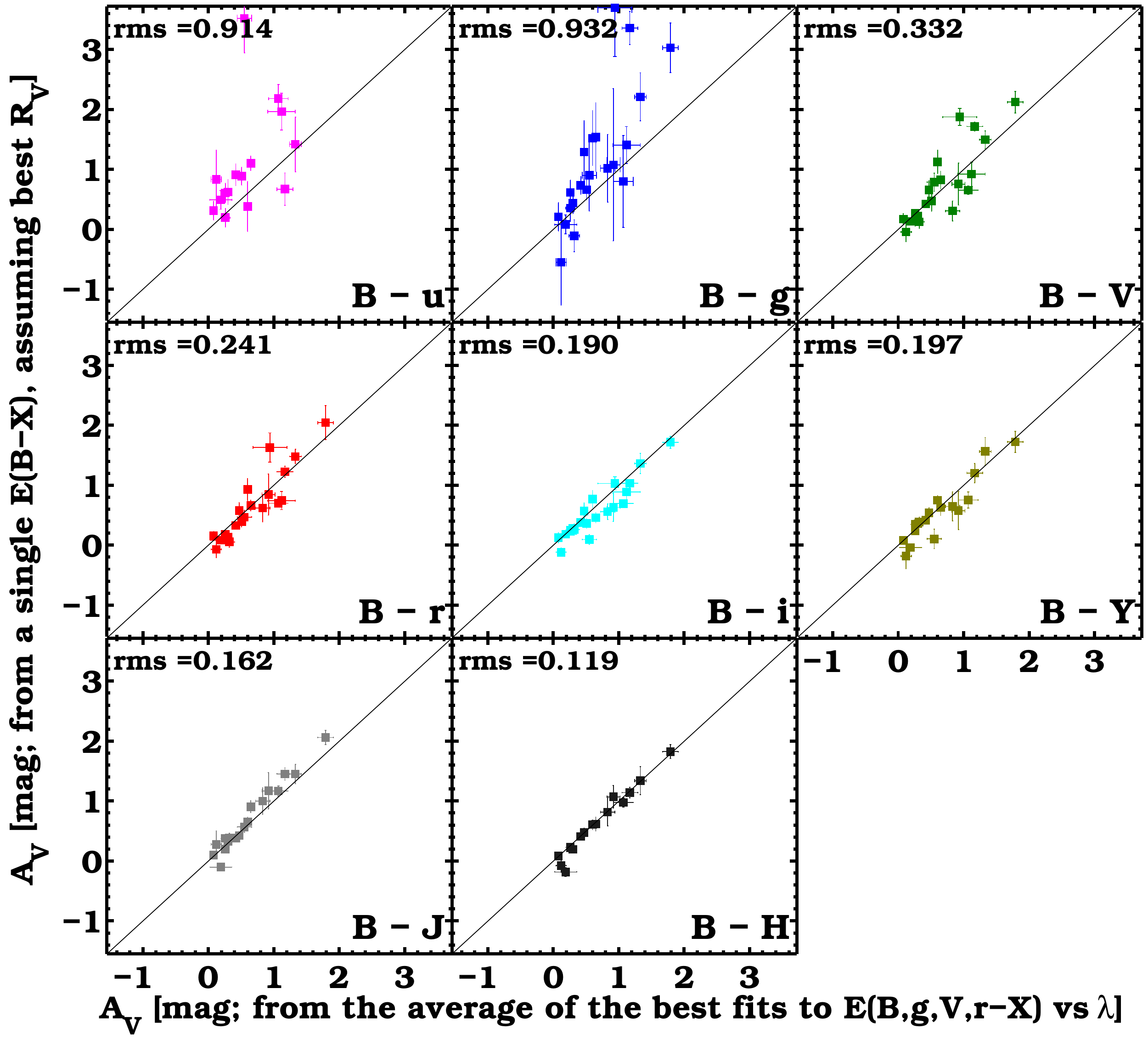}\\
\includegraphics[width=8cm]{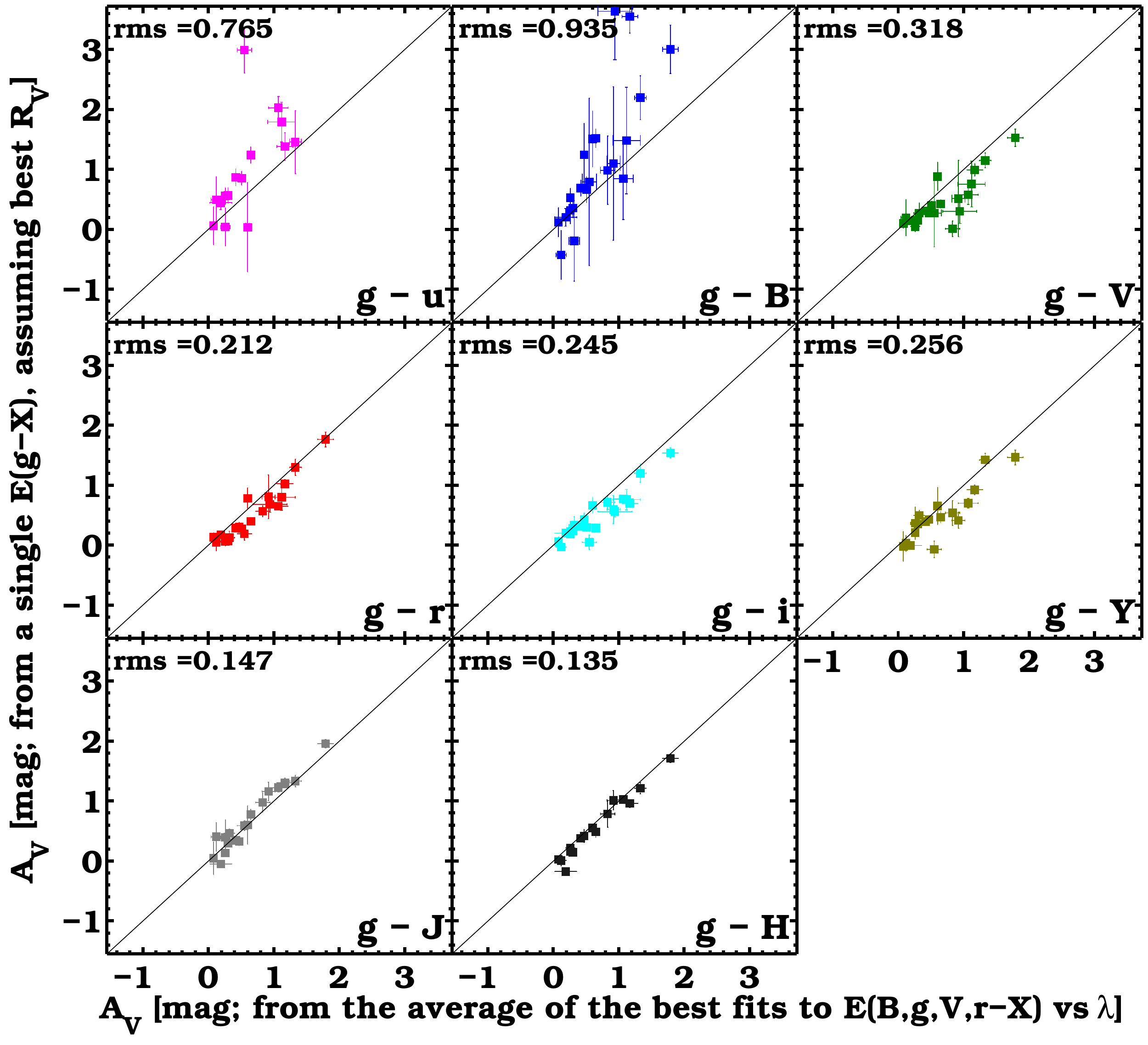}&
\includegraphics[width=8cm]{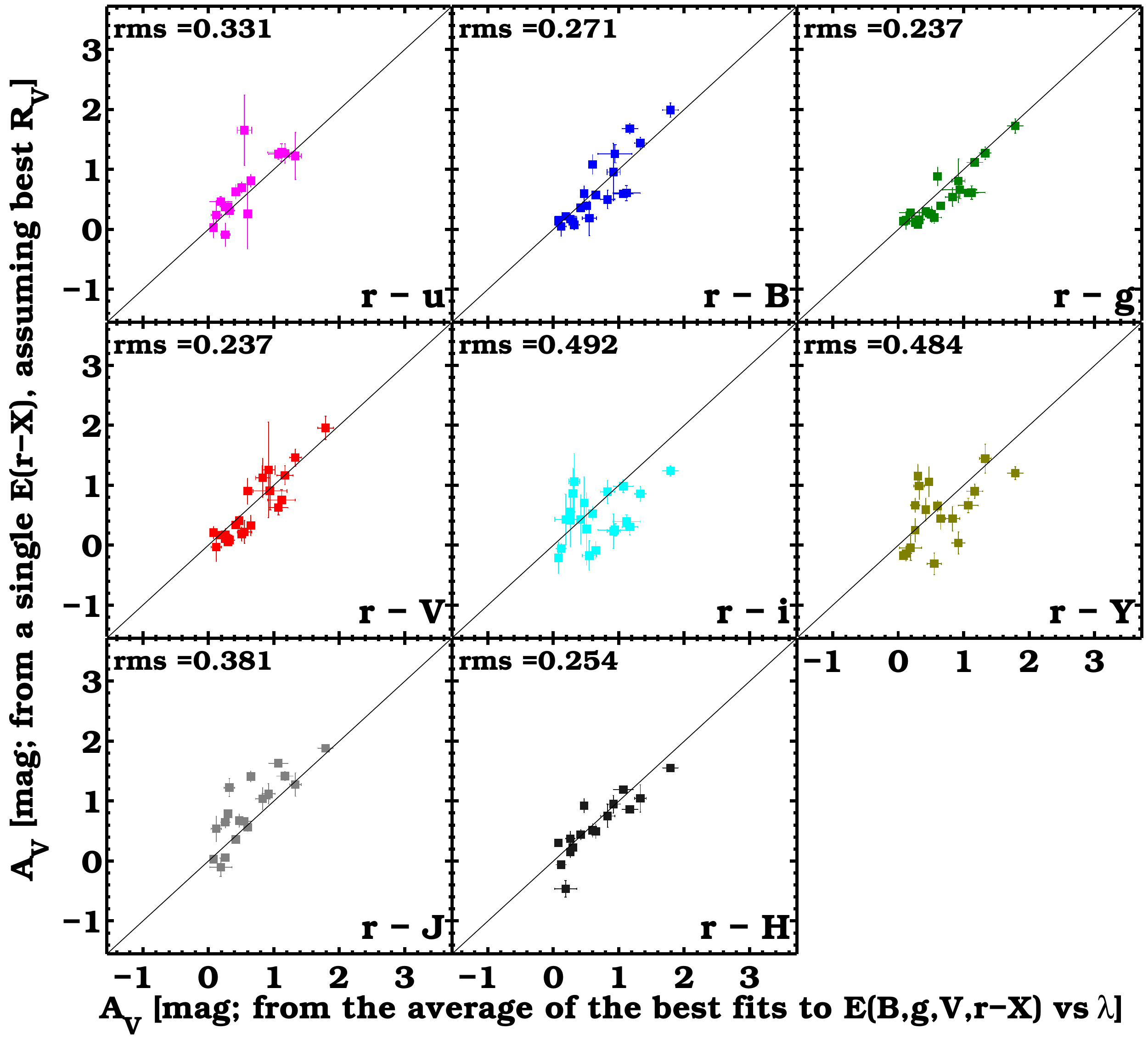}\\
\end{array}$
\caption[]{ $A_V^{host}$ values for 21 SE SNe inferred from the average of the best F99 fits to $E(V-X_{\lambda})_{host}$, $E(B-X_{\lambda})_{host}$, $E(g-X_{\lambda})_{host}$ and $E(r-X_{\lambda})$ as a function of wavelength (see last column in Tables~\ref{tab:veryred} and \ref{tab:EBMV})  vs. $A_V^{host}$ estimated from each $E(V, r, B, g - X_{\lambda})$ color excess value (assuming universal $R_V$ values). The root-mean-square of the difference between the two measurements is given in each sub-panel. \label{fig:checkEBMV}}
\end{figure*}

\clearpage
\begin{figure*}[h]
\centering
\includegraphics[width=12cm]{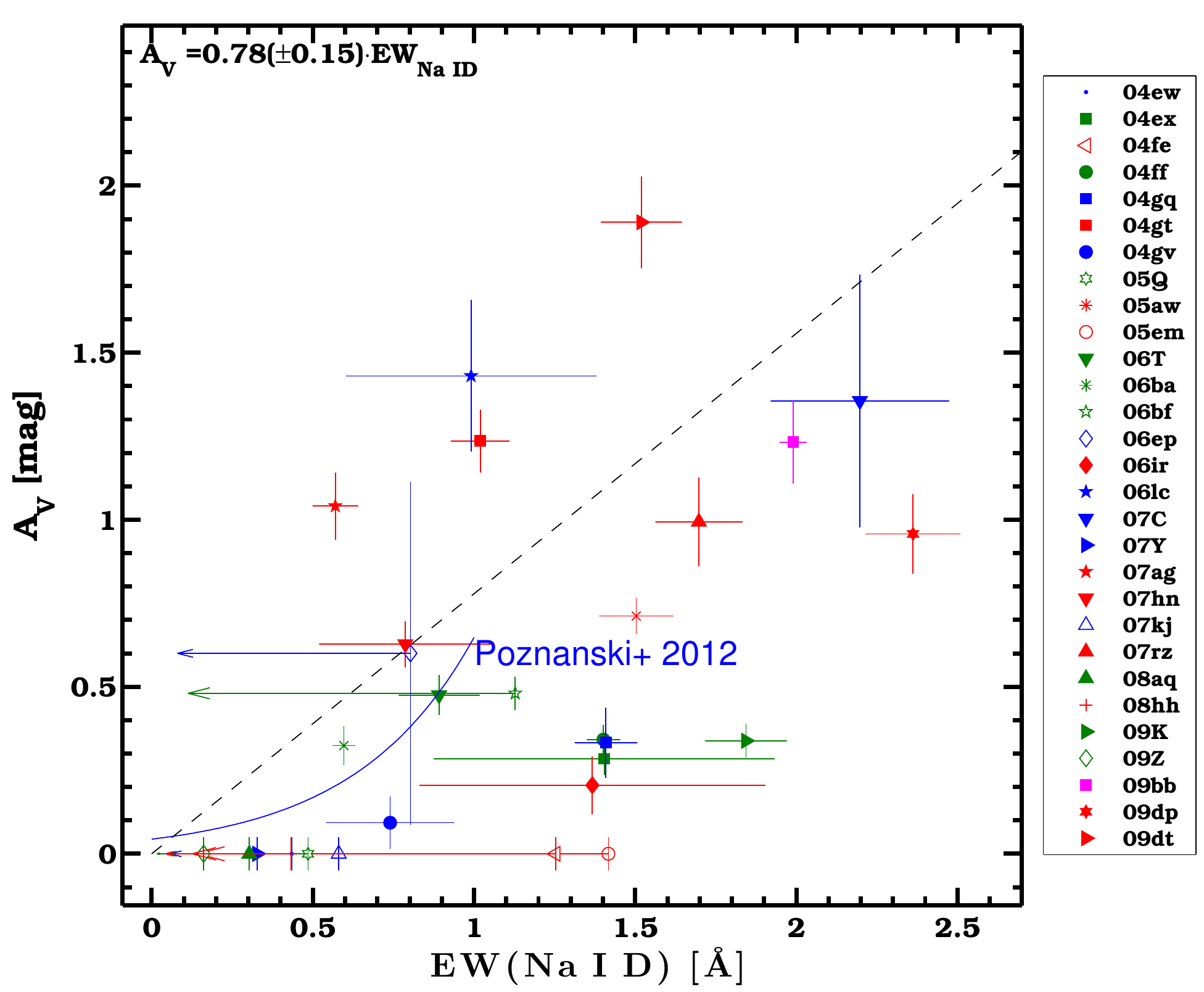}
\caption[]{ $A_V^{host}$ from $E(V-X)$ as given in Tables~\ref{tab:veryred} and \ref{tab:EBMV} versus $EW_{\ion{Na}{i}~D}$.  The  \citet{poznanski12} relation is over-plotted as a solid blue line. Empty symbols indicate upper limits on $EW_{\ion{Na}{i}~D}$. A linear fit between the two quantities (with intercept fixed to zero) is drawn as a black dashed line and the corresponding fit expression is reported  in the top-left corner of the figure. This expression has been obtained by linearly fitting all the points including the limits.}
\label{fig:NAID_EBMV}
\end{figure*}

\clearpage
\begin{figure*}
\centering
\includegraphics[width=9cm]{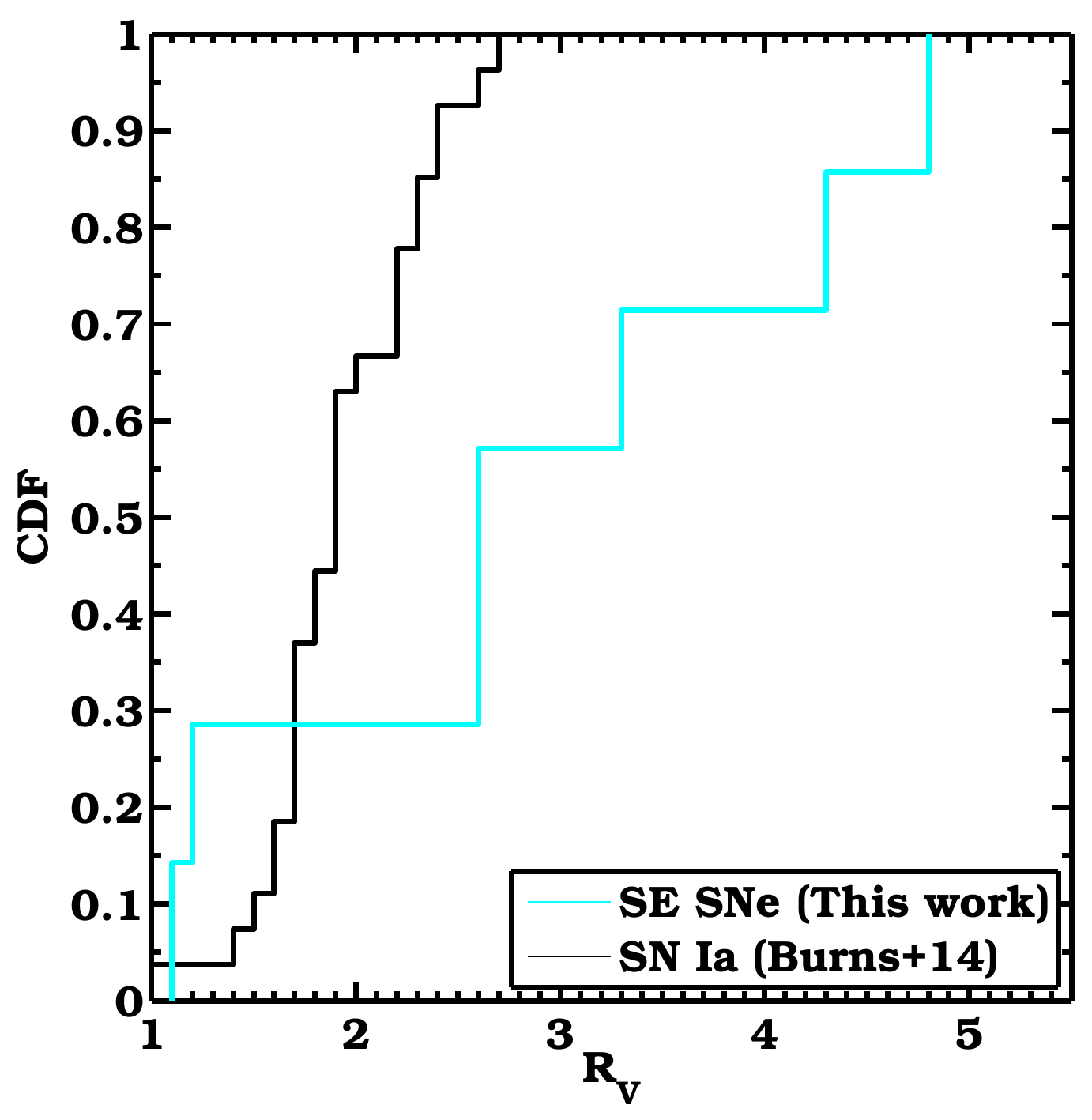}
\caption{\label{RV_cdf}$R_V$ distribution for our most reddened events as compared to those of SNe~Ia from \citet{burns14}, where we have selected objects   characterized by  $E(B-V)_{host} \geq 0.2$ mag and $R_V$ obtained from fitting a F99 reddening law. On average, SNe~Ia  prefer smaller values of $R_V$.}
\end{figure*}

\clearpage

\begin{appendix}

\section{Average effective filter wavelengths}
\label{appendixA}

Plotted in the various panels of Fig.~\ref{effectivewavelength} is the effective wavelength vs. days relative to $V$-band maximum for each of the CSP-I passbands. To compute the black curve the effective wavelength of each filter is computed using the Nugent spectral templates \citep[cf.,][]{levan05}. 
The average of these values is indicated with a dashed horizontal line and this value is adopted as the effect wavelength for each passband. 
\begin{figure*}
\centering
\includegraphics[width=16cm]{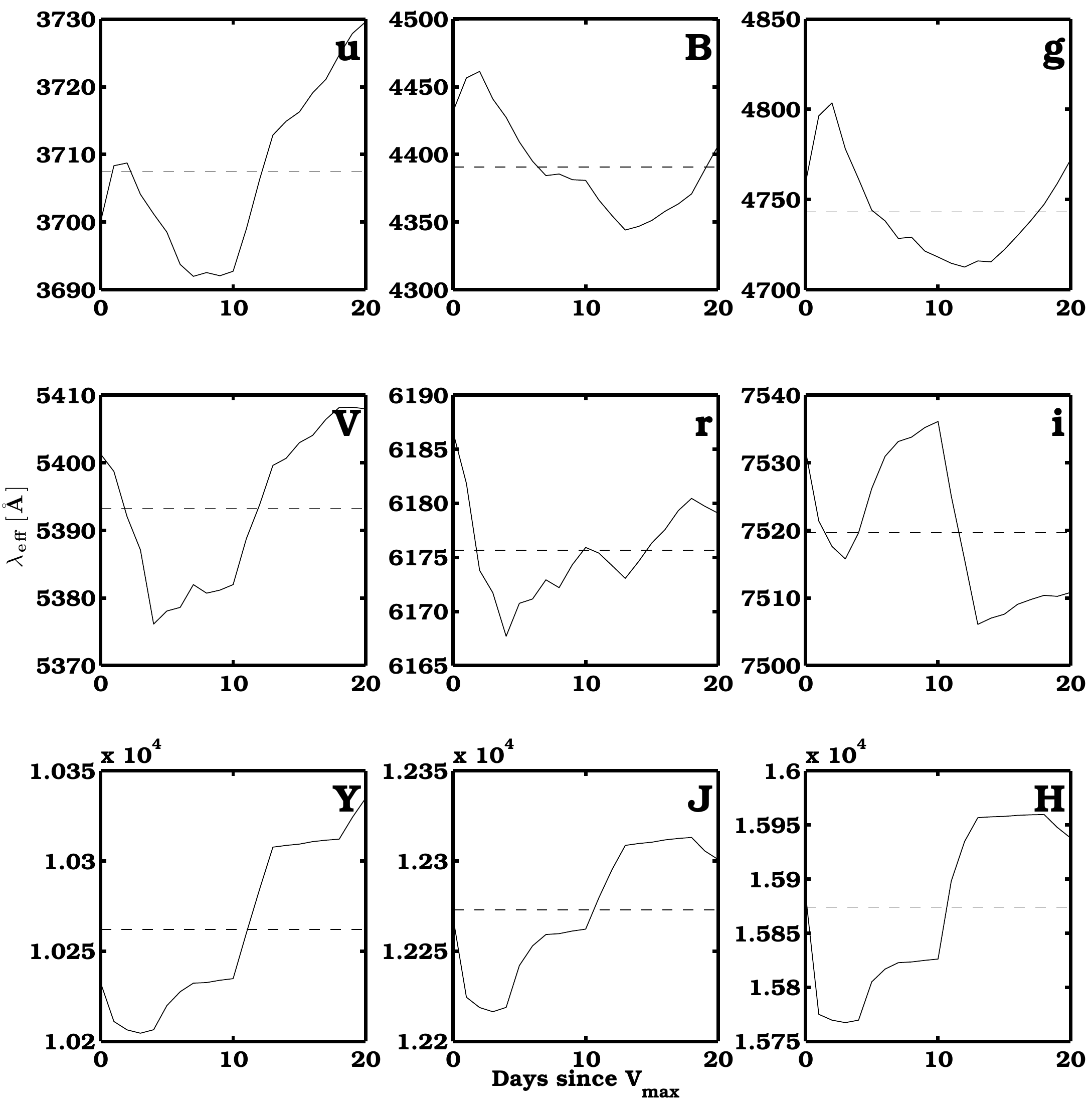}
\caption{Temporal evolution of each passband's effective wavelength, which is driven by the temporal evolution of the SED. Effective wavelengths were measured each CSP-I passbands using the Nugent SN~Ib/c spectral templates. 
\label{effectivewavelength}}
\end{figure*}

\section{Deriving $R_{V}^{host}$ via color excess measurements}
\label{appendixB}

 Here we describe  the $a_x$ and $b_x$ reddening law coefficients are computed for  the $X=ugriBVYJH$ bands. It is these coefficients that  enable us  to  estimate  $R_{V}^{host}$ based on the slope(s) obtained by plotting  $E(V-X)_{+10d}$ vs. $E(B-V)_{+10d}$. 
 Our approach follows directly the method employed by \citet{folatelli10} to determine the   $a_x$ and $b_x$ coefficients  applicable to dust studies related to  Type~Ia supernovae.
 In  our case however we use template spectra computed from the CSP-I SE SN spectroscopy sample \citep{holmbo17}. 

Following \citet{cardelli89}, we first parameterized the reddening law as a function of wavelength in the following form:

\begin{eqnarray}
\frac{A_{\lambda}}{A_{V}} = a_{\lambda} + \frac{b_{\lambda}}{R_V}.
   \label{eq:coefficients}
\end{eqnarray}

\noindent The optical coefficients  are computed by first multiplying a range of $A_{V}^{host}$ values with  $+$10d and $+$20d spectral templates, while for the NIR coefficients we resorted to direct extrapolation from $\sim$ 1.0 to 2.5 microns.
Synthetic photometry is then computed from the  raw spectral templates and the modified spectral templates using the CSP-I system response functions. 
Taking the difference between the resulting synthetic magnitudes then provides values for the 
  observed absorption $A_X$ and $A_V$, corresponding to bands $X$ and $V$. 
  Finally,  estimates of the $a_{\lambda}$ and $b_{\lambda}$ coefficients are obtained for all the bands by fitting linear relations between $A_X$ and $R_V$, as parameterized in 
 Equation~(\ref{eq:coefficients}).
 
 These calculation were performed using the \citet{fitzpatrick99} reddening law and the results are summarized in Table~\ref{tab:appendix}. 
 Inspection of the various coefficients reveals quite consistent results for each passband among each of the SE SN subtypes.  
  As explained in \citeauthor{folatelli10} these coefficients allow us to convert an {\em observed}  color excess measurements, as presented in Sect.~\ref{sec:colorexcesses}, to a {\em true} value of  $R_{V}^{host}$ as defined in Equation~(\ref{eq:coefficients}).
  
\clearpage

\begin{deluxetable}{ccc}
\tablewidth{0pt}
\tablecaption{Reddening-law coefficients for SE SNe at $+$10d.\label{tab:appendix}}
\tablehead{
\colhead{Filter}&
\colhead{$a_{\lambda}$}&
\colhead{$b_{\lambda}$}}
\startdata
\hline
\multicolumn{3}{c}{Type IIb}\\
\hline
$u$  &   0.938  &     1.737  \\   
$B$  &   1.001  &     0.859 \\  
$g$  &   1.004  &     0.532  \\ 
$r$  &   0.970  &    $-$0.467  \\ 
$i$  &   0.781  &    $-$0.533  \\ 
$Y$  &   0.401  &    $-$0.142 \\ 
$J$  &   0.259  &    $-$0.001 \\
$H$  &   0.157  &     0.039 \\
\hline
\multicolumn{3}{c}{Type Ib}\\
\hline
$u$  &   0.944  &     1.689  \\
$B$  &   1.002  &     0.887  \\
$g$  &   1.002  &     0.556  \\
$r$  &   0.975  &    $-$0.455  \\
$i$  &   0.786  &    $-$0.535  \\
$Y$  &   0.402  &    $-$0.139  \\
$J$  &   0.259  &     0.002  \\
$H$  &   0.157  &     0.042  \\
\hline
\multicolumn{3}{c}{Type Ic}\\
\hline
$u$  &    0.929 &      1.789  \\
$B$  &    1.000 &      0.879  \\ 
$g$  &    1.001 &      0.519  \\ 
$r$  &    0.974 &     $-$0.458  \\  
$i$  &    0.784 &     $-$0.532  \\  
$Y$  &    0.402 &     $-$0.139  \\  
$J$  &    0.259 &      0.003  \\
$H$  &    0.156 &      0.042   \\  
\enddata                                                        
\end{deluxetable}

\end{appendix}

\end{document}